\documentclass{aa}  

\usepackage{graphicx}
\usepackage{txfonts}
\usepackage{siunitx}
\usepackage[colorlinks=true,linkcolor=blue,citecolor=blue,urlcolor=blue]{hyperref}
\newcommand{\new}[1]{#1}

\begin{document} 

\title{\textsc{AMICO} galaxy clusters in KiDS-1000: cosmological sample}
\author{
  M. Maturi \inst{1,2}\thanks{\email{maturi@uni-heidelberg.de}}
  \and
  M. Radovich \inst{3}
  \and
  L. Moscardini \inst{4,5,6}
  \and
  G. F. Lesci \inst{4,5}
  \and
  G. Castignani \inst{5}
  \and
  F. Marulli \inst{4,5,6}
  \and
  E. A. Puddu \inst{7}
  \and
  M. Romanello \inst{4,5}
  \and
  M. Sereno \inst{5,6}
  \and
  C. Giocoli \inst{4,5,6}
  \and
  L. Ingoglia \inst{8}
  \and
  S. Bardelli \inst{4}
  \and
  B. Giblin \inst{9}
  \and
  H. Hildebrandt \inst{10}
  \and
  S. Joudaki \inst{11,12}
}
\institute{
  Zentrum f\"ur Astronomie, Universitat\"at Heidelberg, Philosophenweg 12, D-69120 Heidelberg, Germany 
  \and
  Institute for Theoretical Physics, Philosophenweg 16, D-69120 Heidelberg, Germany 
  \and
  INAF - Osservatorio Astronomico di Padova, via dell'Osservatorio 5, I-35122, Padova, Italy 
  \and
  Dipartimento di Fisica e Astronomia ``Augusto Righi'' - Alma Mater Studiorum
  Universit\`{a} di Bologna, via Piero Gobetti 93/2, I-40129 Bologna, Italy 
  \and
  INAF - Osservatorio di Astrofisica e Scienza dello Spazio di Bologna, via Gobetti 93/3, I-40129, Bologna, Italy  
  \and
  INFN - Sezione di Bologna, viale Berti Pichat 6/2, I-40127 Bologna, Italy 
  \and
  INAF - Osservatorio Astronomico di Capodimonte, Salita Moiariello 16, Napoli 80131, Italy 
  \and
  INAF, Istituto di Radioastronomia, Via Piero Gobetti 101, 40129 Bologna, Italy 
  \and
  Institute for Astronomy, University of Edinburgh, Blackford Hill, Edinburgh, EH9 3HJ, UK 
  \and
  Ruhr University Bochum, Faculty of Physics and Astronomy, Astronomical Institute (AIRUB), German Centre for Cosmological Lensing, 44780 Bochum, Germany 
  \and
  Centro de Investigaciones Energéticas, Medioambientales y Tecnológicas (CIEMAT), Av. Complutense 40, E-28040 Madrid, Spain 
  \and
  Institute of Cosmology \& Gravitation, Dennis Sciama Building, University of Portsmouth, Portsmouth, PO1 3FX, United Kingdom 
}
\date{}
   
  \abstract
  {Galaxy clusters provide key insights into cosmic structure formation, galaxy formation and are essential for cosmological studies.}
  {We present a catalog of galaxy clusters detected in the Kilo-Degree Survey (KiDS-DR4) optimized for cosmological analyses and investigations of cluster properties. Each detection includes probabilistic membership assignments for the KiDS-DR4 galaxies within the magnitude range $15<r'<24$}
  {Using the Adaptive Matched Identifier of Clustered Objects (\textsc{AMICO}) algorithm, we identified $\num{23965}$ clusters over an effective area of about $839$ deg$^2$ in the redshift range $0.1 \le z \le 0.9$, with a signal-to-noise ratio $S/N>3.5$. The sample is highly homogeneous across the entire survey thanks to the restrictive galaxy selection criteria we adopted. Spectroscopic data from the GAMA survey were used to calibrate the clusters photometric redshift and assess their uncertainties. We introduced algorithmic enhancements to \textsc{AMICO} to mitigate border effects among neighbor tiles. Quality flags are also provided for each cluster detection. The sample purity and completeness assessments have been estimated using the \textsc{SinFoniA} data driven approach, thus avoiding strong assumptions embedded in numerical simulations. We introduced a blinding scheme of the selection function meant to support the cosmological analyses.}
  {Our cluster sample includes $321$ cross-matches with the X-ray eRASS1 "primary" sample and $235$ matches with the ACT-DR5 cluster sample. We derived a mass-proxy scaling relation based on intrinsic richness, $\lambda_*$, using masses from the eRASS1 catalog.}
  {The KiDS-DR4 cluster catalog provides a valuable data set for investigating galaxy cluster properties and contributes to cosmological studies by offering a large, well-characterized cluster sample.}
  \keywords{galaxies -- galaxy evolution -- clusters of galaxies -- cosmic structure formation -- cosmology}
  \maketitle

%
%
\section{Introduction}

Galaxy clusters serve as powerful laboratories for acquiring new insights into the physics of galaxy formation, feedback mechanisms, distribution of dark matter, properties of dark energy, and cosmology in general. Such power derives from the fact that they represent the culminating phase of cosmic structure formation in the highly nonlinear regime. \new{Over the past decade, a large effort has been dedicated to assembling galaxy cluster catalogs optimized for cosmological analyses. In this effort, the redMaPPer algorithm has been successfully applied to SDSS and DES data sets, providing extensive optically selected cluster samples that have been used to derive cosmological constraints from abundance and clustering statistics \citep{Rykoff14,rykoff16,Costanzi13,costanzi21}. In parallel, Sunyaev-Zel'dovich (SZ) and X-ray surveys, such as Planck, ACT, SPT, and more recently eROSITA's eRASS1, have yielded independent constraints on cosmological parameters as well \citep{bleem15,planckSZ16,bocquet19,abbott20,garrel22,bulbul24,ghirardini24,seppi24,artis25}. These multi-wavelength efforts highlight the efficiency of galaxy clusters as cosmological probes.}

\begin{figure}
  \includegraphics[width=0.48\textwidth]{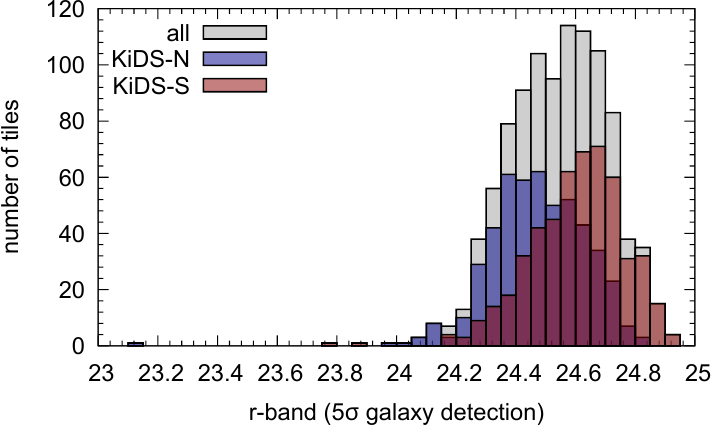}
  \includegraphics[width=0.48\textwidth]{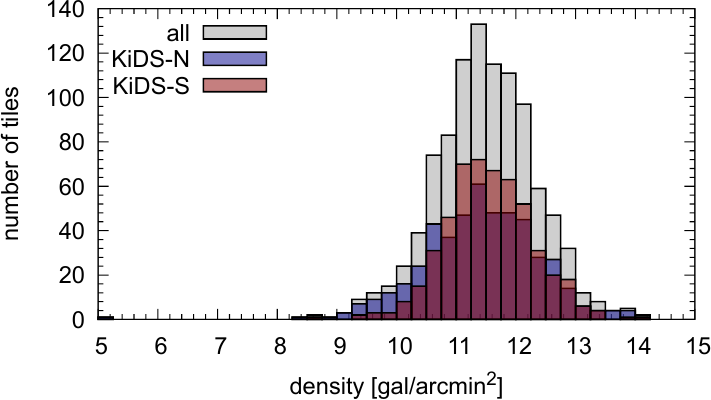}
  \includegraphics[width=0.495\textwidth]{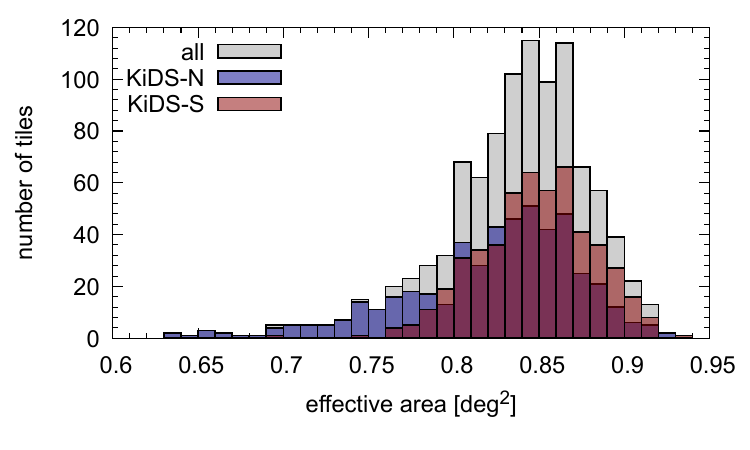}
  \caption{Key statistical properties for each individual tile of the KiDS-DR4 data set. Gray, blue and red histograms refer to the entire data set, to the KiDS-N and KiDS-S stripes, respectively. The distribution of the survey depth in the $r$-band, the galaxy number density per arcmin$^{2}$, and the effective area of each tile are shown in top, central and bottom panels, respectively.}
  \label{fig:galStat}
\end{figure}

\begin{figure}
  \centering
  \includegraphics[width=0.48\textwidth]{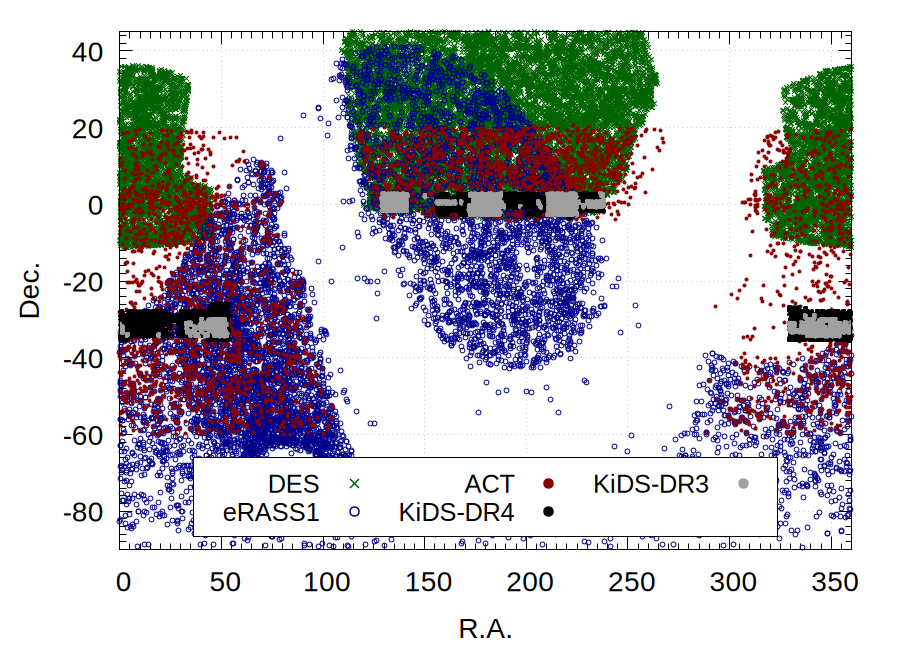}
  \caption{\new{Footprint of KiDS-DR4 data (black area) overlapped by KiDS-DR3 data (gray area), showing the increase in area coverage. For comparison, the RedMaPPer-DES (green crosses), eRASS1 (blue circles) and ACR-DR5 (red points) cluster samples are also shown.}}
    \label{fig:footprint}
\end{figure}
  
\new{In this context, the KiDS-DR4 cluster sample presented in this work contributes a valuable and independent data set, leveraging deep, high-quality optical and near-infrared data with well-calibrated photometric redshifts across a wide area. For the clusters detections we used} algorithms to detect galaxy clusters in optical surveys, the Adaptive Matched Identifier of Clustered Objects (\textsc{AMICO}) algorithm \citep{2005A&A...442..851M,Bellagamba18,maturi19} has proven to be particularly effective. It enabled the creation of comprehensive galaxy cluster samples based on KiDS-DR3, miniJPAS, and COSMOS data sets \citep{maturi19, maturi23, toni24} and it has been selected to be  implemented in the scientific pipeline of the ESA mission Euclid after a dedicated challenge \citep{adam19}. The AMICO-KiDS-DR3 catalog, for instance, has been the basis for numerous studies, including investigations of the properties of Brightest Cluster Galaxies  \citep[BCGs,][]{radovich20,Castignani2022,Castignani2023}, the redshift evolution of the luminosity function of cluster galaxies \citep{puddu21}, the weak lensing mass-richness scaling relations \citep{bellagamba18b, sereno20, lesci22, lesci22b}, luminosity scaling relations \citep{smit22}, the splashback radius measurements \citep{giocoli24}, the large-scale stacked weak lensing profiles \citep{giocoli21}, the halo bias \citep{ingoglia22}, and the clustering of galaxy clusters \citep{lesci22b, romanello24}. 

In this paper, we present and characterize the galaxy cluster sample and probabilistic cluster memberships produced with \textsc{AMICO} from the fourth data release of the Kilo-Degree Survey \citep[hereafter KiDS-DR4,][]{Kuijken2019}. Compared to KiDS-DR3, the KiDS-DR4 data set offers high-quality optical imaging across a wider area (1000 vs. 400 deg$^2$), complemented by near-infrared coverage from the VIKING survey \citep{edge13}. For brevity, we refer to both data sets collectively as KiDS-DR4. The survey's homogeneity, depth ($r<25$), high-quality lensing measurements (\new{with mean seeing in r band of $0.''70$ acquired exclusively during dark time}), and broad spectral coverage ($ugriZYJHKs$) ensure optimal photometric redshifts, making it ideal for studies in cosmology, galaxy evolution, and galaxy cluster science.

In the present analysis, we have introduced several enhancements to the \textsc{AMICO} algorithm and extended our methodology to evaluate the sample purity and completeness using the \new{Selection Function Extractor (\textsc{SinFoniA}) as described in \cite{maturi19}}. Since one of our main scientific drivers is to use the cluster sample to constrain cosmological parameters, we also implemented and applied a blinding strategy to the selection function of the sample. \new{The unblinding was performed following the completion of the cosmological analysis based on cluster counts presented in \cite{lesci25}}.

The GAMA spectroscopic survey, \new{augmented with higher-redshift spectroscopic data from \citet{vandenbusch22},} has been used to calibrate the cluster photometric redshifts and to assign spectroscopic redshifts whenever available. Additionally, we explore the cross-matching of our cluster sample with the Dark Energy Survey catalog created with RedMaPPer \citep{Rykoff14}, the X-ray eRASS1 \citep{bulbul24} and Sunyaev-Zel'dovich (SZ) ACT-DR5 \citep{hilton20} cluster catalogs, providing mass-proxy scaling relations and investigating the most significant mismatches, with a specific focus on X-ray detections lacking an optical counterpart and vice versa.

The paper is organized as follows. Sect.~2 describes the KiDS-DR4 data set. Sect.~3 details the improvements made to the \textsc{AMICO} algorithm and presents the galaxy cluster sample. In Sect.~4, we discuss the mock galaxy catalogs generated with \textsc{SinFoniA}, which are used to estimate the sample purity and completeness, and to evaluate the uncertainties of the associated observables. 
The cross-matching of our cluster sample with X-ray and SZ data is presented in Sect.~5. Finally, Sect.~6 draws our conclusions. In Appendix~(\ref{sec:blinding}) we present the blinding strategy adopted for the cosmological analyses. \new{In this work, we assume a flat $\Lambda$CDM cosmological model with parameters $\Omega_m = 0.3$, $\Omega_\Lambda = 0.7$, and $h = 0.7$. Unless otherwise stated, all uncertainties are quoted at the 1$\sigma$ level, and all magnitudes are given in the AB system.}

%
%
\section{The KiDS-DR4 data set}\label{sec:data-kids}

\begin{figure}
  \includegraphics[width=0.49\textwidth]{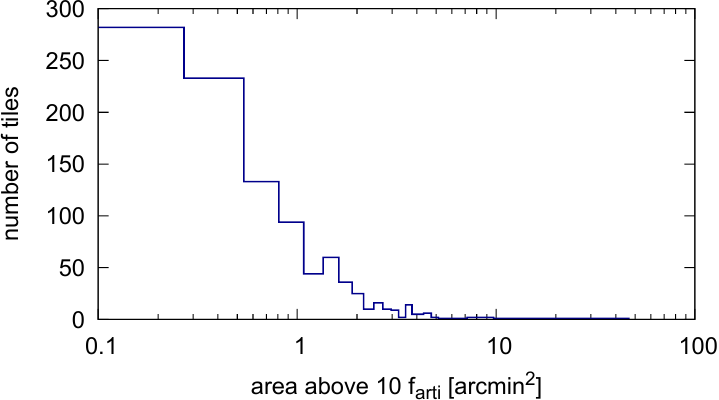}
  \caption{Distribution of the area of each tile affected by unmasked image artifacts mainly caused by spurious reflections and ghosts from bright stars located outside the field of view.}
  \label{fig:diagnostic1}
\end{figure}

\begin{figure}
  \centering
  \includegraphics[width=0.49\textwidth]{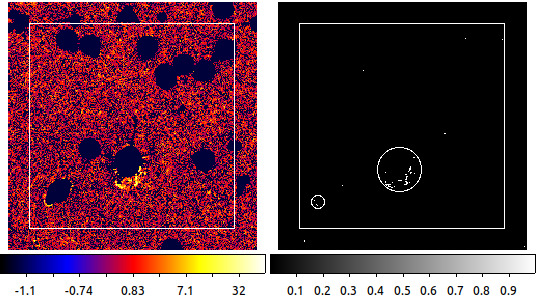}
  \includegraphics[width=0.49\textwidth]{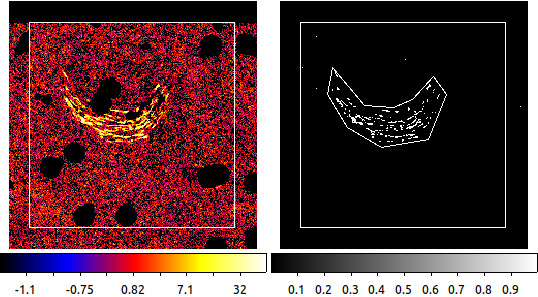}
  SNR density r.m.s.
  \caption{Fluctuations of the  spatial number density of galaxy catalog entries relative to its background r.m.s, $f_{arti}$, for the \detokenize{KiDS_DR4.0_185.0_-0.5} (top left) and \detokenize{KiDS_DR4.0_218.6_2.5} (bottom left) tiles, which are the best and worst tiles listed in Table~\ref{tab:arti_list}, respectively. The values are shown using a logarithmic scale, shown below the panels. Pixels with $f_{arti} > 10$ are used to create additional masks to identify image artifacts, as displayed in the corresponding right panels. The displayed areas are $1.4$ deg on each side. The white square box enclosing most of the area marks the boundary of the survey tile, while the other white contours indicate examples of the handmade polygons enclosing the largest artifacts.}
  \label{fig:arti_worst}
\end{figure}

A detailed description of the KiDS-ESO-DR4 release (DR4 hereafter) was given in \citet{Kuijken2019}. In summary, it consists of a northern patch (KiDS-N) and a southern patch (KiDS-S), covering a 1000 deg$^2$ area in total. $ugri$ photometry was obtained with the OmegaCAM camera attached on the ESO VLT Survey Telescope (VST), on tiles covering approximately 1 deg$^2$ each \new{\citep{Kuijken11}}. In DR4, mean limiting magnitudes (5$\sigma$ in a 2\arcsec aperture) in the $ugri$ bands are 24.23 $\pm$ 0.12, 25.12 $\pm$ 0.14, 25.02 $\pm$ 0.13, 23.68 $\pm$ 0.27, respectively. In addition to the optical bands, DR4 also includes near-IR ($ZYJHKs$) photometry observed with the VIKING survey \citep{edge13} on the ESO VISTA telescope. In DR4, a reprocessing of the VIKING images was done \citep{wright19}, to ensure that photometry is consistent across the nine bands. Photometric redshifts were computed using the Bayesian template fitting code BPZ \citep{benitez00}, which derives best-fitting SEDs and the photometric redshift posterior probability distribution (PDF), based on a prior redshift probability distribution. In DR4, the prior described in \citet{Kuijken2019} was chosen, since it was found to best reduce uncertainties on photometric redshifts and catastrophic failures for faint, high-redshift galaxies: the reason for that was that the photometric catalog needed to be optimized for weak lensing analyses, which is the main science driver of the KiDS survey. However, it was known that this prior was not optimal for bright, low-redshift galaxies, which are instead fundamental for our cluster search. Therefore, for this work, we decided to use instead the prior adopted in KiDS-DR3 \citep{deJong15}  as already done for the \textsc{AMICO}-KiDS-DR3 cluster sample \citep{maturi19}.

KiDS is a very homogeneous survey in terms of exposure time. Despite this, the $r$-band $5\sigma$ detection limit of galaxies shows some scatter, as displayed in the top panel of Fig.~\ref{fig:galStat} where gray, blue and red represent the entire data set, the KiDS-N and KiDS-S stripes, respectively. Therefore, to ensure nearly perfect homogeneity and optimize the final cluster sample for cosmological analyses, we select only galaxies with total $r$-band magnitude brighter than the depth limit of the worst tile (i.e. $r<24$) as in the previous analysis by \cite{maturi19}. We also exclude the outlier tile KiDS\_DR4.1\_196.0\_1.5 from the analysis, as it has an $r$-band $5\sigma$ galaxy depth of $23.1$ only. After the application of this magnitude cut, the number density of galaxies in the KiDS-N and KiDS-S stripes peaks at the same value of the number density, as shown in the central panel of Fig.~\ref{fig:galStat}. \new{In Fig.~\ref{fig:footprint}, we show the footprints of the KiDS-DR3 and KiDS-DR4 data sets, compared to the RedMaPPer-DES, eRASS1 and ACR-DR5 cluster samples that will be discussed in Sect.~\ref{sec:match}.}

To avoid regions affected by cosmetic artifacts, we drop galaxies with a MASK flag in the KiDS-DR4 photometric catalog corresponding to manually masked regions (MASK=3), halos around bright stars in the $r$-band (MASK=12) or outside the area covered by all 9 bands (MASK=14). These restrictive criteria are intended to deliver a final sample of clusters that is as homogeneous and pure as possible across the entire survey. The effective area of the survey is $838.783$ deg$^2$. The bottom panel of Fig.~\ref{fig:galStat} shows the distribution of the effective area in deg$^2$ of each individual survey tile. 

\begin{table}
  \centering
  \caption {List of the tiles where the cosmetic artifacts cover an area larger than $5$ arcmin$^2$. Tile ID identifies the tile name, whereas $A_{arti}$ is the area covered by artifacts in arcmin$^2$.}
  \begin{tabular}{lr}
  \hline
  tile ID & $A_{arti}$\\
  \hline
\texttt{\detokenize{KiDS_DR4.0_185.0_-0.5}} & 5.31\\
\texttt{\detokenize{KiDS_DR4.0_343.7_-31.2}} & 5.49\\
\texttt{\detokenize{KiDS_DR4.0_230.0_-0.5}} & 5.85\\
\texttt{\detokenize{KiDS_DR4.0_138.4_2.5}} & 6.03\\
\texttt{\detokenize{KiDS_DR4.0_178.0_1.5}} & 6.66\\
\texttt{\detokenize{KiDS_DR4.0_351.8_-32.1}} & 7.65\\
\texttt{\detokenize{KiDS_DR4.0_211.6_2.5}} & 7.74\\
\texttt{\detokenize{KiDS_DR4.0_215.0_1.5}} & 9.18\\
\texttt{\detokenize{KiDS_DR4.0_343.9_-30.2}} & 9.36\\
\texttt{\detokenize{KiDS_DR4.0_214.6_2.5}} & 10.17\\
\texttt{\detokenize{KiDS_DR4.0_345.2_-29.2}} & 10.26\\
\texttt{\detokenize{KiDS_DR4.0_213.0_1.5}} & 17.01\\
\texttt{\detokenize{KiDS_DR4.0_212.0_1.5}} & 17.73\\
\texttt{\detokenize{KiDS_DR4.0_344.1_-29.2}} & 17.82\\
\texttt{\detokenize{KiDS_DR4.0_212.6_2.5}} & 20.52\\
\texttt{\detokenize{KiDS_DR4.0_216.0_1.5}} & 22.23\\
\texttt{\detokenize{KiDS_DR4.0_215.6_2.5}} & 23.04\\
\texttt{\detokenize{KiDS_DR4.0_219.6_2.5}} & 23.76\\
\texttt{\detokenize{KiDS_DR4.0_218.6_2.5}} & 38.97\\
  \hline
  \end{tabular}
  \label{tab:arti_list}
\end{table}

Additionally, we identified a few regions affected by reflections within the optical system, caused by bright stars outside the field of view or by other cosmetic defects not fully captured by the masks mentioned above. We identified these regions as significant spatial 2-dimensional overdensities of entries in the KiDS-DR4 galaxy catalog, $\rho$, with density values exceeding ten times the r.m.s of the background density fluctuations, i.e. $f_{arti}=\rho/\sigma_{bkg}>10$. To be sure that we are not rejecting true galaxy overdensities, we adopted relatively small pixels in the evaluation of the density maps, $l_{\text{pix}}=0.3$ arcmin, we considered overdensities produced by the entire line-of-sight (thus discarding redshift clustering), and used a large threshold to exclude these areas.

In Fig.~\ref{fig:diagnostic1}, we show the distribution of the number of survey tiles as a function of the area affected by these artifacts. As visible in the figure, there is a flat tail extending beyond 5 arcmin$^2$ containing 19 tiles that we list in Table~\ref{tab:arti_list}. As examples, the left panels of Fig.~\ref{fig:arti_worst} show the aforementioned fluctuations, $f_{arti}$,  of the best and worst tiles listed in Table~\ref{tab:arti_list} (top and bottom panels, respectively), while in the right panels we show the corresponding masked pixels exceeding the $f_{arti}>10$ threshold. Note that the intensity and values of the color maps are in logarithmic scale.

Since the area of the most affected tiles is small (at most $1\%$ of the total area), we do not exclude these tiles from our analysis. Instead, we discard the galaxies that fall within these regions when applying AMICO to identify clusters. Additionally, we assign to each cluster detection a flag, \texttt{ARTIFACTS\_FLAG}, to facilitate a potential subsequent rejection. This flag takes the value 0 if the detection occurs in a tile with no artifacts or with a total affected area smaller than 5 arcmin$^2$, the value 1 for detections in tiles with a total affected area larger than 5 arcmin$^2$, and the value 2 for those located in the manually created masks enclosing the largest contiguous compromised regions.

The latest and final KiDS data release (DR5) has recently been published \citep{wright24}. This release offers expanded coverage, twice the depth in the $i$-band, and improved photometric redshifts. \new{Cluster detection with \textsc{AMICO} and the corresponding cosmological analysis will also be carried out using this new dataset.}

\section{The \textsc{AMICO} galaxy cluster sample}

\subsection{Basics of the detection algorithm}\label{sec:amico}

We detect galaxy clusters using the \textsc{AMICO} algorithm \citep{2005A&A...442..851M, Bellagamba18, maturi19}. We refer to these papers for a more complete description of the mathematical derivation, and of the cluster model implemented in this code. Here, we summarize the core concepts and discuss in detail the specific setup, the differences, and new features implemented for this specific study.

\textsc{AMICO} employs an unbiased linear optimal matched filter specifically designed to maximize the signal-to-noise ratio of cluster detections. This optimization assumes that the data $D$ can be modeled as $D(\vec{x}, z, \vec{m}) = A \tau(\vec{x}, z, \vec{m}) + N(z, \vec{m})$, where $\tau$ is a cluster template characterizing the observational properties of clusters, $N$ represents the noise from field galaxies, and $A$ is an amplitude factor that scales with the cluster's richness. The amplitude $A$ is derived using a linear estimator obtained through the constrained optimization procedure detailed in \cite{maturi19}. \new{To characterize the luminosity distribution of cluster members, as in the previous KiDS-DR3 data analysis, the cluster model used as a template, $\tau$, assumes a Schechter luminosity function $\Phi(m)$ \citep{schechter76}, with a faint-end slope $\alpha=-1.06$ from \cite{Zenteno16}, and $m_*$ in the r-band from \cite{Hennig17}.}

\begin{figure}
  \includegraphics[width=0.45\textwidth]{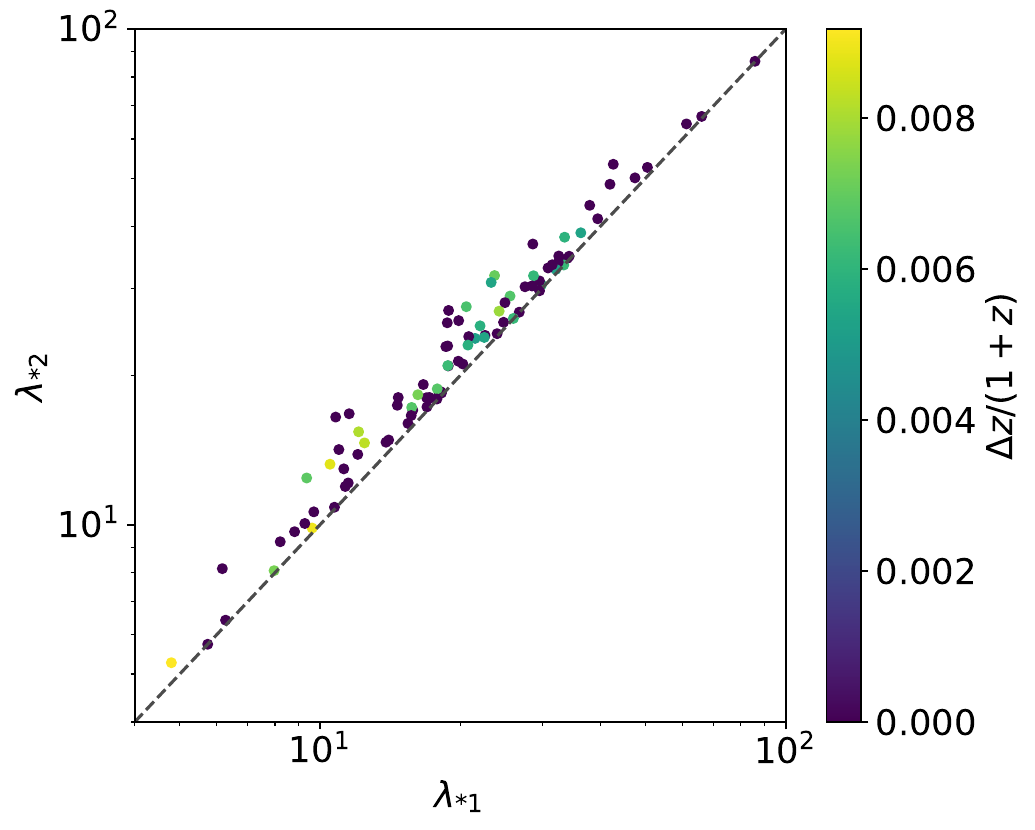}
  \caption{Duplicated detections occurring when merging overlapping survey tiles. Duplicates are identified by a maximum angular separation of $\Delta R <1.2$ arcmin and a redshift difference of $\Delta z = 0.015(1+z)$. The strong correlation between their intrinsic richness 
  ($\lambda_{*1}$ 
  and $\lambda_{*2}$) confirms that these detections originate from the same structure. The color scale represents their redshift separation.}
  \label{fig:repetitions}
\end{figure}

Additionally, it describes the projected density of cluster members assuming a Navarro-Frenk-White (NFW) radial profile \citep{1997ApJ...490..493N} scaled to match the radius $R_{200}=1$ Mpc\footnote{Here, $R_{200}$ is the radius within which the average density of a halo is 200 times the critical density of the universe.}.
The weight of each galaxy at a given redshift is given by its own photo-$z$ probabilistic redshift distribution, $P(z)$, as provided by BPZ. Beyond the list of cluster detections, the algorithm calculates the probability of each galaxy being a member of a detected cluster. This probability is then used to estimate the apparent richness $\lambda$, i.e. the number of observable members in a cluster, and the intrinsic richness $\lambda_*$, defined as the number of cluster members with an $r$-band magnitude brighter than $m_*+1.5$ within a radial distance $R_{200}$. Here, the radius $R_{200}$ and the characteristic magnitude in the Schechter luminosity function, $m_*$, are inherited from the model used to construct the optimal filter. Two different signal-to-noise ratios for the amplitude $A$, are returned: one accounts for the noise induced by field galaxies only (\texttt{SN\_NO\_CLUSTERS}), while the other additionally includes the shot noise due to the cluster members ($S/N$).

\begin{figure}
  \includegraphics[width=0.241\textwidth]{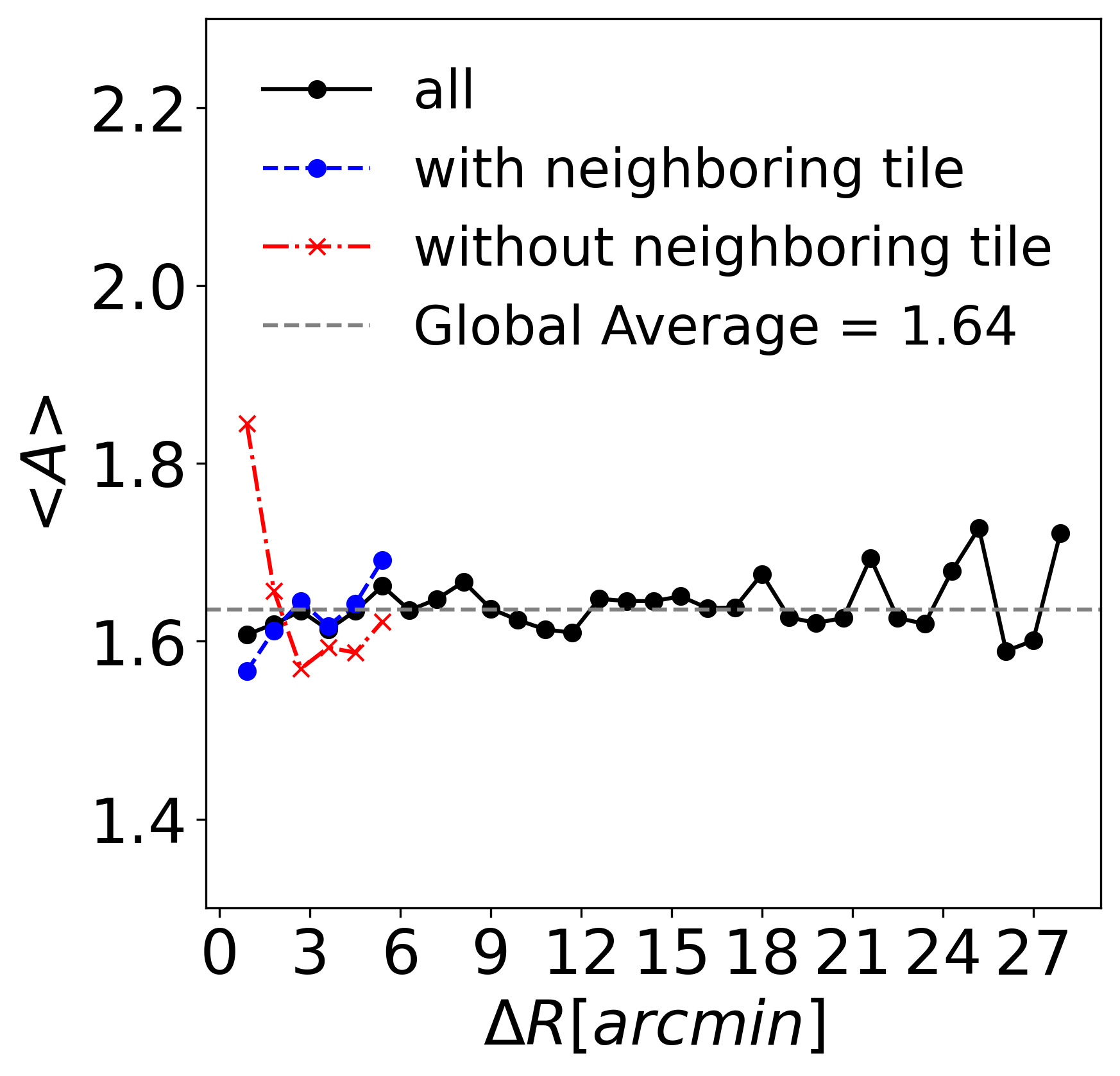}
  \includegraphics[width=0.235\textwidth]{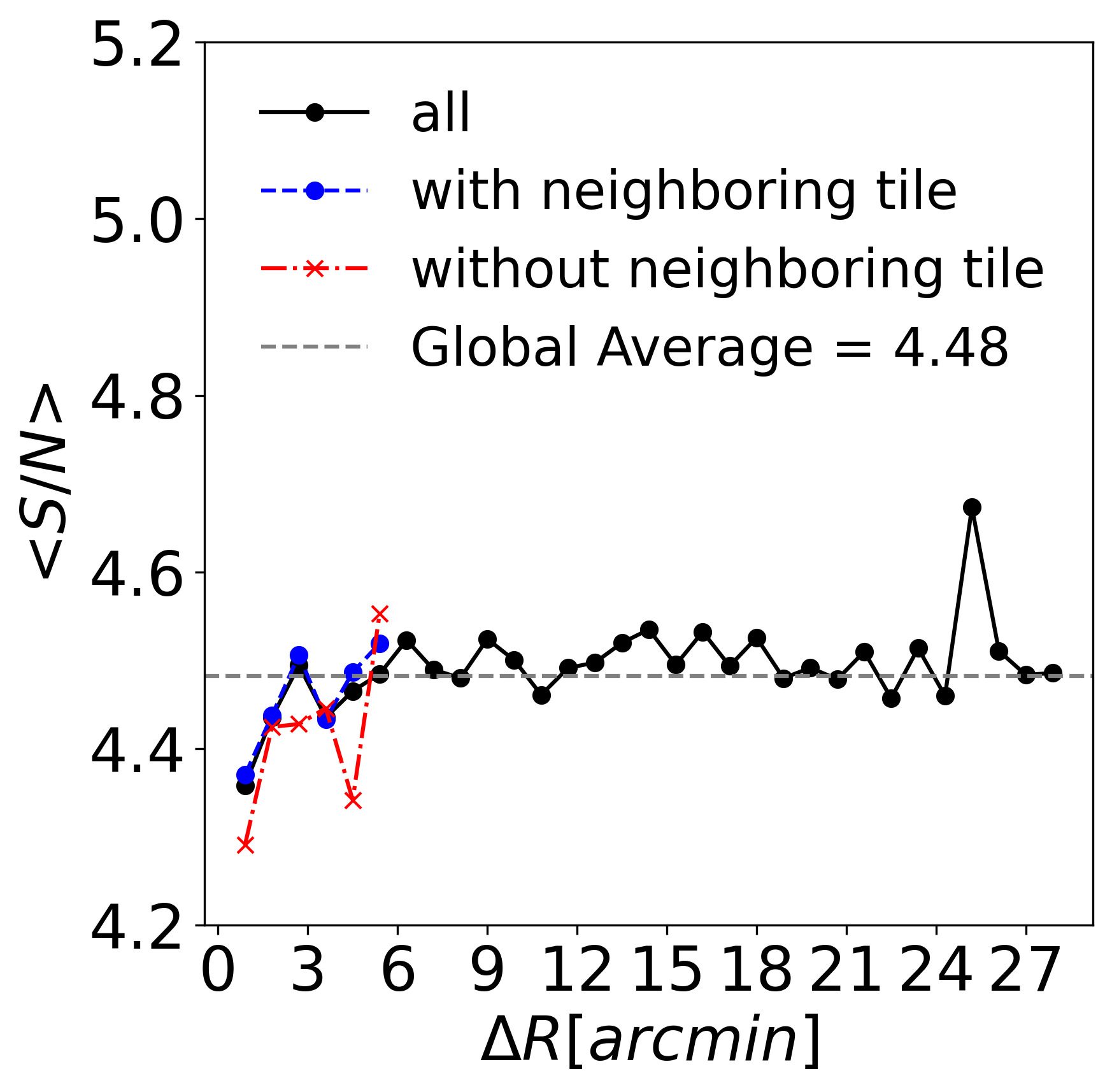}\\
  \includegraphics[width=0.241\textwidth]{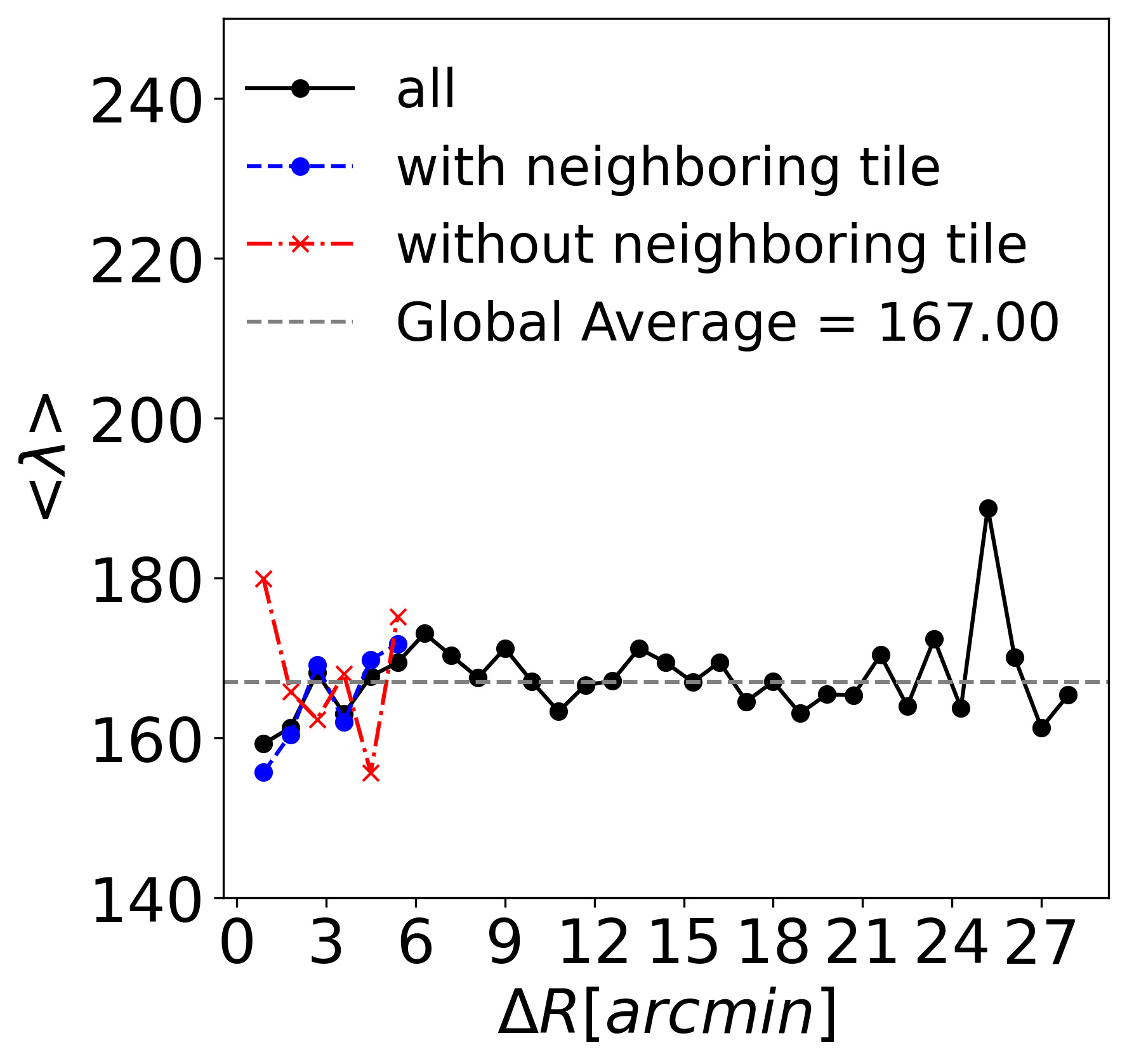}
  \includegraphics[width=0.235\textwidth]{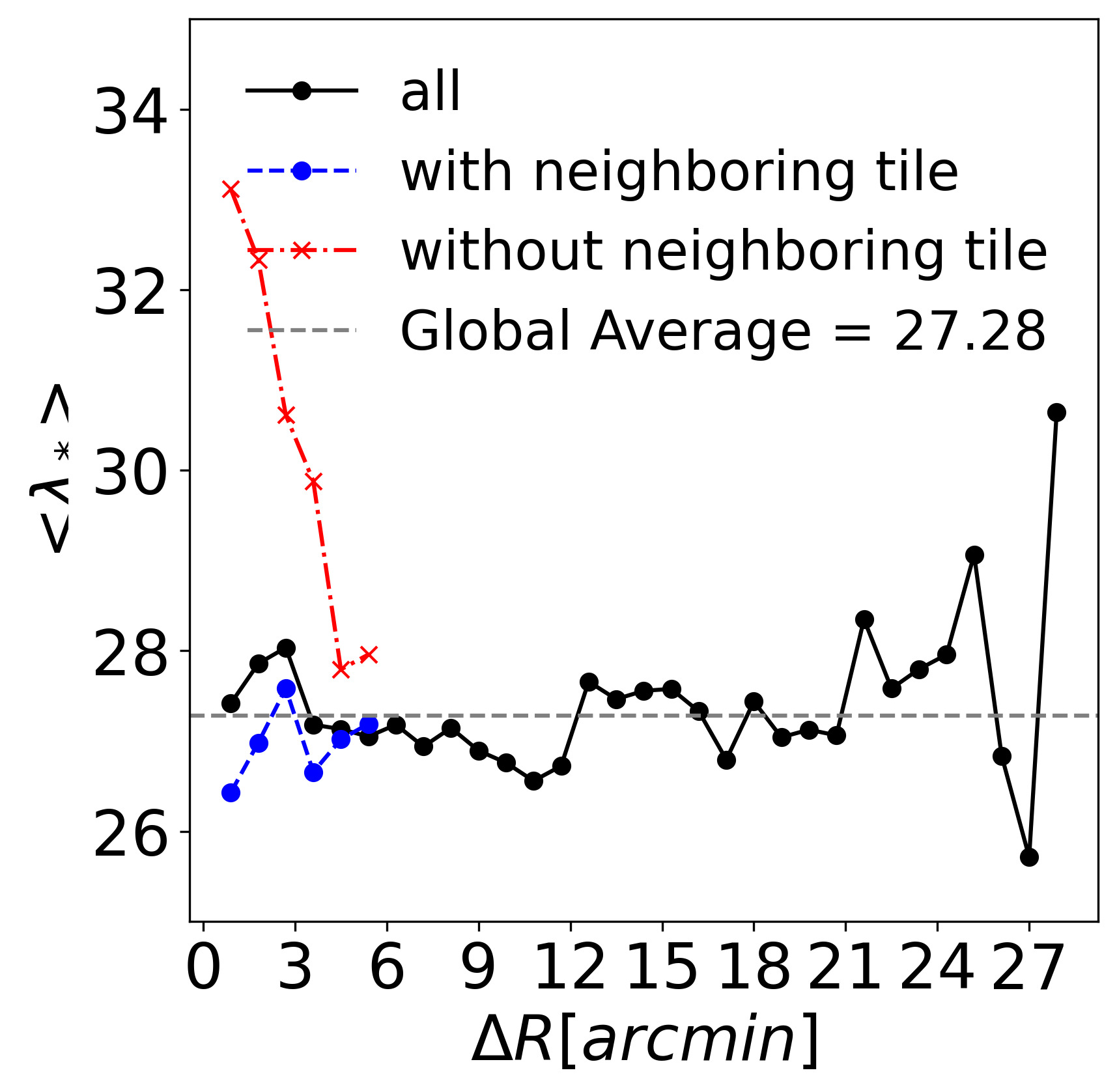}
  \caption{Average properties of detections as a function of their distance from the tile border $\Delta R$:
  amplitude $A$ (top left panel), signal-to-noise ration $S/N$ (top right panel), apparent richness $\lambda$ (bottom left) and intrinsic richness $\lambda_*$ (bottom right). The results for all detections are shown in black, those within $5$ arcmin of a tile border having a neighboring tile are shown in blue, and those with a missing neighboring tile are shown in red. Most values are insensitive to the tiling scheme, except for $\lambda_*$ near the survey's external borders and in regions with missing tiles within the overall area.}
  \label{fig:borders_stat}
\end{figure}

\subsection{Application to the KiDS-DR4 data set}\label{sec:application}

We considered galaxies within the magnitude range $15<r'<24$. Unlike in the KiDS-DR3 analysis, where noise properties $N$ were estimated on a tile-by-tile basis, in this work we derived them across the entire survey area to leverage the survey homogeneity and apply the same filter throughout the survey. This approach enhances the statistical robustness of the noise estimate and simplifies both the analysis and interpretation of the results. The sky was sampled in steps of $0.3$ arcmin and redshift intervals of $\Delta z=0.01$. The search was conducted over the redshift range $0.05<z<1.2$, with the final cluster sample restricted to $0.1<z<0.9$. The initial broader redshift range was adopted to mitigate border effects at high redshifts due to the possible truncations of the galaxy $P(z)$. To maintain sensitivity near the edges of each survey tile, we included a buffer of $0.1$ deg on each side. Consequently, \textsc{AMICO} was run on overlapping square areas of $1.2$ deg per side.

\new{A word of caution must be raised for detections at $z > 0.8$, as the value of $m_* + 1.5$ used to define the $\lambda_*$ mass proxy approaches the sample’s magnitude limit of $r = 24$. As a result, $\lambda_*$ becomes strongly redshift-dependent and, given the small number of galaxies at those redshifts, increasingly unstable. Nevertheless, we chose to include these high-redshift detections, even though they are less robust, as they represent promising candidates that may be confirmed with deeper or complementary multi-wavelength data.}

\begin{figure}
  \includegraphics[width=0.241\textwidth]{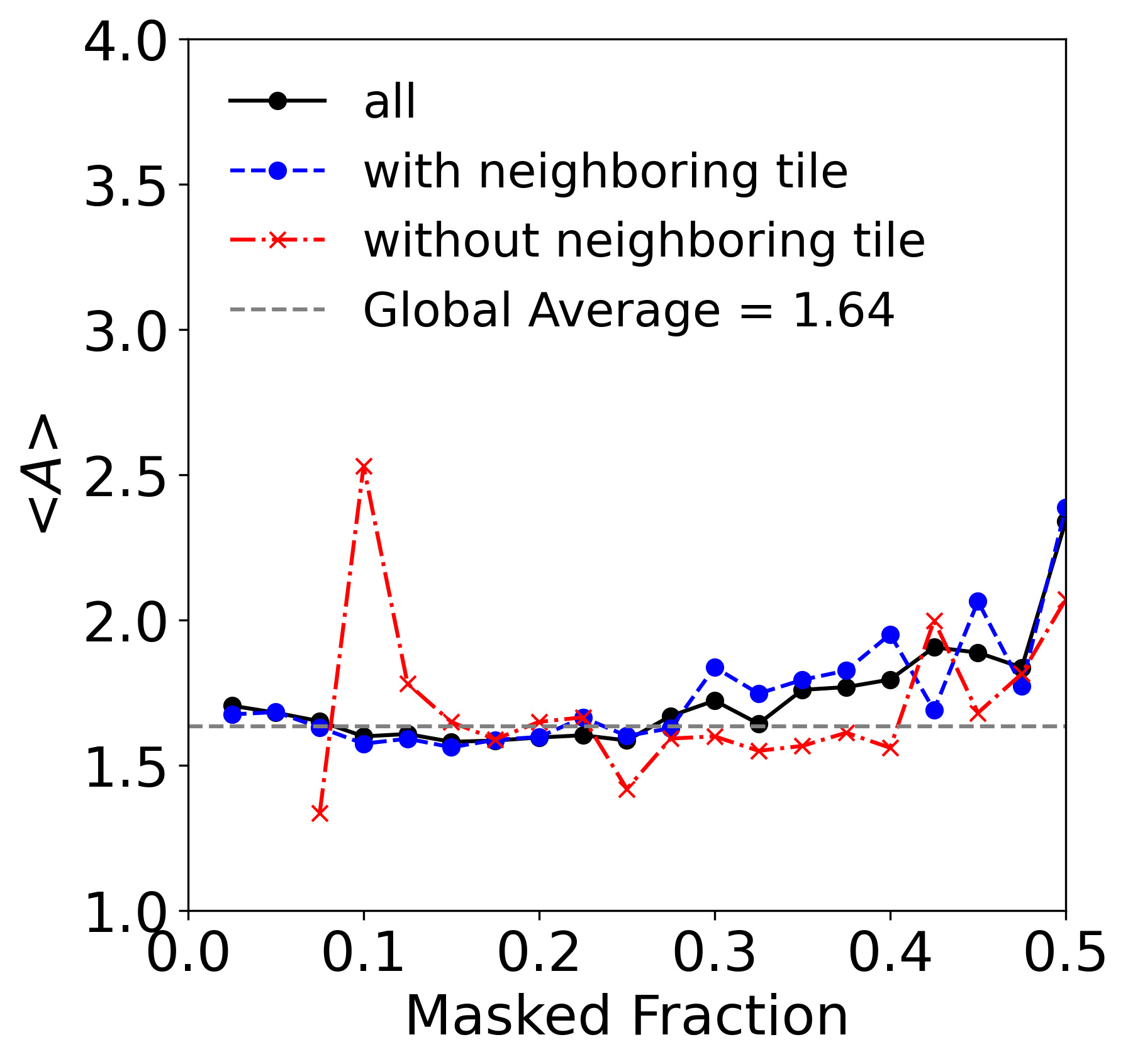}
  \includegraphics[width=0.235\textwidth]{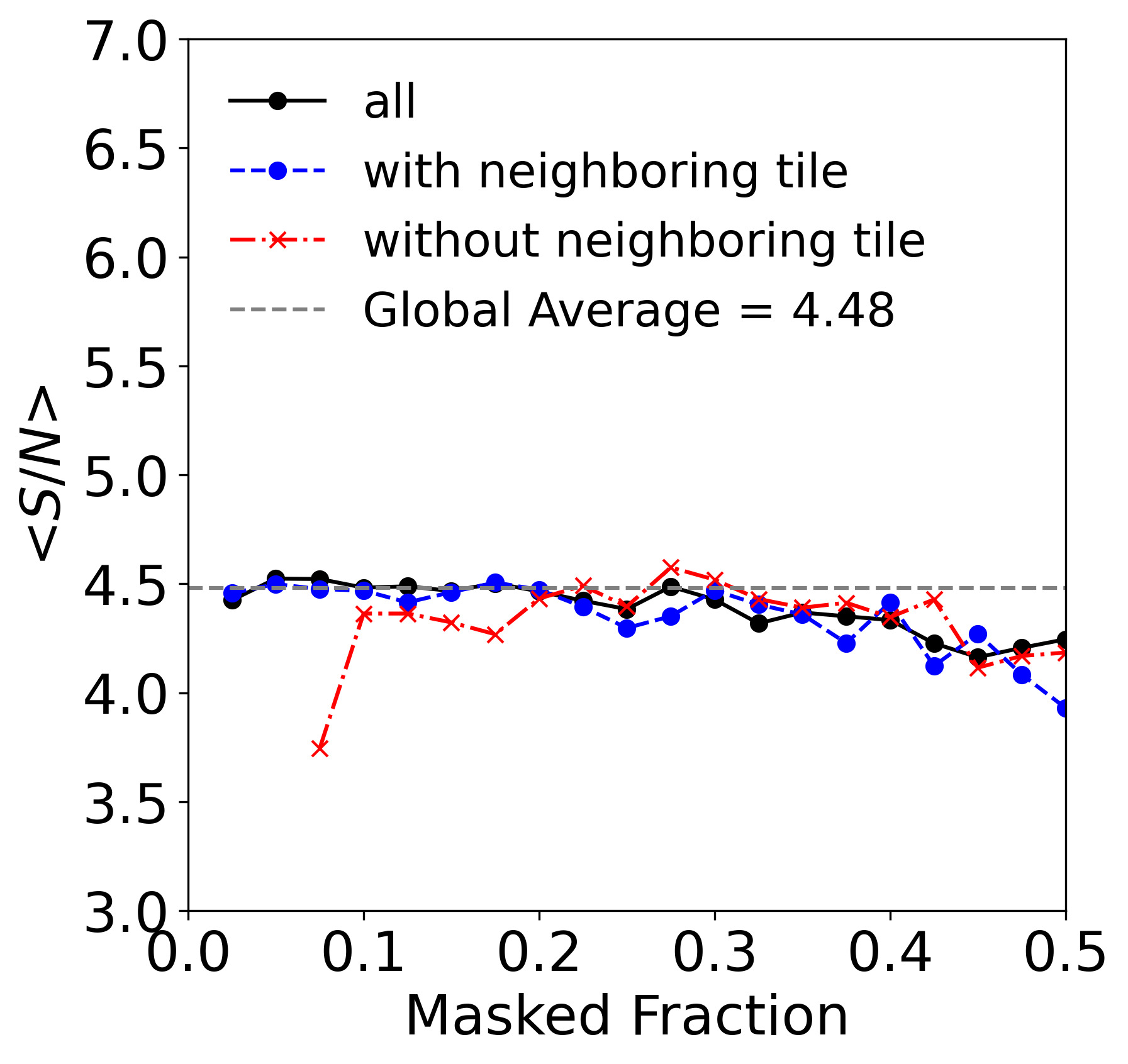}\\
  \includegraphics[width=0.241\textwidth]{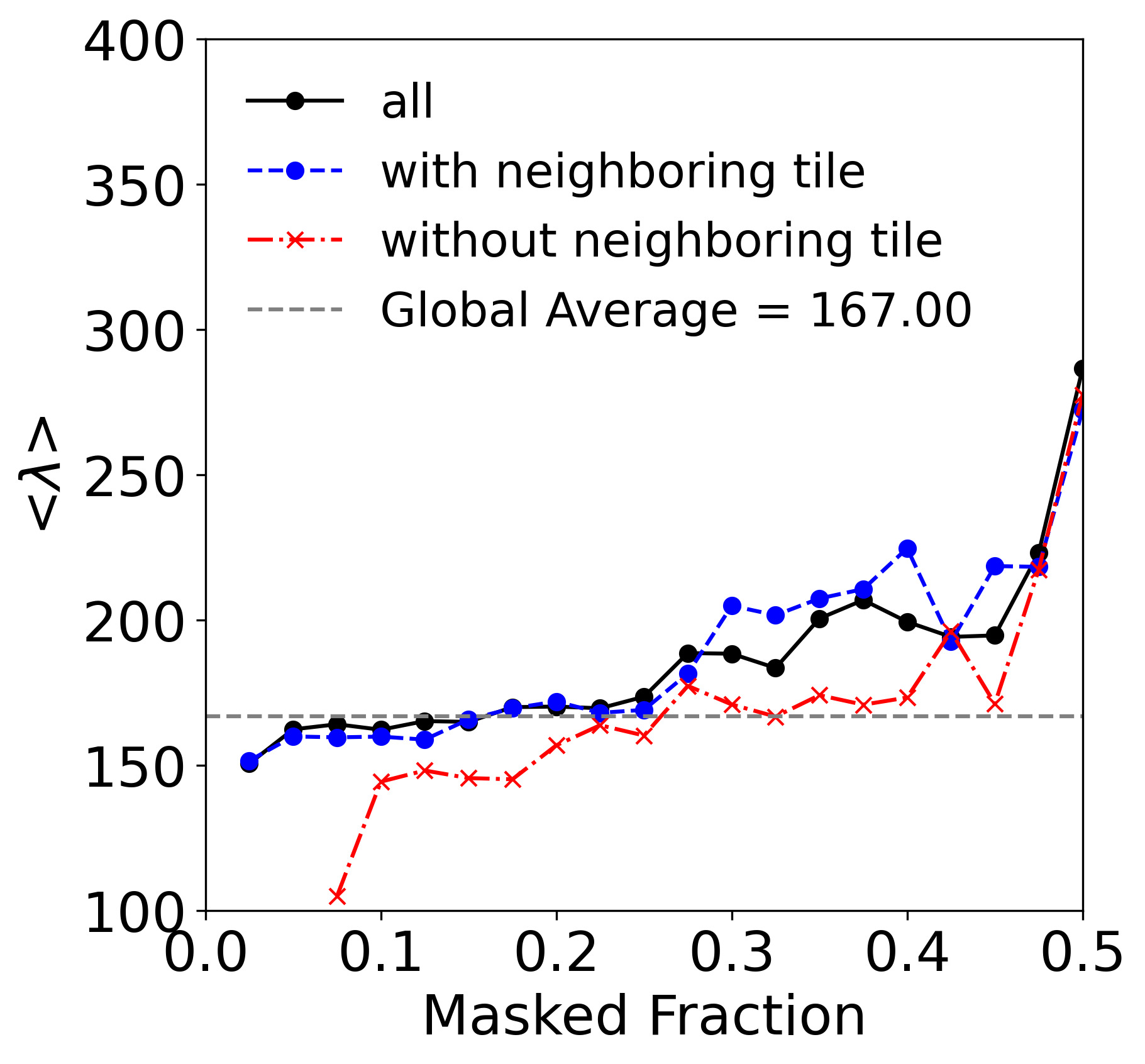}
  \includegraphics[width=0.235\textwidth]{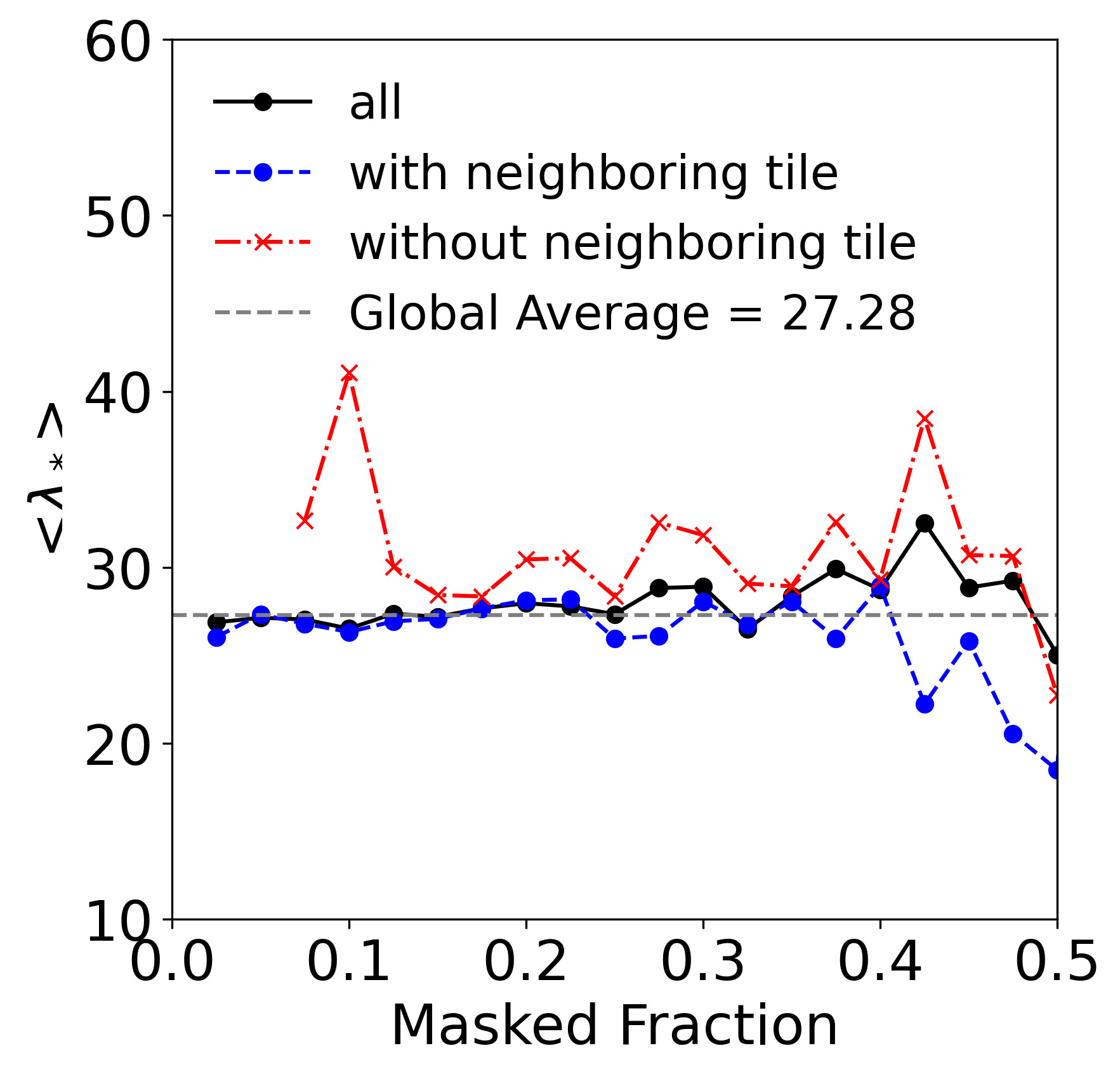}
  \caption{The same as Fig.~\ref{fig:borders_stat},
  but as a function of the masked fraction of each detection. All quantities remain stable up to a masked fraction of $0.3$, with $\lambda_*$ remaining stable up to $0.4$.}
  \label{fig:maskfrac_stat}
\end{figure}

The catalogs of cluster detections obtained from each survey tile are then combined into a master catalog. To avoid duplicate entries due to tile overlaps, we apply a geometric cut dictated by the tiles geometry and identify pairs with differences in angular position and redshift smaller than $\Delta R<1.2$ arcmin (equivalent to 2 pixels of the \textsc{AMICO} amplitude maps) and $\Delta z<0.015(1+z)$, respectively. We retain both detections if they are located within the same survey tile; otherwise, we keep only the detection closest to the center of its respective tile. The repeated objects are shown in Fig.\ref{fig:repetitions} as a function of their intrinsic richness $\lambda_*$. The clear correlation between their intrinsic richness (which is not used for their identification) demonstrates that distinct nearby clusters are not mistaken as duplicates. Indeed, after the removal of the duplicates, the probabilistic membership of galaxies has been re-adjusted to ensure consistency with the purged catalog. The number of rejected duplicates is minimal, specifically 31 entries, because the buffer areas between tiles are small.

\subsection{Testing and flagging detections in the proximity of areas with missing data}

\begin{figure}
  ~~~~~\includegraphics[width=0.455\textwidth]{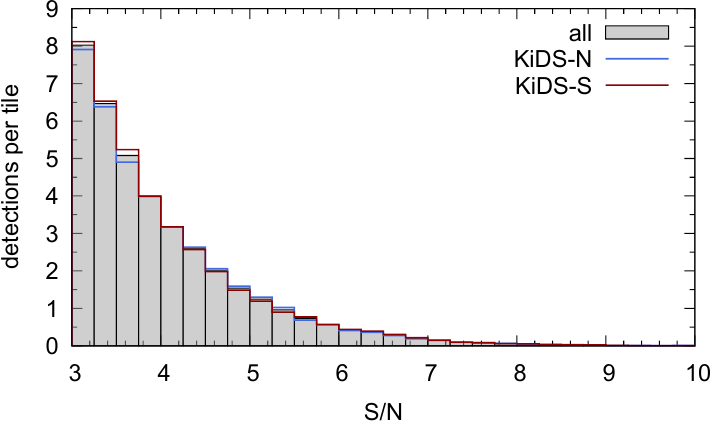}\\
  \includegraphics[width=0.48\textwidth]{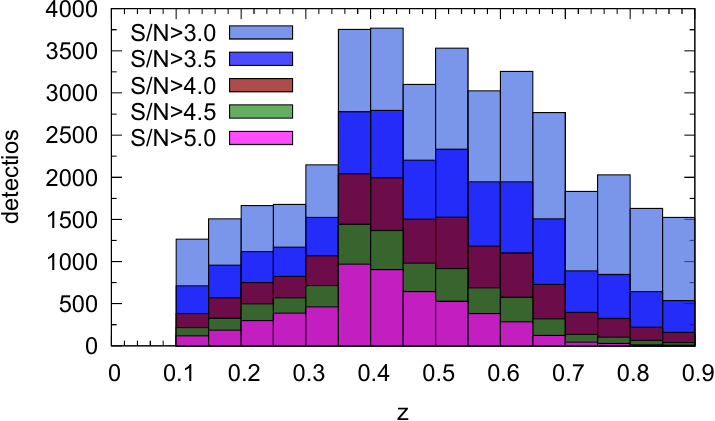}\\
  \includegraphics[width=0.5\textwidth]{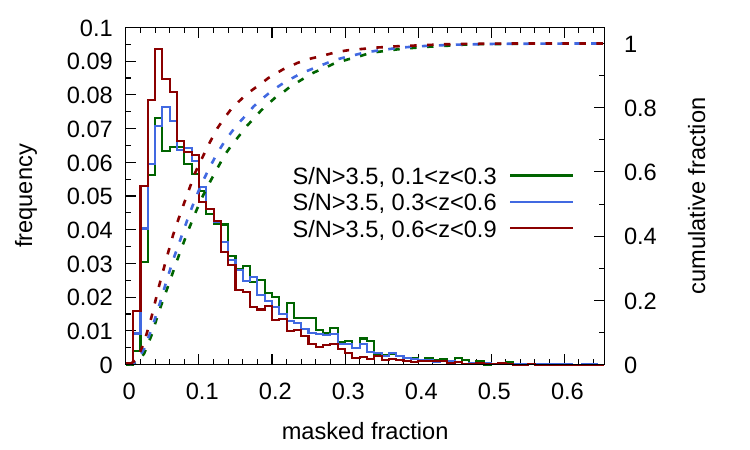}
  \caption{The main properties of the cluster sample. Top panel: the average number of detections per survey tile as a function of the signal-to-noise ratio $S/N$ for the entire sample (gray histogram), and for the KiDS-N and KiDS-S stripes (blue and red lines, respectively). Central panel: the redshift distribution for five different $S/N$  cutoffs. Bottom panel: distribution of the masked fraction of detections in different redshift intervals.}
  \label{fig:detStat}
\end{figure}

To verify the behavior of the detections at the interface between tiles, we show in Fig.~\ref{fig:borders_stat} the average values of the signal-to-noise ratio $S/N$, amplitude, apparent and intrinsic richness as a function of their distance from the edge of the survey tile to which they belong. We distinguished between all detections (black lines) and those within $\Delta R<5$ arcmin, further divided into two categories: those having a neighboring tile (blue lines), where border effects should be minimal due to the overlapping tile scheme we implemented, and those without a neighboring tile, where border effects are expected (red lines). The latter case occurs at the survey's edge or where tiles within the survey area had not yet been observed in KiDS-DR4. It is clear from the figure that the overall average values of $A$, $S/N$, and $\lambda$ are not affected by border effects. This is because the filter construction accounts for the presence of missing data when evaluating the amplitude $A$ and the signal-to-noise ratio $S/N$. Moreover, the mask fraction used to correct the apparent richness $\lambda$ is accurate. However, the intrinsic richness $\lambda_*$ of detections within $3$ arcmin of the tile edge is overestimated. The reason for this overestimation is that we apply for $\lambda_*$ the same correction factor used for $\lambda$, which does not account for the magnitude selection criteria imposed in the definition of $\lambda_*$. This results in an over-correction of its value. In fact, the brightest galaxies used in the evaluation of $\lambda_*$ are preferentially located closer to the cluster center, and their number is less affected by masks, which generally tend to `erode' the outer regions of clusters.

To enable in the catalog the selection of clusters based on their proximity to tile edges, we include two parameters: the minimum distance of each detection from the edge of its host survey tile, denoted as \texttt{TILE\_EDGE\_DISTANCE}, and a flag, \texttt{TILE\_EDGE\_FLAG}, which can assume the following three possible values:
\begin{itemize}
\item 0: assigned to detections located more than $\Delta R>5$ arcmin away from the tile edge.
\item 1: assigned to detections within $\Delta R<5$ arcmin of the tile edge, in cases where an adjacent neighboring tile is present. These detections are minimally impacted by edge effects, as the data are not truncated unless the cluster's radius exceeds $0.1$ deg (the width of the buffer region used for tile overlap).
\item 2: assigned to detections within $\Delta R<5$ arcmin of the tile edge, in cases where no adjacent neighboring tile exists. This scenario typically occurs at the survey's boundaries or near unobserved regions. Although similar to \texttt{TILE\_EDGE\_FLAG}=1, the absence of a neighboring tile increases the likelihood of edge effects or anomalies, particularly due to minor image artifacts at the tile boundaries.
\end{itemize}

\begin{figure}
  \centering
  \vspace{0.2cm}
  \includegraphics[width=0.49\textwidth]{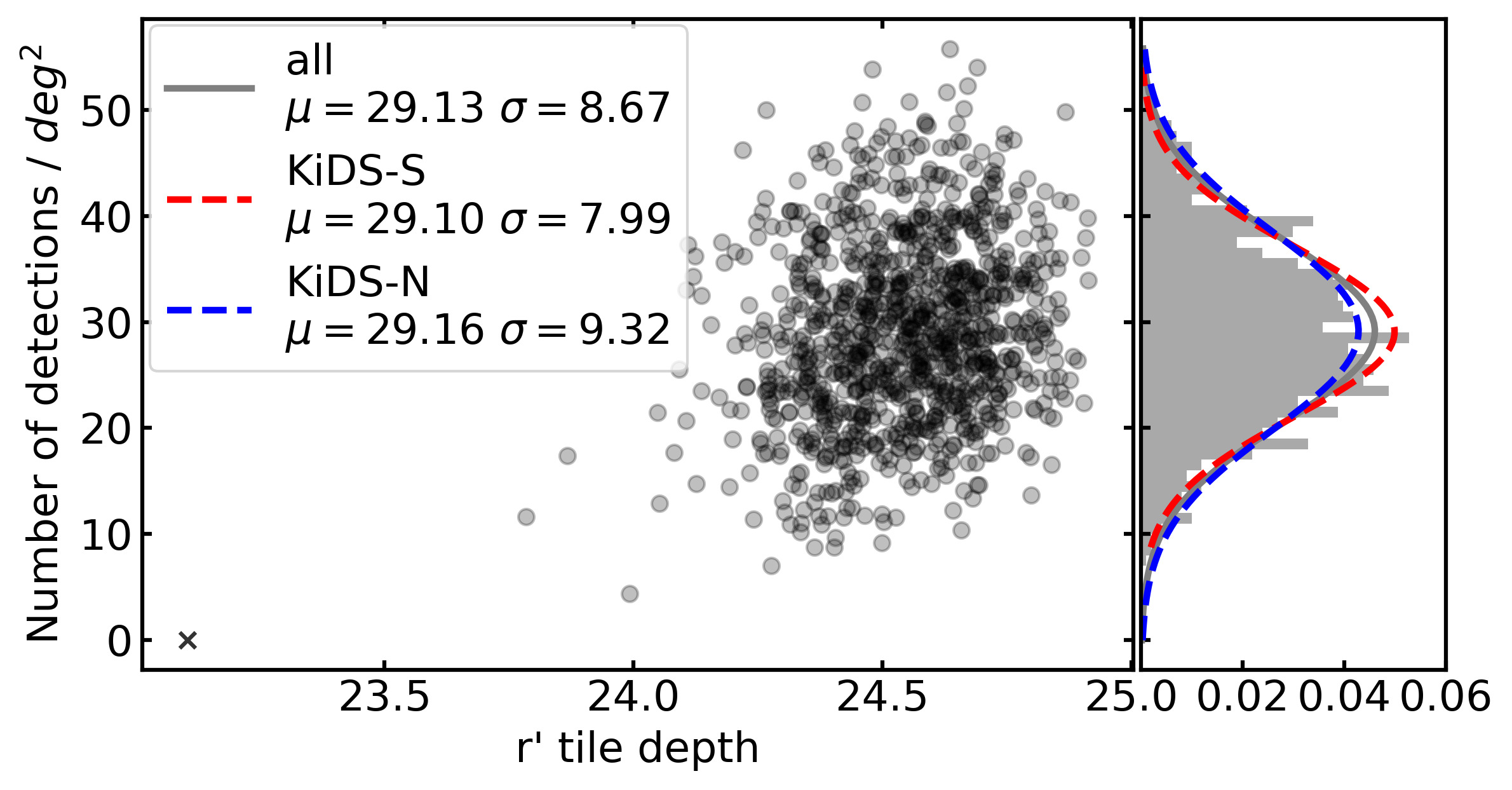}
  \caption{Left panel: number density of detections with $S/N>3.5$ as a function of the survey tile depth in the $r$-band. Each point represent an individual tile. Side panel: probability distribution of the number of detections per tile for the entire sample (gray histogram), and for the KiDS-N and KiDS-S stripes (blue and red lines, respectively). The corresponding values of the mean $\mu$ and RMS $\sigma$ are reported in the legend. The absence of any significant correlation and the Gaussian shape of the number density distributions indicate the homogeneity of the survey clusters detection efficiency when a strict cutoff of $r < 24$ is applied in the selection of the input galaxy sample. A few outliers are visible, corresponding to tiles with problematic photometry (see their list in Table~\ref{tab:arti_list}). We excluded from the analysis the tile marked with an "x" \new{(see bottom left corner)}.}
  \label{fig:galStatDet}
\end{figure}

\begin{figure}
    \centering
    \includegraphics[width=0.48\textwidth]{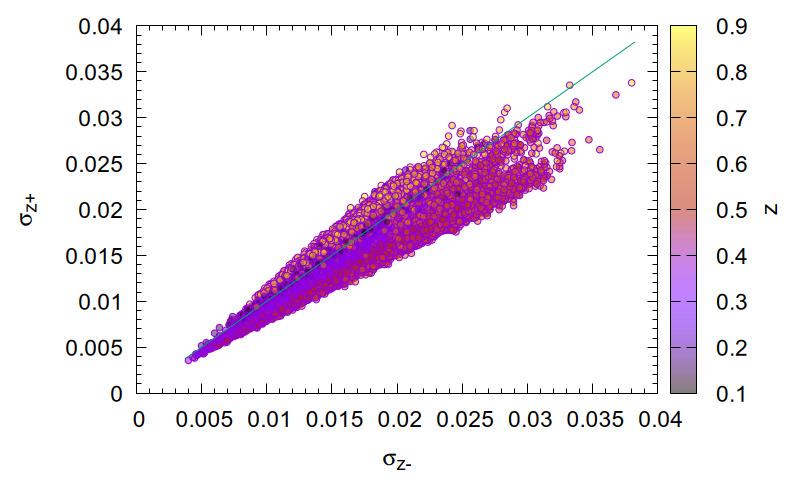}
    \includegraphics[width=0.48\textwidth]{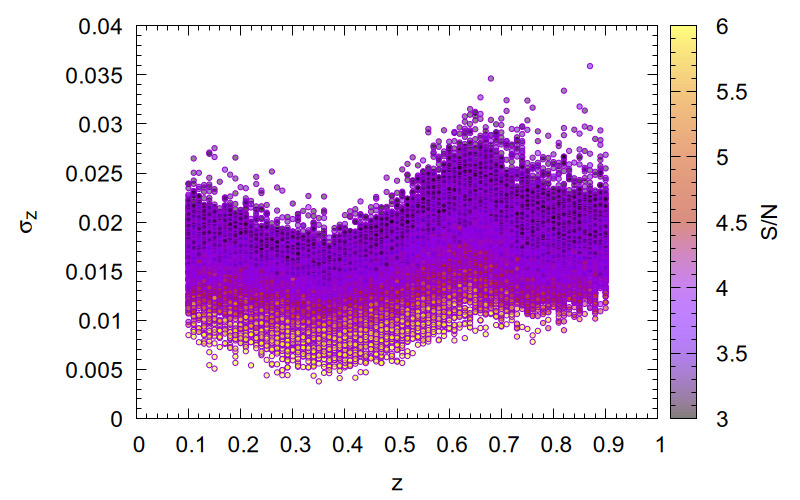}
    \includegraphics[width=0.45\textwidth]{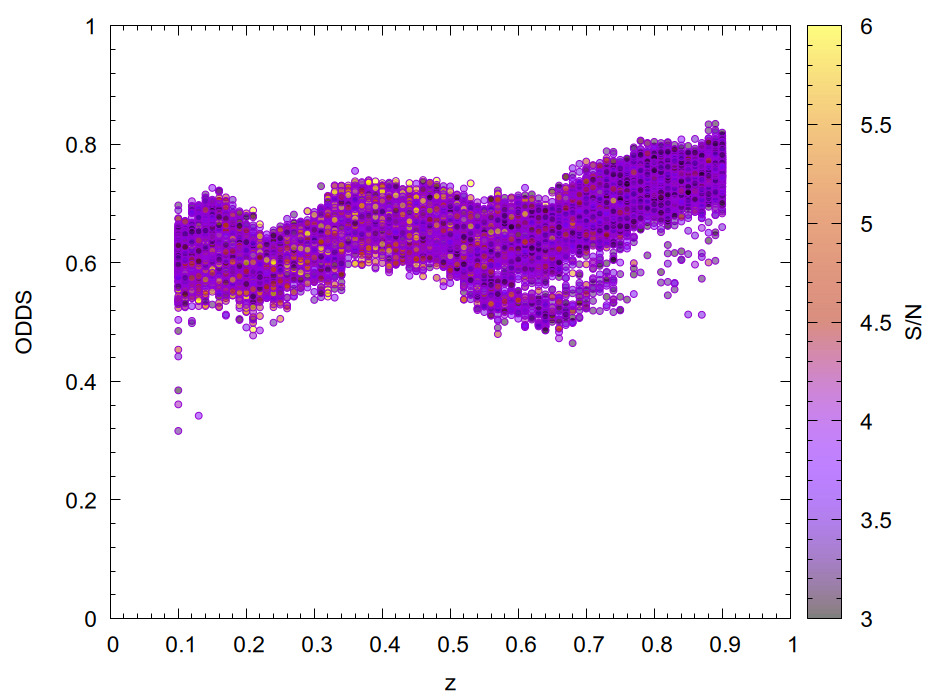}
    \caption{Top panel: upper and lower $67\%$ percentile limits of the redshift probability distribution, $P_{det}(z)$, for each cluster detections. Central panel: redshift uncertainty derived from the previous percentiles. Bottom panel: ODDS values, where it is possible to notice that a sub-population of detections in the redshift range $0.5 < z < 0.75$ exhibits significantly lower ODDS values compared to the other detections. Data are colored according to their redshift in the upper panel, and to $S/N$ in the other two panels.}
    \label{fig:odds}
\end{figure}

We conducted a similar analysis, this time examining how detections are influenced by the masked fraction, defined as the proportion of cluster members located within masked regions. This evaluation is based on the density profile used in the filter model described in Sect.~\ref{sec:amico}. Unlike the edge-proximity analysis, this study considers the entire survey area, not just regions near tile edges. As in the previous analysis of border effects, we categorize detections according to the three values of \texttt{TILE\_EDGE\_FLAG}. Fig.~\ref{fig:maskfrac_stat} presents the results for the same set of average quantities as in the previous figure, but analyzed as a function of the masked fraction of detections. The average values of $A$, $S/N$, $\lambda$, and $\lambda_*$ remain unbiased for masked fractions up to $30\%$. Beyond this threshold, deviations from the global average value become apparent, indicating the presence of a bias. Among these quantities, the intrinsic richness $\lambda_*$ is the most robust, showing a good consistency up to a masked fraction of $40\%$. In any case, the number of detections with masked fractions exceeding $0.3$ is notably small, as illustrated in the central panel of Fig.~\ref{fig:detStat}.

\subsection{The cluster sample}\label{sec:clustersample}

The resulting catalog comprises $23\,965$ detections with $S/N>3.5$. Fig.~\ref{fig:detStat} provides an overview of the principal properties of the identified cluster sample.
In the top panel, we display the average number of detections per tile as a function of the signal-to-noise ratio for the entire sample (gray histogram) and for the KiDS-N and KiDS-S stripes (blue and red lines, respectively). The distributions for both survey regions are very similar, despite the KiDS-S stripe being, on average, slightly deeper than the KiDS-N one. Again, this similarity shows the effectiveness of the conservative cut of $r < 24$ applied to the input galaxy sample, as discussed in Sect.~\ref{sec:data-kids}.
The central panel displays the redshift distribution. A discontinuity is visible at $z \sim 0.35$, due to issues with photometric redshifts occurring at the transition between the $g$ and $r$ bands of the $4000\AA$ break. This feature, which was much more prominent in the KiDS-DR3 analysis, is here significantly alleviated thanks to the incorporation of the NIR data from the VIKING survey.
The bottom panel illustrates the distribution \new{and the cumulative} of the masked fraction for detections with $S/N > 3.5$. Among these detections, $96\%$, $85\%$, $53\%$, and $19\%$ have masked fractions below $30\%$, $20\%$, $10\%$, and $5\%$, respectively. Higher-redshift detections tend to have slightly lower masked fractions compared to those at lower redshifts, due to their smaller angular extent.

Due to their definitions, the amplitude $A$ and intrinsic richness $\lambda_*$ serve as nearly redshift-independent mass proxies. For the amplitude, this redshift independence arises because the cluster template used in the \textsc{AMICO} filtering process is built to incorporate the survey's magnitude limits. For the intrinsic richness, it is achieved thanks to the embedded magnitude cut, which is never below $r<24$ at all redshifts. As expected, the minimum detectable cluster mass increases with redshift, leading to a corresponding rise in the lower values of both amplitude and intrinsic richness. Conversely, the lower limit of the apparent richness, $\lambda$, which represents the number of observable galaxies, remains relatively constant with redshift. This behavior is expected, as the survey's sensitivity to galaxy cluster detection primarily depends on the number of member galaxies that can be observed, as quantified by the apparent richness $\lambda$.

\begin{figure}
  \centering
  \includegraphics[width=0.49\textwidth]{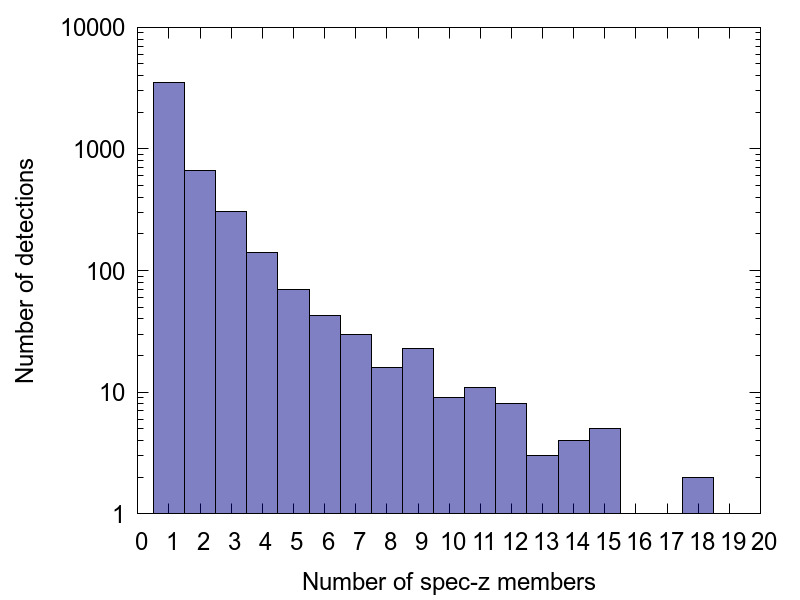}
  \caption{Distribution of 
  the number of detections as a function of the number  
  of spectroscopic members with a probabilistic membership association greater than $90\%$.}
  \label{fig:spec-z_stat}
\end{figure}

Fig.~\ref{fig:galStatDet} displays the number density of cluster detections as a function of the $r$-band magnitude depth for each tile, with each point representing an individual tile. The outlier tile that we excluded is visible in the bottom-left corner, marked with an "x". For the remaining tiles, there is no apparent correlation between the survey depth and the number density of detections. Furthermore, the local survey depth and the number density of detections follow a Gaussian distribution with the same mean for both the KiDS-N and KiDS-S stripes (see the side panel). The only notable difference is that the scatter is slightly larger for the KiDS-N stripe. This confirms that the initial galaxy selection, with the magnitude cut at $r<24$, effectively homogenized the properties of the survey to a very high degree.

Finally, the analysis has been enhanced with new features recently introduced in \textsc{AMICO}, including:
\begin{itemize}
\item an estimate of the probabilistic redshift distribution, $P_{det}(z)$, for each detection. This is calculated as the weighted sum of the $P(z)$ distributions of individual galaxies, where the weights correspond to the probabilistic association of each galaxy with the given detection;
\item the 16th and 84th percentiles of $P_{det}(z)$, which represent the $1\sigma$ lower and upper bounds of the redshift uncertainties for each detection, denoted as $\sigma_{z,\mathrm{min}}$ and $\sigma_{z,\mathrm{max}}$;
\item an ODDS value for each detection, \new{defining the probability that the true redshift lies within $\pm 0.1$ of the best-fit photo-$z$. It is computed as} $ODDS=\int_{z_-}^{z+} P_{det}(z) dz$ with integration limits $z_-$ and $z_+$ corresponding to a redshift interval of width $0.2\,(1+z_{det})$, centered on the mode of $P_{det}(z)$ \citep{benitez00}.
\end{itemize}

The upper and lower bounds, the symmetrized redshift uncertainty, and the ODDS values for each detection are shown (with colors depending on their $S/N$) in the different panels of Fig.~\ref{fig:odds}, from top to bottom. As expected, detections with higher signal-to-noise ratios exhibit smaller redshift uncertainties. More specifically, the average redshift uncertainties, $\sigma/(1+z)$, are $0.01$, $0.009$, $0.008$ and $0.007$ for $S/N>3.5$, $4.0$, $4.5$ and $5.0$, respectively. These uncertainties are smaller than those estimated using GAMA spectroscopic redshifts, as it will be discussed in Sect.~\ref{sec:gama}. This difference may be attributed to the input $P(z)$ of the galaxies photometric redshifts being slightly too optimistic, i.e. narrower than what they should be. In the distribution of ODDS values, a population of detections between redshift $z=0.5$ and $z=0.8$ with very low ODDS values appears. However, this population of detections is not of concern, as no other quantities characterizing them exhibit any peculiar or anomalous behavior. Additionally, visual inspection of the $gri$ color composite images of these objects reveals no issues that would suggest problems with their reliability. Therefore, we decided not to flag them as the anomaly seems to be related to the ODDS estimate itself.

\subsection{Redshift calibration with GAMA spectroscopic redshifts}\label{sec:gama}

The Galaxy and Mass Assembly (GAMA) survey consists \citep{Driver2011,Liske2015} of spectroscopic observations carried out with the Anglo-Australian Telescope’s AAOmega spectrograph across four equatorial and one southern field. In its Data Release 4 \citep{Driver2022}, GAMA incorporated KiDS-DR4 photometry within an overlapping area of approximately 210 deg$^2$ \citep{Bellstedt2019}. \new{From this data set, we selected galaxies with reliable redshifts (NQ > 2; \citealt{Driver2022}) \new{and further augmented the spectroscopic sample with higher-redshift estimates from \citet{vandenbusch22}.} We used this combined spectroscopic sample} to evaluate the redshift uncertainties in our cluster sample and to investigate the presence of potential biases.

\begin{figure*}[h!]
  \includegraphics[width=0.99\textwidth]{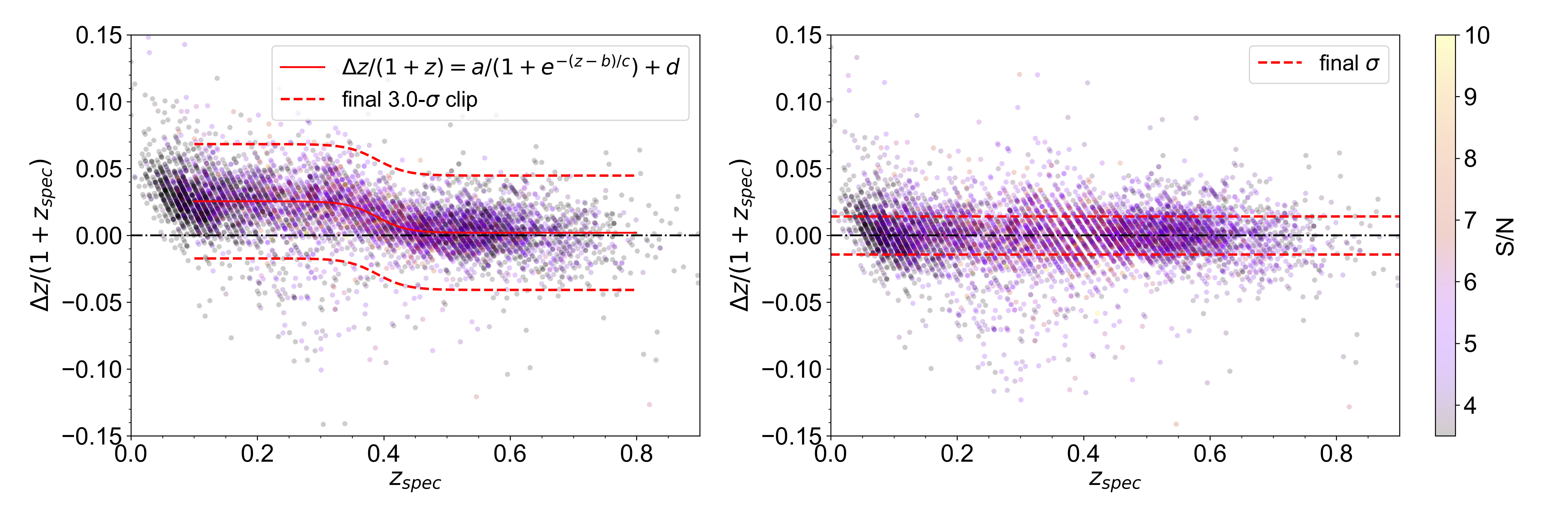}
  \caption{Scatter of the detections' redshifts obtained with \textsc{AMICO} (based on galaxy photo-$z$s) with respect to the corresponding spectroscopic redshifts. Left and right panels show the values before and after redshift calibration, respectively. The left panel also displays the associated bias model (red solid line) and the final limits used for the k-sigma clipping (dashed red line). In the right panel the dashed red lines show the final 1$\sigma$ uncertainty.}
  \label{fig:spec-z_calib}
\end{figure*}

We consider GAMA spectroscopic redshifts for cluster members with an \textsc{AMICO} probabilistic membership exceeding $90\%$. This high probability threshold ensures that the selected members can be confidently associated with their respective detections. We do not impose a minimum requirement on the number of spectroscopic galaxies per cluster detection: a cluster is assigned a spectroscopic redshift even if it contains only one member with a spec-$z$. However, we provide information on the number of spectra used for each cluster detection along with additional parameters to access the reliability of the spectroscopic evaluation and facilitate further selections based on the specificity of the science goals. These additional quantities, all based on members with a probabilistic association greater than $P>90\%$, are included in the final catalog for each cluster detection. They are:
\begin{itemize}
\item \texttt{SPEC\_NGAL}: the number of members with an associated spectroscopic redshifts.
\item \texttt{SPEC\_SUMP}: the sum of the \textsc{AMICO} membership probabilities for galaxies having a spectroscopic redshift.
\item \texttt{SPEC\_ZAVE}: the weighted average of the spectroscopic redshifts, where the weights are given by the \textsc{AMICO} membership probabilities.
\item \texttt{SPEC\_ZMED}: the median spectroscopic redshift of members with spectroscopic redshifts.
\item \texttt{SPEC\_RMS}: the RMS scatter around the spectroscopic weighted average.
\end{itemize}

A total of $4883$ detections have been assigned a spectroscopic redshift. In Fig.~\ref{fig:spec-z_stat}, we display the distribution of the number of cluster members with spectroscopic information. We found that $1343$, $237$ and $50$ detections have more than $2$, $5$ and $10$ members with $P>0.9$ and a spectroscopic redshift, respectively. 

In the left panel of Fig.~\ref{fig:spec-z_calib} we show the deviation between the cluster redshifts computed by \textsc{AMICO} using the galaxies' photometric redshifts, and the corresponding spectroscopic redshift estimates.
A clear bias is observed in the photo-$z$ estimates for clusters at $z<0.4$, which we model using a sigmoid function:
\begin{equation}
  \frac{\Delta z_{bias}}{1+z} = a / (1 + e^{-(z-b) / c}) + d \;.
\end{equation}
Here $a$, $b$, $c$, and $d$ are free parameters that are determined using Orthogonal Distance Regression (ODR) combined with an iterative approach to handle outliers in an effective way. \new{This functional form effectively captures the discontinuity caused by the transition of the $4000\AA$ break between the $g$ and $r$ bands at $z \sim 0.34$ already discussed in Sect.~\ref{sec:clustersample}.} The fit is restricted to cluster detections with $S/N>3.5$ within the redshift range $0.1<z<0.8$, as spectroscopic data are sparse at higher redshifts. The aforementioned iterative procedure begins with an initial fit of the bias model to all data points, providing a preliminary estimate of the RMS for bias-corrected photometric redshifts. In each subsequent iteration, we apply the bias correction and clip all data points deviating by more than 3$\sigma$ from their corresponding spectroscopic redshifts. The model parameters are then re-fitted, and the RMS is recomputed for the following iteration. Five iterations ensure convergence. 
The final model and the boundaries for the k-sigma clipping are reported in the left panel of Fig.~\ref{fig:spec-z_calib} as solid and dashed red lines, respectively. The best-fit parameter values for the bias model are
\begin{equation}
    \begin{aligned}
    a &=-0.0235\pm0.0006, \;  b =0.389\pm0.003,\\
    c &=0.024\pm0.003, \; \mbox{and} \; d =-0.0255\pm-0.0004.
    \end{aligned}
\end{equation}

The right panel of the same figure shows the scatter of the photometric redshifts of the cluster detections after the bias correction, together with their RMS. 
The RMS for bias-corrected photometric redshifts, determined using this k-sigma clipping method, is $\sigma_z/(1+z)=0.014$ for the entire sample, while for detections with $S/N>4.0$, $S/N>4.5$ and $S/N>5.0$, the corresponding RMS values are $0.013$, $0.012$ and $0.0118$, respectively. These scatters are approximately 1.5 times larger than those estimated using the percentiles of the detection probability redshift distribution, $P_{det}(z)$, which is derived from the weighted $P(z)$ of cluster members, as discussed above. This discrepancy may arise from an overestimation of errors based on the spectroscopic redshifts, potentially due to member misassociations. Alternatively, $P_{det}(z)$ might be narrower than the true distribution. Although the difference is not substantial, we consider safer to rely on the errors derived from the spectroscopic redshift estimates, as they are based on external data rather than an internal evaluation.

\begin{table*}
    \centering
    \caption{Descriptions of the columns present in the \textsc{AMICO}-KiDS-DR4 galaxy cluster catalog.}
    \begin{tabular*}{\textwidth}{l@{\extracolsep{\fill}}llr}
        \hline
        \multicolumn{1}{l}{Column} & \multicolumn{1}{@{\extracolsep{\fill}}l}{Unit} & \multicolumn{1}{l}{Description} & \multicolumn{1}{r}{Example} \\
        \hline
 \texttt{NAME} (1) & & ESO format object name & AK4 J000523.41-284831.0 \\
 \texttt{UID}  (2) & & unique identification number & 1 \\
 \texttt{TILE} (3) & & survey tile name & KiDS\_DR4.0\_1.1\_-29.2 \\ 
 \texttt{TID}  (4) & & identification number within a survey tile& 10 \\
 \texttt{XPIX} (5) & pixel & pixel (x axis) amp. map with buffer: [0,nx-1] & 156\\
 \texttt{YPIX} (6) & pixel & pixel (y axis) amp. map with buffer: [0,nx-1] & 194\\
 \texttt{ZPIX} (7) & pixel & pixel (z axis) amp. map with buffer: [0,nx-1] & 50\\
 \texttt{RA}   (8) & deg & right ascension of XPIX & 1.3475235 \\
 \texttt{DEC}  (9) & deg & declination of YPIX & -28.808636 \\
 \texttt{Z}    (10) & & redshift of ZPIX& 0.55 \\
 \texttt{ZFIX} (11) & & calibrated redshift based on GAMA spec-$z$ & 0.55 \\
 \texttt{ZFIX\_ERR} (12) & & ZFIX RMS around GAMA spectroscopic redshifts & 0.030225 \\
 \texttt{SPEC\_NGAL} (13) & & number of members with $P>0.9$ and GAMA spec-$z$ & -99 \\
 \texttt{SPEC\_SUMP} (14) & & sum of memberships with $P>0.9$ and GAMA spec-$z$ & -99 \\
 \texttt{SPEC\_ZAVE} (15) & & weighted average with $P>0.9$ and GAMA spec-$z$ & -99 \\
 \texttt{SPEC\_ZMED} (16) & & median of the GAMA spec-$z$ with $P>0.9$ & -99 \\
 \texttt{SPEC\_RMS}  (17) & & RMS of the weighted average spec-$z$ & -99 \\
 \texttt{SN}              (18) & & S/N: with N = background and cluster contrib. & 4.4 \\
 \texttt{SN\_NO\_CLUSTER} (19) & & S/N: with N = cluster contribution only & 15.3 \\
 \texttt{MSKFRC}          (20) & & masked fraction of the detection & 0.285 \\
 \texttt{AMP}             (21) & & signal amplitude returned by the filter& 1.88 \\
 \texttt{LAMBDA}          (22) & & apparent richness & 181.89 \\
 \texttt{LAMBDA\_STAR}    (23) & & intrinsic richness & 39.35 \\
 \texttt{PZ} (24) & & probability redshift distribution & (0.0126, 0.0017, ..., 0.0)\\
 \texttt{PZ\_ODDS} (25) & & ODDS photo-$z$ based on the cluster $P(z)$ & 0.65999776 \\
 \texttt{PZ\_ZPIX\_SIGM} (26) & & ZPIX $16\%$ percentile of the detection $P(z)$ & 1.55 \\
 \texttt{PZ\_ZPIX\_SIGP} (27) & & ZPIX $84\%$ percentile of the detection $P(z)$ & 1.26 \\
 \texttt{PZ\_Z\_SIGP}    (28) & & Z $16\%$ percentile of the detection $P(z)$ & 0.015 \\
 \texttt{PZ\_Z\_SIGP}    (29) & & Z $84\%$ percentile of the detection $P(z)$ & 0.016 \\
 \texttt{TILE\_EDGE\_DISTANCE}   (30) & arcmin & distance from the border of the survey tile & 6.84 \\
 \texttt{TILE\_EDGE\_FLAG}       (31) & & flag: proximity to border of the survey tile & 0 \\
 \texttt{ARTIFACTS\_FLAG}   (32) & & flag: possible issues of photometry in survey tile & 0 \\
 \texttt{M500\_eRASS1\_SCAL} (33) & $10^{13} M_{\odot}$ & $M_{500}$ from eRASS1 scaling relation & 12.28\\
 \texttt{M500\_eRASS1\_SCAL\_MIN} (34) & $10^{13} M_{\odot}$ & upper bound (1$\sigma$) of $M_{500}$ eRASS1 scaling relation & 6.16 \\
 \texttt{M500\_eRASS1\_SCAL\_MAX} (34) & $10^{13} M_{\odot}$ & lower bound (1$\sigma$) of $M_{500}$ eRASS1 scaling relation & 25.71 \\
 \hline
    \end{tabular*}
    \label{tab:catentries}
\end{table*}

To test the stability of this approach, we also evaluated $\sigma_z$ using two alternative robust estimators: (1) the Interquartile Range (IQR) clipping, which is similar to k-sigma clipping but based on data distribution quartiles, and (2) the Median Absolute Deviation (MAD), a method that relies on the median and the absolute deviations from the median, thus avoiding the dependency on arbitrary parameters, like the k-value used in the other methods. Both these methods yielded $\sigma_z=0.014(1+z)$, further validating the robustness of our approach. Additionally, repeating this analysis by considering members with a probabilistic association to clusters greater than $P>0.95$ (instead of $P>0.9$) produced consistent results, with a redshift uncertainty of $\sigma_z=0.0138(1+z)$.

Table~\ref{tab:catentries} lists all entries of the \textsc{AMICO}-KiDS-DR4 cluster catalog. \new{For convenience, the column number of each entry is indicated between parentheses. Missing values are labeled as ``-99''}. A gallery of postage-stamp $gri$-color composite images of newly discovered clusters, which allows to assess the data quality, is presented in Appendix~\ref{sec:gallery}.

\section{Quality assessment of the cluster sample}\label{sec:pur-sel}

The sample purity and selection function are derived from the analysis of mock data using the data-driven approach implemented in the Selection Function extrActor (\textsc{SinFoniA}) code. This method has been previously employed by \cite{maturi19} to characterize the \textsc{AMICO}-KiDS-DR3 cluster sample and is currently used as a part of the Euclid cluster pipeline. The core methodology of \textsc{SinFoniA} relies on a Monte Carlo approach, using the probabilistic galaxy memberships provided by \textsc{AMICO} to generate a mock galaxy catalog that distinguishes between field galaxies and cluster members. This is done by reproducing the survey properties, such as the actual masks, galaxy density, photometry and photo-$z$s.

After generating the mock galaxy sample, the procedure involves running the detection algorithm on the mock data and comparing the resulting detections to the "true" clusters in the mock galaxy catalog. This comparison is used to derive the statistical properties of the sample, such as purity, completeness, and the uncertainties associated with the key quantities characterizing the cluster detections. This data-driven approach avoids the need for numerical simulations, which often fail to capture the full complexity of the data and can introduce biases due to the cosmological and astrophysical assumptions they rely on.

In this work, we have significantly enhanced the \textsc{SinFoniA} code, achieving an acceleration by nearly a factor of $10$. Additionally, we introduced new features, including mock clusters with an ellipsoidal shape and an additional Monte Carlo extraction method used to select the properties of the mock clusters to be generated (see Sect.~\ref{sec:CDF} and Sect.~\ref{sec:mockClusters} for details).

\subsection{Stacks of mock cluster members}

Before generating the actual mock clusters, we randomly extract galaxies based on their probabilistic membership association, $P_i(j)$. A higher value of $P_i(j)$ indicates a larger probability for the $i$-th galaxy to be selected. Here the $i$ index refers to galaxies, while $j$ refers to the detection with which the galaxy is associated. During this extraction process, we only consider members associated with detections that meet the criteria $S/N>2.5$ and \texttt{MASK\_FRAC}$<0.25$.

The extracted potential cluster members are grouped into bins defined by the redshift and the apparent richness of the parent detection $(z, \lambda)$. Additionally, the radial distance of each galaxy from the parent detection center is recorded. This setup will allow us to populate a mock cluster at a specific redshift $z$ and richness $\lambda$ by selecting galaxies from the overall population of members belonging to the corresponding $(z, \lambda)$ bin. The large pool of available members enables the use of fine-grained bins with $\Delta\lambda=5$ and $\Delta z=0.01$.

In contrast to our previous analysis of KiDS-DR3 data, where bins were defined solely by redshift and richness, the current study introduces an additional refinement by grouping detections into sets of 50 tiles having a similar survey depth. This refinement ensures that the galaxies within each bin exhibit uniform data quality, particularly in terms of photometry and photometric redshifts (deeper tiles have better photometry and photo-$z$s). Consequently, these bins gain a third dimension, representing the depth of the tiles. These will be the bins containing the members that will be used to generate individual realizations of mock clusters.

\subsection{Monte Carlo extraction of the properties of the mock clusters}\label{sec:CDF}

Before extracting the members to generate individual realizations of mock clusters we need first to define their key properties, such as number, position, redshift and richness. There are various way to do that depending on the goal. For instance, if the intention is the quantification of detection blending as a function of angular and/or redshift separation, pairs of clusters with various separations could be generated in the data. Here, we aim at characterizing the sample purity and completeness of the detected clusters across the range of richness they exhibit while preserving the three-dimensional auto-correlation of clusters and their cross-correlation with the field galaxies. To achieve this, the number of mock clusters, visible cluster members ($\lambda$), redshifts, and angular positions of the mock clusters are directly derived from the sample of cluster detections identified in KiDS-DR4.

\begin{figure}
 \includegraphics[width=0.49\textwidth]{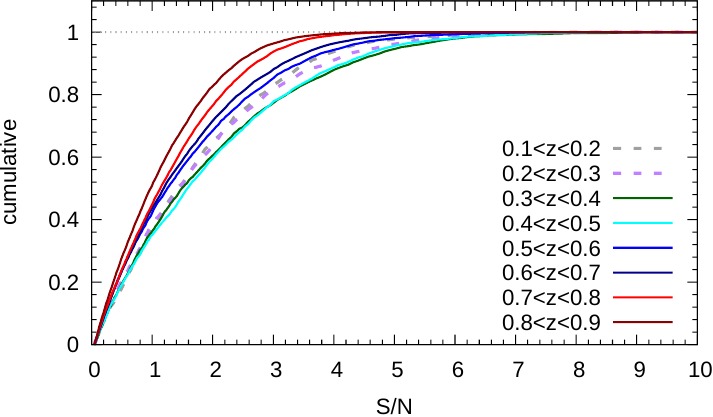}
  \caption{Cumulative distribution function of the number of detections as a function of the signal-to-noise ratio $S/N$, computed excluding the cluster shot noise contribution. Different colors represent distinct redshift intervals, as labelled. Solid and dashed lines refer to redshift intervals above and below $z\sim 0.3$, respectively, near which a noticeable transition in the photo-$z$ properties occurs, as shown in Fig.~\ref{fig:spec-z_calib}.}
  \label{fig:mockCDF}
\end{figure}

Since not all detections correspond to real clusters, particularly at lower signal-to-noise ratios, we first assess the probability of each detection being a true cluster: this will be used to perform the Monte Carlo extraction required to populate the mocks with clusters. To achieve this, we derive the cumulative distribution function (CDF) of detections, sorted according to a quality indicator: the signal-to-noise ratio provided by \textsc{AMICO}. The CDF is computed in distinct redshift intervals to account for potential redshift-dependent variations in the statistics. The resulting CDFs for all considered redshift intervals are shown in Fig.~\ref{fig:mockCDF}. A clear redshift-dependent trend is observed, with the CDFs shifting to higher values of $S/N$ at lower redshifts. However, this trend inverts below $z=0.3$, corresponding to the threshold where the properties of photometric redshifts change due to the transition of the $4000\AA$ break between the $g$ and $r$ bands.

The CDFs, one for each redshift interval $z_{bin}$, are then employed for the random sampling of detections, with higher signal-to-noise ratios corresponding to higher probabilities,
$P_{extraction} = CDF(z,S/N)$.
of being selected. The properties of the extracted detections - such as position, redshift, and richness - serve as the basis for defining the properties of the mock clusters. Conversely, detections that are not selected are labeled as spurious, and the probabilistic memberships of all galaxies associated with them are removed accordingly. At this stage, the probabilistic membership of each galaxy reflects the likelihood that the galaxy belongs to a specific detection, conditional on the probability of that detection being a true cluster. This approach integrates our best understanding of the sample while eliminating any ambiguity associated with applying a hard signal-to-noise ratio threshold. By considering all detections, including those with very low signal-to-noise ratio values (approaching zero), we ensure a comprehensive and unbiased treatment avoiding any arbitrary cut off in the sample selection.

\subsection{Realization of individual mock galaxy clusters}\label{sec:mockClusters}

In the previous step the list of the key properties of clusters and cluster members have been defined. Now we describe the generation of individual realizations of mock clusters by finalizing their final properties such as position, redshift, richness, actual members, and ellipsoidal shape.

The apparent richness $\lambda$ remains unchanged to ensure consistency with the probabilistic memberships used in the process, maintaining a statistically compatible number of galaxies (field and cluster members combined) with the original catalog.
Only the angular sky positions and redshifts are randomly perturbed, with displacements drawn from flat distributions: a maximum displacement of $\Delta R = 250$ kpc/h (consistent with the typical core radius of detected clusters) and a maximum of $\Delta z = 0.03$ (three times the redshift resolution of \textsc{AMICO}). These limits allow nearby clusters to nearly overlap without erasing their spatial correlations. The angular diameter distance used to calculate physical separations on the sky assumes the $\Lambda$CDM cosmology defined in the Introduction. The exact cosmology is not critical here since the perturbations serve only to slightly reshuffle cluster positions while preserving their auto-correlation and the correlation with the large scale structure. Throughout the whole process, survey masks are carefully preserved.
If the redshift reshuffling results in an empty redshift-richness bin, \textsc{SinFoniA} begins an alternate search, prioritizing redshift first and then richness, to find a bin populated with cluster members. If no populated bin is found after $5$ iterations per dimension, resulting in a total of $25$ attempts, the cluster is not generated. This occurred in only $9\%$ of mock tiles and for few clusters per mock tile.

\begin{figure}
  \includegraphics[width=0.49\textwidth]{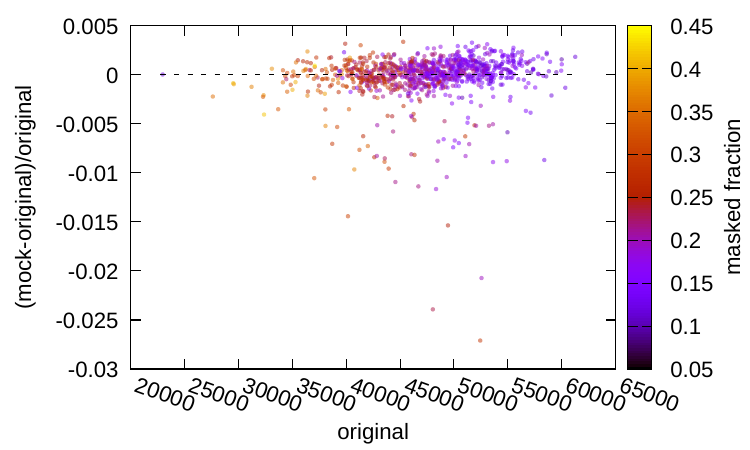}
  \caption{Deviation with respect to the original data set of the total number of galaxies (field and member galaxies combined) in each mock tile (including the surrounding buffer area), plotted against the corresponding number of galaxies in the original KiDS data. Colors represent the masked fraction of each tile.}
  \label{fig:mockNumberGal}
\end{figure}

\begin{figure*}
  \includegraphics[width=0.49\textwidth]{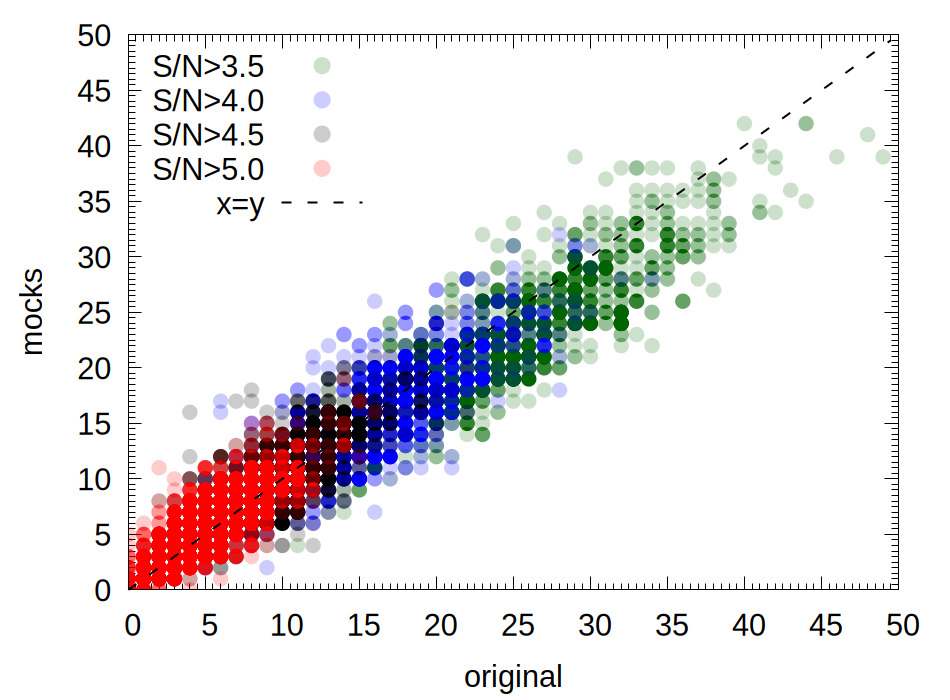}
  \includegraphics[width=0.49\textwidth]{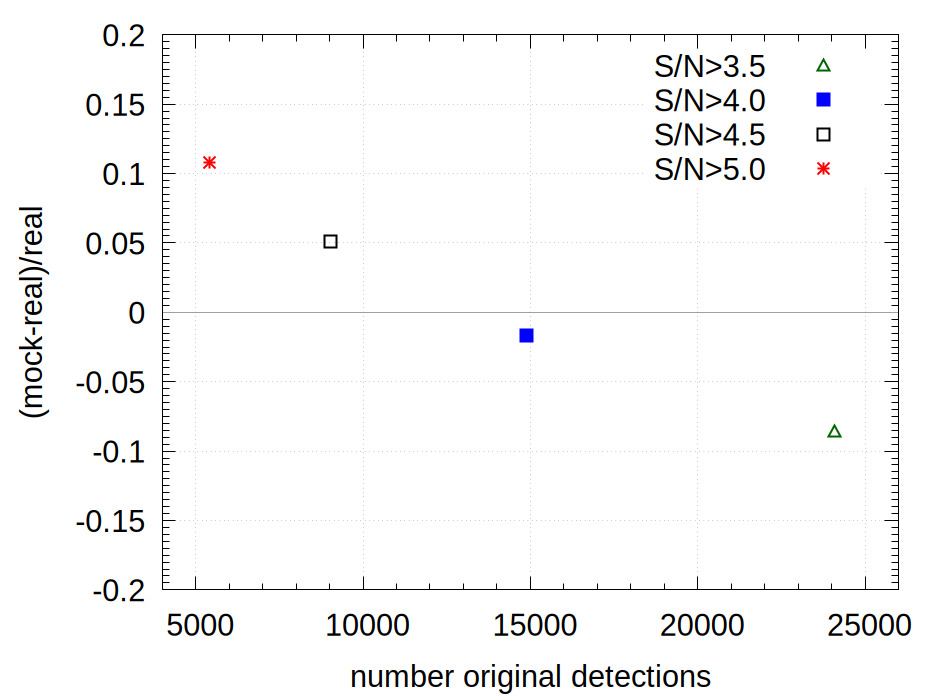}
  \caption{Comparison of the number of detections per tile (including the buffer area) in the original and in the mock catalogs. Colors refer to different signal-to-noise ratio thresholds. The left panel displays the absolute numbers, while the right panel shows the relative deviations.}
  \label{fig:mockNumberDet}
\end{figure*}

In Fig.~\ref{fig:mockNumberGal}, we show the deviation in the number of galaxies per tile between the original data set and the corresponding mocks. The fluctuations around zero are due to the Monte Carlo process used to extract both field galaxies and cluster members. The tiles with the largest negative deviations correspond to the cases mentioned above where clusters could not be generated. Overall, the maximum deviation is well below $1\%$, with only three tiles showing deviations exceeding $1.5\%$. The colors in the figure represent the masked fraction of each tile.

In addition to position and richness, we assign each cluster an ellipticity for the projected density of its members. This is necessary because mock members are generated by extracting them from stacks that comprise several cluster detections. These detections have uncorrelated intrinsic three-dimensional shapes and orientations. Consequently, if no asymmetry is imposed, the spatial distribution of members within a mock cluster would exhibit circular symmetry.
To address this, we apply a coordinate transformation to the members' sky coordinates using the following Jacobian matrix:
\begin{equation}
  \mathbf{J} = \begin{bmatrix}
    1-e\cos(2 \phi) & -e \sin(2\phi) \\
    -e \sin(2\phi)  & 1+e\cos(2 \phi)
  \end{bmatrix} \;.
\end{equation}
This transformation imparts an "ellipsoidal" shape to clusters by adding a trace-free component to a diagonal matrix, analogous to the Jacobian used in gravitational lensing with zero magnification. The position angle $\phi$ of the major axis is randomly selected, while the projected ellipticity, $e\equiv (a-b)/(a+b)$, is drawn from a distribution that mimics the properties of dark matter halos in N-body simulations \citep{despali17} \new{and are therefore not based on direct measurements from our observational data.} Here, $a$ and $b$ represent the major and minor axes of the ellipsoid, respectively.

\subsection{Mock field galaxies}

Field galaxies are randomly selected from the full KiDS-DR4 galaxy sample based on heir probability of not being associated with any mock cluster. The field probability for the $i$-th galaxy is calculated as $P^{\text{field}}_i = 1 - \sum_j P_i(j)$, where the index $j$ runs over all detections to which the galaxy is probabilistically associated. A higher value of $P^{\text{field}}_i$ corresponds to a larger probability for the galaxy to be extracted and classified as a field galaxy.

Unlike in our previous analysis of KiDS-DR3 data, in this study we do not perturb the angular positions of field galaxies, thus preserving their intrinsic clustering properties. As a result, the three-dimensional correlation of the noise is maintained throughout the process. The randomization of field galaxies naturally emerges from the Monte Carlo extraction based on galaxy memberships, and from the sampling of mock clusters discussed in Sect.~\ref{sec:CDF}, which influences the field galaxy population as well.

Note that each galaxy can only be extracted once during the generation of field galaxies. However, it can, in principle, be extracted again during the generation of mock clusters.
In any case, the galaxy's final location, when extracted as cluster members, will differ from its original position, as it will depend on the location of the mock cluster it is assigned to, along with the additional spatial angular randomization applied within each mock cluster.

\begin{figure*}
  \includegraphics[width=0.33\textwidth]{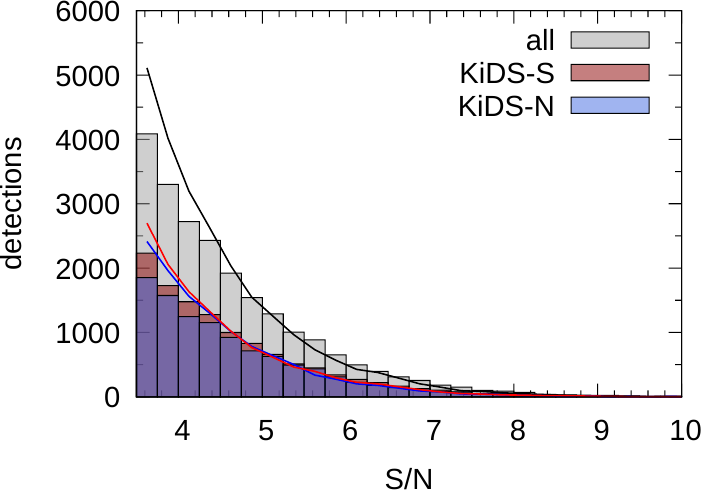}
  \includegraphics[width=0.33\textwidth]{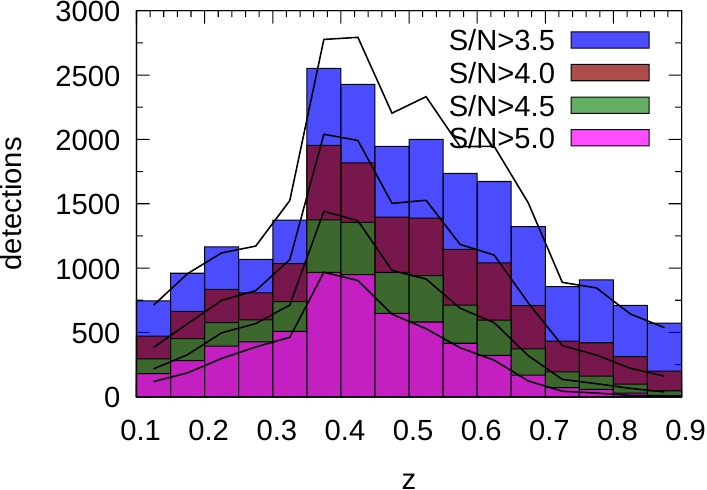}
  \includegraphics[width=0.33\textwidth]{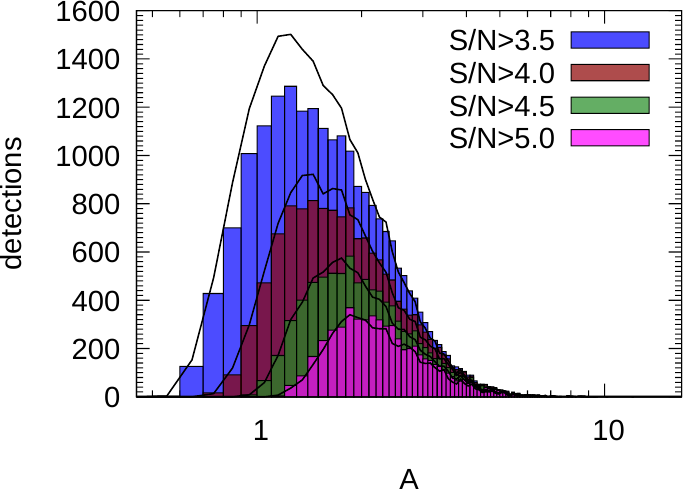}\\
  ~\\
  \includegraphics[width=0.34\textwidth]{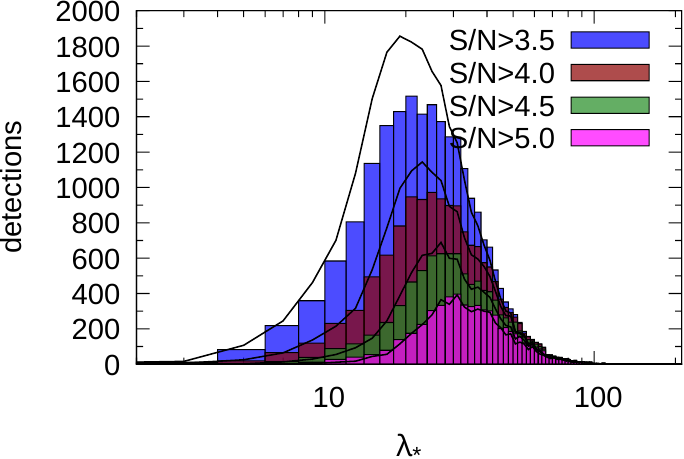}
  \includegraphics[width=0.32\textwidth]{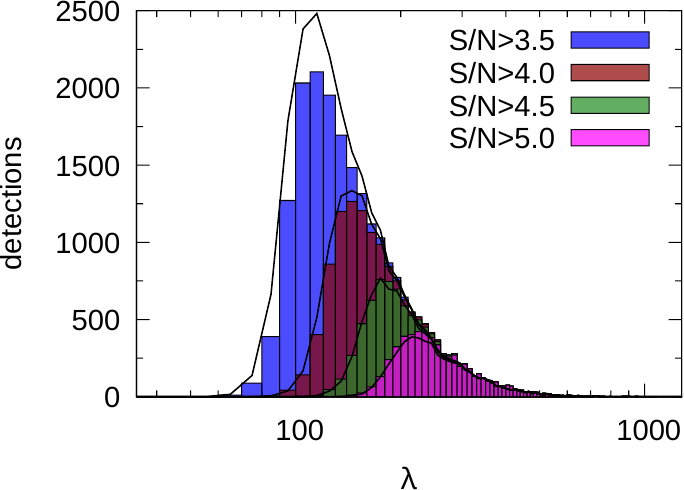}
  \includegraphics[width=0.33\textwidth]{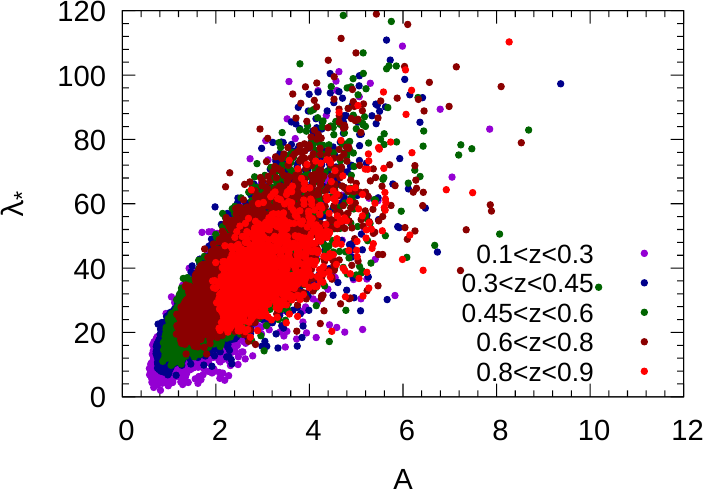}
  \caption{
  Comparison of the key properties computed in the original \textsc{AMICO}-KiDS-DR4 sample (solid lines) and in the mock sample (solid bars).
  Different panels show the distributions of signal-to-noise ratio for the entire sample and for both KiDS-N and KiDS-S stripes (top left), redshift $z$ (top center), amplitude $A$ (top right), intrinsic richness $\lambda_*$ (bottom left), and apparent richness $\lambda$ (bottom center). The bottom right panel shows the correlation between $A$ and $\lambda_*$ in the mocks. }
  \label{fig:mockStat}
\end{figure*}

\subsection{Detection of clusters in the mock data}

To evaluate sample purity, completeness, and the scatter of all observable quantities, we ran \textsc{AMICO} on the mock galaxy catalog using the exact same methodology as for the real data. The left panel of Fig.~\ref{fig:mockNumberDet} shows the number of detections per tile for real data and mocks, colored according to their signal-to-noise ratio ($S/N$),  while the right panel displays the relative deviation between mocks and real data. There is a $6\%$ excess in the number of detections with high $S/N$ values ($S/N>5.0$) in the mocks compared to the real data. At lower $S/N$ values, the number of detections in the mock data is lower than in the real data, with a maximum deviation of $12\%$ for detections in the redshift range $0.35<z<0.7$. This pattern suggests that larger mock clusters are somewhat easier to detect, whereas the smaller ones are more challenging with respect to those in the real data. Since spurious detections are largely unaffected due to the scheme described in Sect.~\ref{sec:CDF}, this deviation likely originates from the injected mock clusters. In fact, their generation is influenced by spurious detections and by the overestimation of the radial profile of the $P(z)$ used to extract their members (the probabilistic membership relies on a cluster model assuming a relatively large mass, $10^{14}\,M_{\odot}/h$), resulting in clusters with a spatial density of galaxies lower than expected.

Despite these deviations, the distribution of counts across all observable quantities is well reproduced, and the full range of redshifts and richness values is adequately covered. This is made evident in Fig.~\ref{fig:mockStat}, where we show the distribution of detections as a function of several parameters: signal-to-noise ratio for both KiDS-N and KiDS-S stripes, redshift, amplitude, intrinsic richness, and apparent richness. The solid lines represent the distributions for the real data. For completeness, we also show the correlation between $A$ and $\lambda_*$, which again well matches the one observed in the real data.

It is important to note that cluster catalogs are typically truncated at $S/N$ thresholds where sample purity and completeness are sufficiently high. For instance, cosmological samples often require high purity (e.g., $P>95\%$, corresponding to $S/N \sim 4.5$). Therefore, in this regime, the mock catalogs effectively resample the data while preserving the expected total number of detections, ensuring consistency with the real data.

\subsection{Estimate of the uncertainty of the \textsc{AMICO} observables}\label{sec:mock-matching}

We compare now the properties of the detections in the mocks with the corresponding true values to evaluate the scatter of the observables, as well as the sample purity and completeness.

\begin{figure*}
  \includegraphics[width=0.49\textwidth]{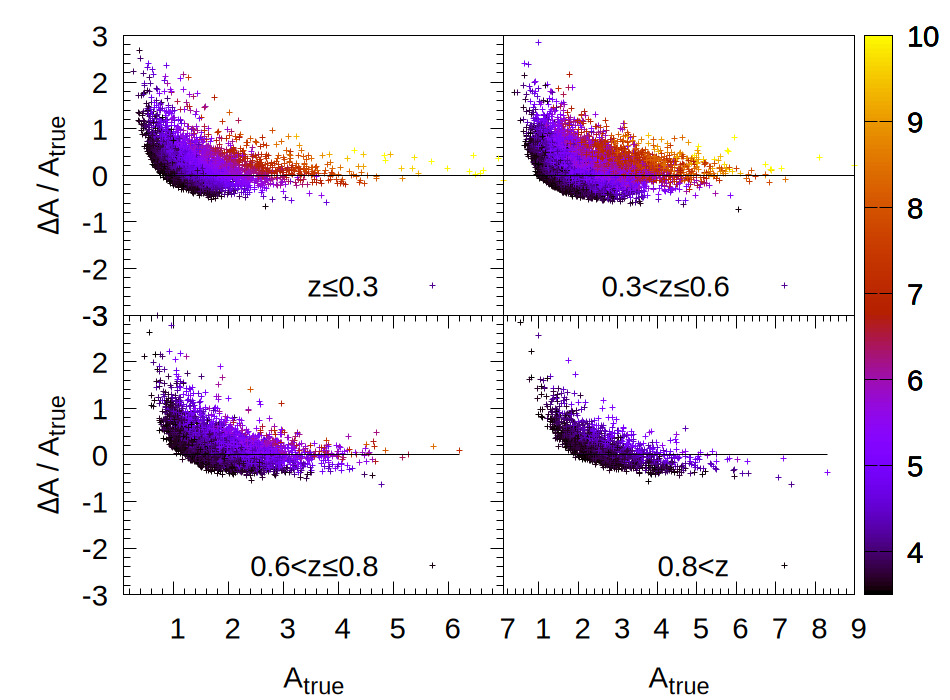}
  \includegraphics[width=0.49\textwidth]{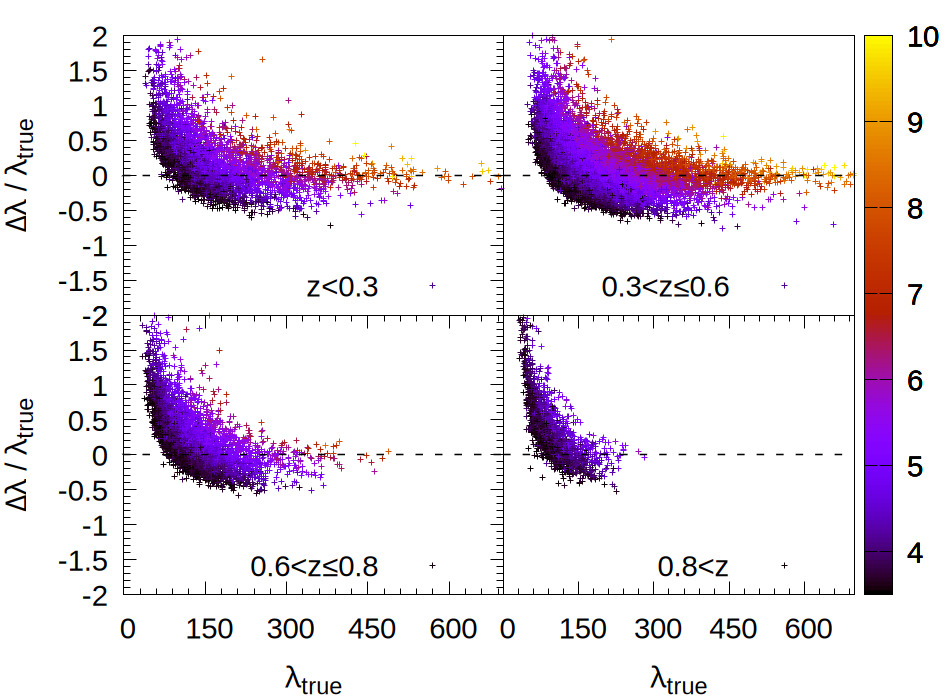}\\
  ~\\
  \includegraphics[width=0.49\textwidth]{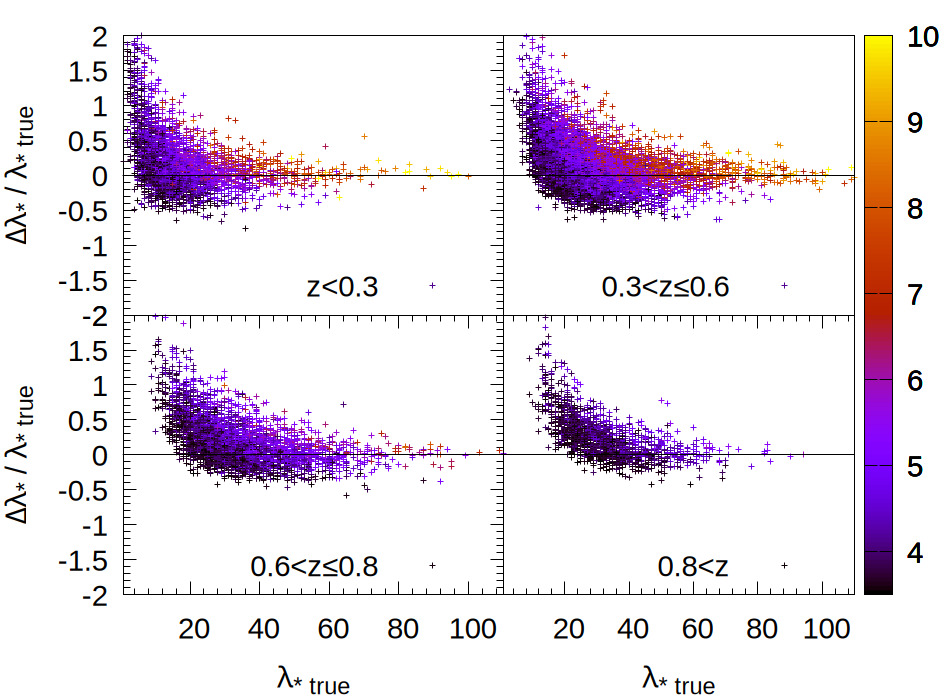}
  \includegraphics[width=0.49\textwidth]{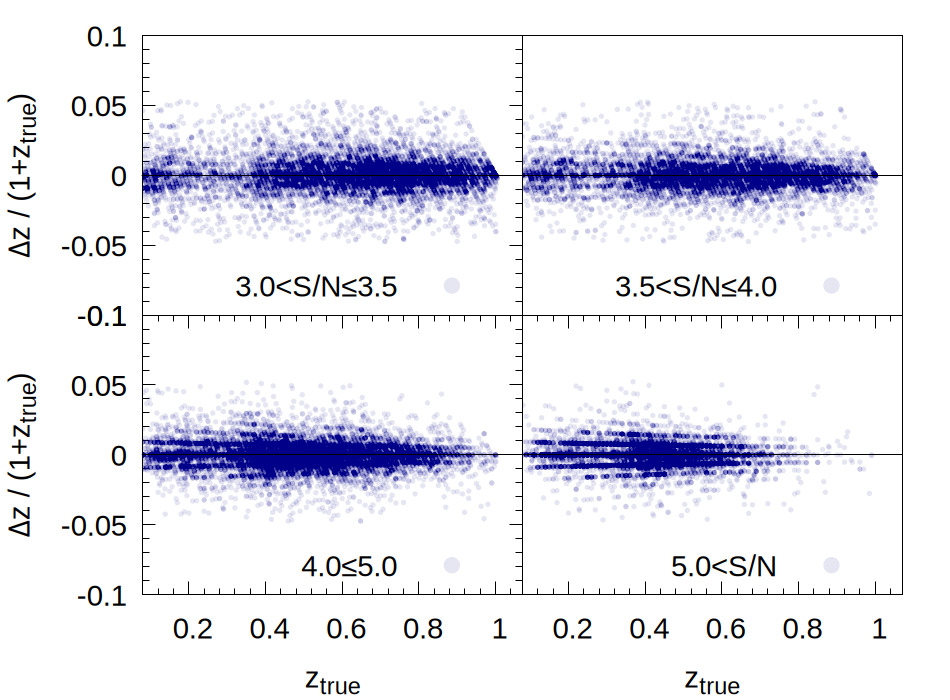}
  \caption{Scatter plots comparing the mock measured versus mock true values (e.g. $\Delta \lambda = \lambda_{obs}-\lambda_{true}$) of amplitude $A$ (top left panels), apparent richness $\lambda$ (top right panels), intrinsic richness $\lambda_*$ (bottom left panels) and redshift $z$ (bottom right panels). Each analysis is presented for 4 different bins in redshift, as labeled in each sub-panels. The colorbars refer to different values of $S/N$.
  }
  \label{fig:mockErrors}
\end{figure*}

To achieve this, we perform a three-dimensional matching between detections and mock clusters. The matching criteria are an angular distance smaller than $\Delta R = 0.5$ Mpc/h and a redshift difference of $\Delta z = 0.05(1+z)$. Additionally, we sort the detections and mock clusters in descending order of apparent richness, prioritizing larger detections for matching with larger clusters. This strategy minimizes confusion with the more abundant low-richness objects and effectively implements an implicit "soft" matching based on richness, following \cite{maturi19}. The scatter in redshift, amplitude, intrinsic richness and apparent richness, is presented in the different panels of Fig.~\ref{fig:mockErrors}. \new{For the lowest values of $A$, $\lambda$, and $\lambda_*$, the sample selection bias, typically affecting samples in the low signal-to-noise regime (where preferentially noise-scattered high estimates are selected), is evident. Where the sample is complete,} the intrinsic richness $\lambda_*$ appears unbiased across the full sample, \new{confirming the overall good behavior of the estimates.} The apparent richness $\lambda$ shows instead a slight negative bias at redshifts $z>0.6$, while the amplitude $A$ exhibits a trend transitioning from a slight positive to a slight negative bias. In the redshift range $0.6<z<0.8$, the amplitude bias is negligible. The Malmquist bias, which is caused by the sample selection and not by the detection algorithm, affects estimates close to the detection limit. As expected, all bias contributions are smaller for detections with higher signal-to-noise ratios, where the measurements are inherently more robust. Note that no intrinsic redshift bias can be measured within this approach because the redshifts of the mock clusters are directly based on the original photo-$z$s and therefore the scatter is inherently defined in the photo-$z$ space. To estimate the bias, an independent reference redshift catalog, such as the GAMA spectroscopic redshifts discussed in Sect.~\ref{sec:gama}, is required.

\begin{figure*}
  \includegraphics[width=0.32\textwidth]{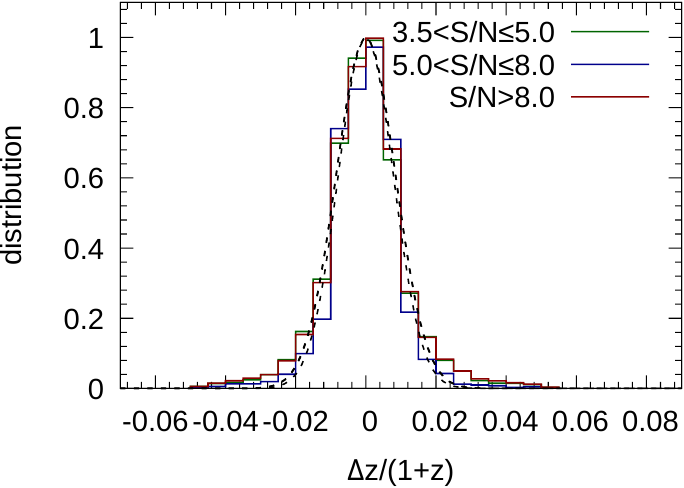}
  \includegraphics[width=0.32\textwidth]{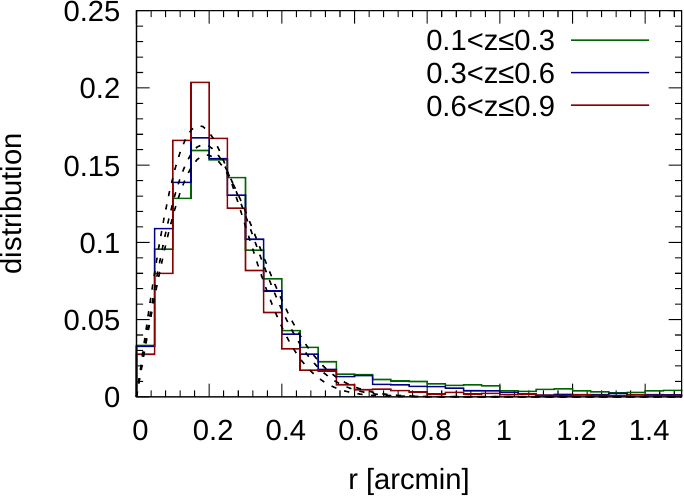}
  \includegraphics[width=0.33\textwidth]{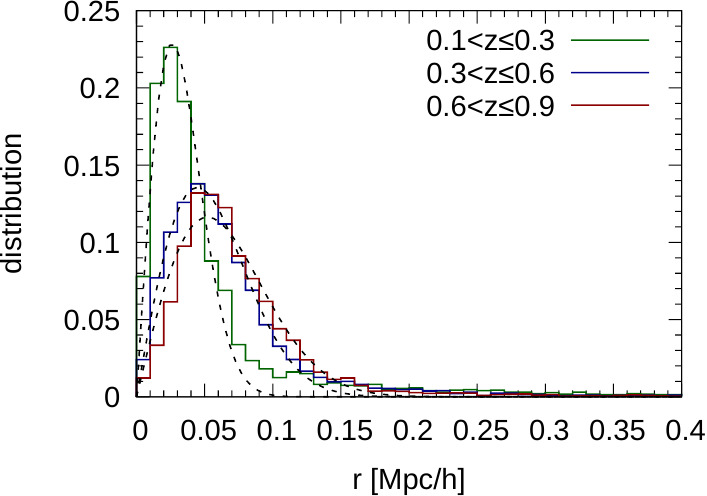}
  \caption{Scatter between the observed and true values for redshift in three different bins of $S/N$ (left panel), and for angular and physical separations (central and right panels, respectively) in three different redshift intervals. The dashed lines represent the corresponding best-fit Gaussian and Rayleigh distributions for redshift and radius, respectively.}
  \label{fig:mockMiscentering}
\end{figure*}

In the left panel of Fig.~\ref{fig:mockMiscentering}, we show the scatter in redshift for three different $S/N$ bins: the scatter turns out not to depend on the detection significance, with $\sigma_z/(1+z) \sim 0.008$. Comparing this result with the spectroscopic measurements given in Sect.~\ref{sec:gama},it appears that this uncertainty is slightly underestimated, by the same amount as those based on the detections $P_{det}(z)$ which directly rely on the photo-$z$. Again, this might be due to an underestimation of the photo-$z$ uncertainties of the input galaxies. Even if the difference is not substantial, we recommend the use of the error estimates based on the spec-$z$ as they rely on independent data.

The central (right) panel of the same figure displays, for three different redshift intervals, the scatter due to angular (physical) spatial miscentering. The angular displacement is largely determined by the angular resolution of the \textsc{AMICO} sky sampling adopted in this study, \new{which is $0.3$ arcmin,} and shows a minimal dependence on the cluster redshift. The scatter in angular displacement is $\sigma_r = 0.19$, $0.18$, and $0.17$ arcmin ($0.026$, $0.045$, and $0.051$ Mpc/h) for detections in the redshift intervals $0.1<z<0.3$, $0.3<z<0.6$, and $0.6<z<0.9$, respectively. The dashed lines represent the best-fit Rayleigh distributions for these scatters. Erroneously matched mock clusters contribute to extending the tail visible in these distributions. To better understand and interpret these results, we recall that the mock clusters are modeled with ellipsoidal shapes, lacking any substructure or irregularity. As a result, their centers are better defined compared to real clusters which can have more complex morphologies. Consequently, the positional uncertainties derived here reflect purely algorithmic errors rather than incorporating these additional complexities present in real clusters and, more importantly, the intrinsic ambiguity in the definition of their center.

\begin{figure*}
  \includegraphics[width=1.0\textwidth]{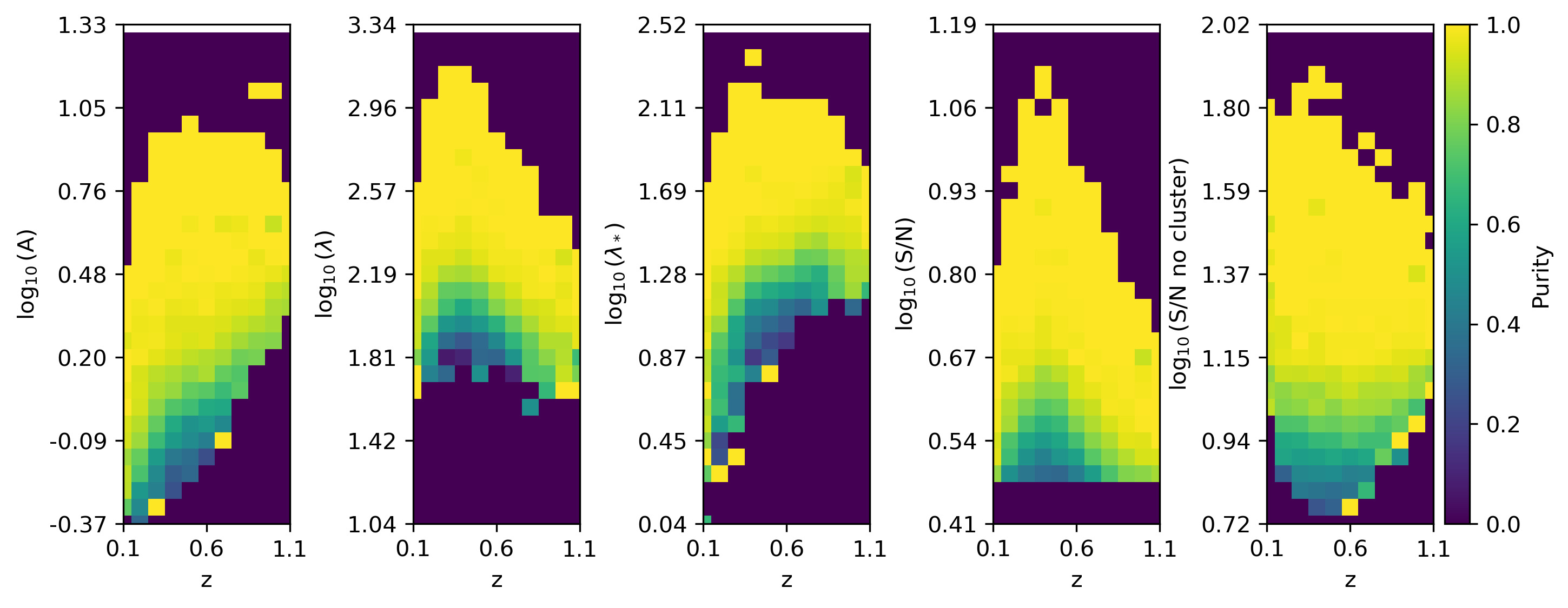}
  \caption{Sample purity derived from the \textsc{SinFoniA} mocks as a function of true redshift and various observables:  the different panels refer to amplitude $A$, intrinsic richness $\lambda_*$, observed richness $\lambda$, signal-to-noise ratio $S/N$, and signal-to-noise ratio excluding the shot noise contribution from clusters, from left to right.}
  \label{fig:mockPurity}
\end{figure*}

\subsection{Sample purity and selection function}\label{sec:pur-sel}

The sample purity is defined as the ratio between the detections with a matched counterpart in the mocks and the total number of detections. This ratio is evaluated in bins of true redshift and different observed properties of the mock clusters. The results are shown in Fig.~\ref{fig:mockPurity} for amplitude $A$, intrinsic richness $\lambda_*$, apparent richness $\lambda$, signal-to-noise ratio $S/N$, and for an alternative definition of the signal-to-noise ratio that excludes the shot noise contribution from cluster members (\texttt{SN\_NO\_CLUSTER}). For a given fixed value of purity, all observables exhibit a redshift dependence except for \texttt{SN\_NO\_CLUSTER}. A few isolated points at low observable values, with apparent $100\%$ purity (see the panels referring to $\lambda_*$ and \texttt{SN\_NO\_CLUSTER}), result from random matches of single detections with mock clusters and should be disregarded. The choice of observable used for selection clearly influences the resulting sample. The choice should be based on the science goal at hand.

The completeness is defined as the ratio between the detections with a matched counterpart in the mocks and the total number of clusters in the mocks. This ratio is evaluated in bins of true redshift and expected input cluster properties. Completeness is a critical quantity that \new{was initially} blinded to preserve the integrity of the cosmological analysis based on this cluster sample \citep{lesci25}. Blinding safeguards against subjective influences and ensures the robustness of the derived cosmological constraints. Details on the blinding procedure are provided in Appendix~\ref{sec:blinding}, and the corresponding selection functions based on $\lambda_*$ for the three blind realizations are shown in Fig.\ref{fig:blind_completeness}. \new{Now that the cosmological analysis has been completed \citep{lesci25}, the unblinding has been carried out, and the true selection function, displayed in the central panel of Fig.\ref{fig:blind_completeness}, is publicly available.}

\section{Matching with external catalogs of clusters}\label{sec:match}

We compared our cluster sample, restricted to $S/N>3.5$, with three external cluster catalogs: the Dark Energy Survey catalog created with RedMaPPer \citep{rykoff16}, the eRASS1 X-ray cluster sample produced by the extended ROentgen Survey with an Imaging Telescope Array \citep[eROSITA,][]{bulbul24}, and the cluster sample identified via SZ effect in the ACTpol data \citep{hilton20}. The comparison is performed over the effective area covered by the KiDS-DR4 data. To match AMICO detections with clusters in the external catalogs, we employed the same geometric matching procedure introduced in Sect.~\ref{sec:mock-matching}. This method includes a preliminary sorting step based on the detection significance to ensure that the most significant detections are prioritized. In this case, a maximum redshift discrepancy of $\Delta z = 0.1(1+z)$ and a spatial separation of $0.5\,\mathrm{Mpc}/h$ are used as matching criteria.

\subsection{RedMaPPer cluster catalog}

\begin{figure*}
\centering
  \includegraphics[width=0.31\textwidth]{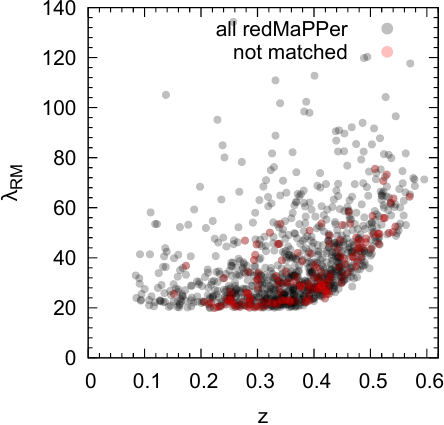}  
  \includegraphics[width=0.325\textwidth]{
  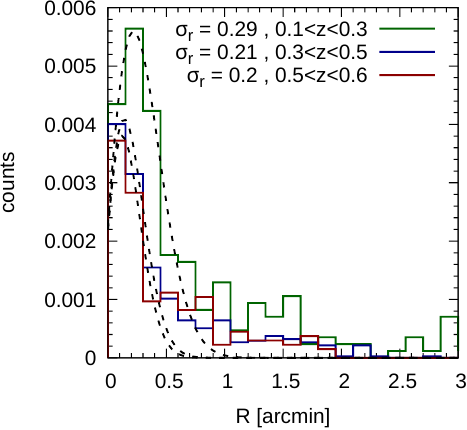}
  \includegraphics[width=0.303\textwidth]{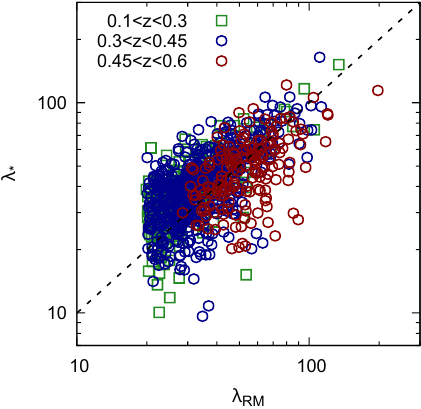}
  \caption{Left panel: the distribution in the richness-redshift plane of all RedMaPPer detections present in the KiDS effective area. Objects with no \textsc{AMICO} counterpart are displayed in red.
  Central panel: the distribution of the angular displacement between the centers of the matched detections between the RedMaPPer and \textsc{AMICO} samples for three different redshift intervals. The best-fit Rayleigh distributions are shown as dashed lines; the values of their corresponding $\sigma_r$ are also reported in the legend. Right panel: comparison between the RedMaPPer richness $\lambda_{RM}$ and the \textsc{AMICO} intrinsic richness $\lambda_*$, for the matched objects. Data are color-coded according to three different redshift intervals. The dashed line indicates the equality line.}
  \label{fig:redmapper_match}
\end{figure*}

In this section we compare our catalog with the cluster sample identified using the redMaPPer algorithm, which exploits the red sequence of galaxy clusters to detect them in photometric data \citep{Rykoff14}. This sample spans both the Dark Energy Survey (DES) Science Verification (SV) and Sloan Digital Sky Survey (SDSS) DR8 photometric data sets, comprising a total of $1055$ entries within the KiDS-DR4 effective area in the redshift range $0.1<z<0.6$. Using the three-dimensional matching approach outlined earlier, we achieve $902$ (i.e. $88\%$) positive matches. The redMaPPer detections with no \textsc{AMICO} counterpart are highlighted in the left panel of Fig.~\ref{fig:redmapper_match} as the red points.

The central panel of the same figure shows the angular displacement between the RedMaPPer and \textsc{AMICO} detections for three distinct redshift intervals. The distributions feature a $\sigma$ value for a Rayleigh distribution ranging from 0.29 (at high redshift) to 0.2 arcmin (at low redshift). A tail extending toward larger separations, which is particularly noticeable at lower redshifts, can be observed. This extended tail is likely a result of a different definition adopted for the cluster center: \textsc{AMICO} centers are determined using the spatial distribution of all member galaxies, whereas RedMaPPer centers are defined by the location of the candidate brightest central galaxy (BCG). A detailed analysis of the BCGs identified in our sample will be provided in a dedicated paper \citep{radovich25}.

In the right panel of Fig.~\ref{fig:redmapper_match}, we compare the richness estimates provided by the two detection algorithms. The results exhibit a strong correlation across the entire redshift range probed by RedMaPPer and agree well despite their different definitions.  This consistency arises because the \textsc{AMICO} intrinsic richness $\lambda_*$ primarily accounts for the brightest member galaxies, which are predominantly the red elliptical galaxies targeted by RedMaPPer.

\begin{figure*}
\centering
  \includegraphics[width=0.322\textwidth]{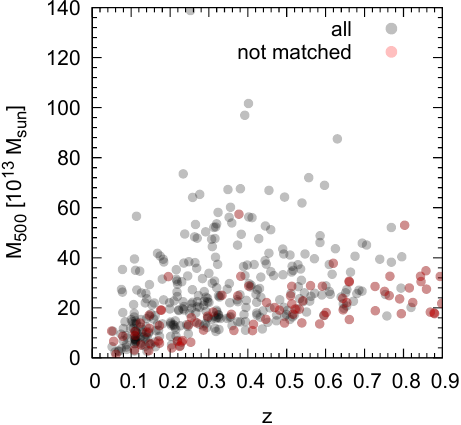}  
  \includegraphics[width=0.325\textwidth]{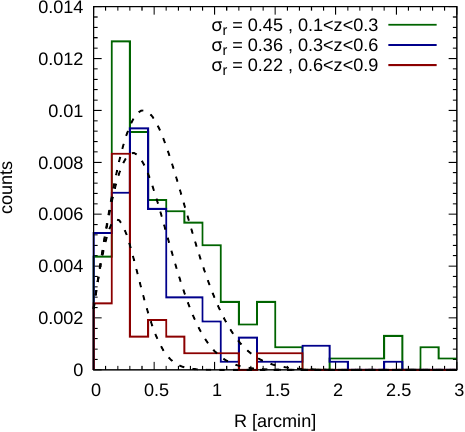}
  \includegraphics[width=0.305\textwidth]{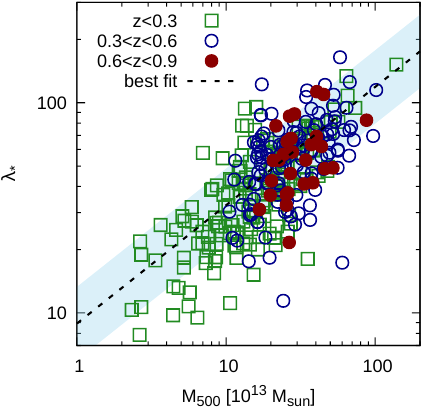}
  \caption{Left panel: the distribution in the mass-redshift plane of all eRASS1 primary clusters in the KiDS effective area. Objects with no \textsc{AMICO} counterpart are displayed in red.
    Central panel: the distribution of the angular displacement between the centers of the matched detections between the eRASS1 primary and \textsc{AMICO} samples for three different redshift intervals. The best-fit Rayleigh distributions are shown as dashed lines; the values of their corresponding $\sigma_r$ are also reported in the legend. Right panel: relation between the \textsc{AMICO} intrinsic richness $\lambda_*$ and the eRASS1 X-ray derived mass $M_{500}$, for the matched objects. Data are color-coded according to three different redshift intervals. \new{The dashed line indicates the best-fit scaling relation with $\gamma=0$, and the shaded region indicates the R.M.S of the points around the best-fit model.}}
  \label{fig:eRASS1_dr}
\end{figure*}

\begin{figure*}
\centering
  \includegraphics[width=0.322\textwidth]{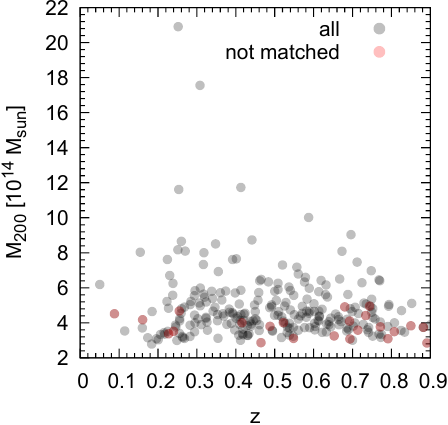}
  \includegraphics[width=0.325\textwidth]{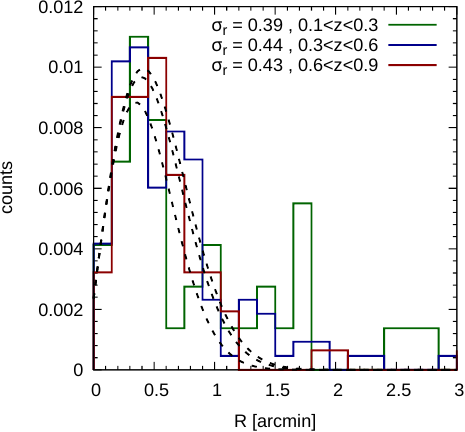}
  \includegraphics[width=0.305\textwidth]{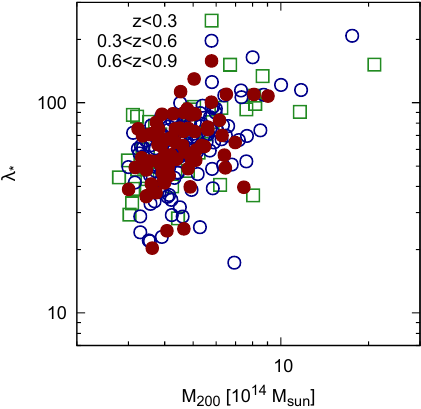}
  \caption{Left panel: the distribution in the mass-redshift plane of all ACT SZ clusters in the KiDS effective area. Objects with no \textsc{AMICO} counterpart are displayed in red.
  Central panel: the distribution of the angular displacement between the centers of the matched detections between the ACT and \textsc{AMICO} samples for three different redshift intervals. The best-fit Rayleigh distributions are shown as dashed lines; the values of their corresponding $\sigma_r$ are also reported in the legend. Right panel: 
  relation between the \textsc{AMICO} intrinsic richness $\lambda_*$ and the ACT SZ-derived mass $M_{200}$, for the matched objects. Data are color-coded according to three different redshift intervals.}
  \label{fig:act_match}
\end{figure*}

\new{To better understand the origin of the mismatch between RedMaPPer and \textsc{AMICO} detections at redshifts $z<0.6$, we visually inspected KiDS $gri$ color-composite images of the 30 richest \textsc{AMICO} detections that lack an apparent RedMaPPer counterpart. These detections have a minimum richness of $\lambda_* \ge 63.9$. Our inspection revealed the following cases:
\begin{itemize}

\item Six RedMaPPer detections are located at nearly identical sky positions to \textsc{AMICO} detections but were not automatically matched due to redshift discrepancies that slighlty exceed the adopted matching tolerance.

\item Four \textsc{AMICO} detections have RedMaPPer counterparts with consistent redshifts, but angular positions that are too distant to satisfy the matching criteria.
This mismatch arises from the differing centering strategies of the two algorithms: RedMaPPer aims to identify the central dominant galaxy, but in these cases, it selects a bright, isolated galaxy that is offset from the main galaxy overdensity traced by \textsc{AMICO}.

\item Fourteen detections exhibit a clear overdensity of elliptical galaxies in the KiDS color images but lack a nearby RedMaPPer detection. All are located at $z>0.35$, and nearly half (six) lie at $z>0.55$, close to the upper redshift limit of the RedMaPPer catalog.
  
\item Two detections show a galaxy overdensity, but not as prominent as the previous ones. In one case, missing photometric bands prevented the creation of a proper color composite image making visual inspection difficult.
 
\item Four detections are clearly caused by imaging artifacts. However, they fall within those faulty areas that we manually masked during post-processing and are therefore excluded from our sample (see Sect.\ref{sec:application}).
\end{itemize}

We also performed a visual inspection of the 30 RedMaPPer detections without an \textsc{AMICO} counterpart. Our analysis revealed the following:
\begin{itemize}
\item Two RedMaPPer detections have a corresponding \textsc{AMICO} counterpart with consistent sky position but a redshift difference slightly exceeding the matching threshold.

\item Nine clusters have \textsc{AMICO} counterparts at similar redshift but with a larger angular separations up to $2.5'$. As in the previous analysis, this is due to the different definition of the cluster centers of the two codes. 

\item Three detections correspond to asymmetric clusters composed by two galaxy clumps. In these cases, RedMaPPer and \textsc{AMICO} appear to select different components of the same overall system.

\item Five detections lie within large masked regions of the KiDS data where \textsc{AMICO} does not search for clusters. An additional one falls in a tile rejected due to significant image artifacts.

\item Three RedMaPPer detections do not correspond to obvious galaxy overdensities in the visual inspection. One is centered on a galaxy extremely close to a bright star (RMJ144349.2+011943.0), another lacks any evident overdensity (RMJ084122.4+020707.5), and the third lies in between three \textsc{AMICO} detections, one of which has the same redshift, suggesting possible miscentering issues (RMJ084122.4+020707.5).

\item Eight detections show no \textsc{AMICO} counterpart. Except in one case, the RedMaPPer detections are not that evident in the images but some galaxy overdensity is visible.
\end{itemize}
In conclusion, the majority of unmatched RedMaPPer detections can be attributed to differences in centering strategies, redshift mismatches near the matching threshold, or the presence of masks and artifacts in the KiDS data which are excluded in our analysis. Only a minority lacks a plausible \textsc{AMICO} counterpart entirely, and in many of those cases, image quality or morphological complexity likely played a role. This analysis reinforces the importance of understanding algorithmic differences when comparing cluster catalogs and accurate sample purity and completeness estimates.
}

\subsection{eRASS1 cluster catalog}

The eRASS1 cluster samples \new{\citep{bulbul24}} were derived from the first six months of operations of the eROSITA X-ray space telescope \new{\citep{predehl21}}, as part of the extended ROentgen Survey with an Imaging Telescope Array (eRASS1), and processed by the German eROSITA Consortium \new{\citep[eROSITA-DE,][]{merloni24}}. eROSITA operates in the energy range 0.2 to 8 keV, focusing on a uniform all-sky flux limit of $F_{0.5-2\,keV} > 5 \times 10^{-14}$ erg s$^{-1}$ cm$^{-2}$. Cluster candidates were primarily identified in the $0.2-2.3$ keV band, which is the eROSITA's most sensitive spectral range. These X-ray detections are limited to half of the sky that is accessible to the eROSITA-DE Consortium: this overlaps nearly all of the KiDS-N stripe and approximately $65\%$ of the KiDS-S stripe. The missing parts fall within the sky area managed by the Russian eROSITA-RU Consortium or in the Galactic equatorial stripe, which was excluded from the X-ray analysis. Two main cluster samples were produced: the "cosmological" and "primary" catalogs.

The "cosmological" sample of eRASS1 contains $5\,258$ detections, of which $224$ overlap the effective area of KiDS-DR4. Using the automated matching criteria described above, $91\%$ of these X-ray detections (corresponding to $204$ objects) were successfully matched to an \textsc{AMICO} cluster. For the remaining X-ray detections without an AMICO counterpart, a visual inspection of KiDS $gri$ color-composite images reveals the following:
\begin{itemize}
\item three detections correspond to regions without any evident overdensity of galaxies, with one X-ray detection possibly caused by the presence of an AGN.
\item two detections contain two distinct visible galaxy overdensities. In both cases, one of the two overdensities is very near the X-ray detection center, but \textsc{AMICO} assigned as the center the other overdensity, located outside the matching radius. Increasing the matching radius to $1 Mpc/h$ would have resolved these mismatches.
\item two detections have clear \textsc{AMICO} counterparts just outside the matching radius.
\item six detections are near the edges of large  regions which are masked because of the presence of bright stars and star halos. For this reason, they are not found by \textsc{AMICO}.
\item seven detections correspond to low-redshift clusters ($0.12<z<0.17$) where the poor photo-$z$ quality led to missed identifications by \textsc{AMICO}.
\end{itemize}
This analysis emphasizes the high correspondence between the X-ray-based eRASS1 clusters and the optically detected \textsc{AMICO} clusters, particularly at higher redshifts, where the photo-$z$ accuracy improves.

We also performed the matching with the eRASS1 'primary' cluster sample, which comprises $12\,247$ clusters. Of these, $10\,020$ objects lie within the redshift interval $0.05<z<1$ and have a non-zero X-ray luminosity. Similarly to the "cosmological" sample, we selected only objects inside the KiDS-DR4 effective area, yielding $409$ clusters. Of these, $321$, i.e. $78\%$, are matched to \textsc{AMICO} detections.

The key properties of the matched clusters (including those visually inspected mentioned above) are shown in Fig.~\ref{fig:eRASS1_dr}. In the left panel we show the matched (gray) and not matched (red) eRASS1 detections falling in the KiDS-DR4 area as a function of redshift and mass. The angular scatter between the centers of the X-ray and \textsc{AMICO} detections, shown in the central panel, decreases with increasing redshift: the values of the RMS are
$\sigma_r=0.45$ arcmin for $0.1<z<0.3$,
$\sigma_r=0.36$ arcmin for $0.3<z<0.6$ and
$\sigma_r=0.22$ arcmin for $0.6<z<0.9$.

The right panel of Fig.~\ref{fig:eRASS1_dr} displays the relationships between the \textsc{AMICO} intrinsic richness $\lambda_*$ and the X-ray derived mass $M_{500}$\footnote{Here, $M_{500}$ is the mass
included inside a radius within which the average density of a halo is 500 times the critical density of the universe.}, showing the existence of a strong correlation at all redshifts. Using orthogonal distance regression, we fit a mass–proxy scaling relation \cite[see also][]{bellagamba18b,lesci22}
\begin{equation}\label{eq:scaling-no-redshift}
  \log_{10}\left(\frac{M}{10^{13}M_\odot}\right) = \alpha
  + \beta \log_{10}\left(\frac{\lambda_*}{\lambda_{*piv}}\right)
  + \gamma \log_{10}\left(\frac{E(z)}{E(z_{piv})}\right) \;.
\end{equation}
Here, $E(z)\equiv H(z)/H_0$ is computed assuming a flat $\Lambda$CDM model with $\Omega_\Lambda=0.7$, $H_0=70$ km/s/Mpc and we set $\lambda_{*piv}=30$ as well as $z_{piv}=0.35$. Assuming no redshift dependence, i.e. $\gamma=0$, the best-fit parameters are \new{$\alpha = 0.94 \pm 0.03$ and $\beta = 1.78 \pm 0.10$}, while leaving $\gamma$ free we obtain $\alpha=0.91 \pm 0.04$, $\beta = 1.85 \pm 0.12$ and $\gamma = -0.7 \pm 0.4$. \new{The fit accounts for the uncertainties in mass, $\lambda_*$, and redshift. In both cases, the R.M.S. around the best-fitting model is $0.31$ dex.}
The mass estimates derived using this scaling relation are reported in the final catalog, and are also shown in the right panel of Fig.~\ref{fig:eRASS1_dr}. \new{For simplicity of representation, we plot the scaling relation without accounting for the mild redshift evolution.} Notice that here we present the results for the ``primary'' eRASS1 sample, but for the "cosmological" eRASS1 sample, which is restricted to higher masses, the amount of scatter for all shown quantities is very similar. Mass scaling relations derived from a subset of clusters using KiDS weak gravitational lensing data is presented by \cite{lesci25b}, where both the cosmological parameters and the scaling relation parameters are simultaneously fitted.

\subsection{Atacama Cosmology Telescope DR5 cluster sample}

The Atacama Cosmology Telescope (ACT) is a single-dish millimetric telescope located in Chile \new{\citep{swetz11,thornton16}}. The SZ-detected cluster sample was obtained using the Advanced ACTPol receiver mounted on ACT, employing the $98$ and $150$ GHz channels, which provide beam FWHM values of $2.2$ arcmin and $1.4$ arcmin, respectively. The detections were identified using a multi-channel optimal matched filter, yielding a sample of $4\,195$ entries having an optical counterpart and a signal-to-noise ratio  $S/N>4$. This sample covers an area of $13\,211$ deg$^2$ and spans the redshift range $0.04<z<1.91$ \citep{hilton20}. Among these SZ cluster detections, $267$ fall within the KiDS-DR4 effective area and are in the redshift range $0.1<z<0.9$ covered by our \textsc{AMICO} catalog. 
Of these SZ detections, $235$ have an \textsc{AMICO} counterpart, corresponding to a matching fraction of $88\%$. Considering separate redshift intervals $0.1<z<0.3$, $0.3<z<0.7$, and $0.7<z<0.9$, the recovery fractions are $83\%$, $92\%$, and $78\%$, respectively. 
By relaxing the angular constraints from $0.5$ Mpc/h to $1$ Mpc/h, the number of matched clusters increases to $244$, resulting in a recovery fraction of $91\%$ over the entire redshift range $0.1<z<0.9$, and $92\%$, $95\%$, and $80\%$ for the intervals $0.1<z<0.3$, $0.1<z<0.7$, and $0.7<z<0.9$, respectively.  \new{The reduction in sensitivity in the highest redshift bin is primarily due to the limited number of galaxies observable beyond $z > 0.8$, where the sample depth reaches $m_* + 1.5$ (see Sect.~\ref{sec:application}). Despite the reduced sensitivity, AMICO remains effective at detecting significant structures thanks to its optimal filtering approach, which assigns higher weights to intrinsically brighter galaxies, such as the brightest group galaxies. These luminous galaxies, detectable even at high redshifts, serve as reliable tracers of galaxy clusters. Their large weight in the filtering scheme enhances the likelihood of identifying genuine overdensities, even under limited sensitivity.}

We show the matched (gray) and not matched (red) ACT detections falling in the KiDS-DR4 area as a function of redshift and mass $M_{200}$\footnote{Here, $M_{200}$ is the mass included inside a radius within which the average density of a halo is 200 times the critical density of the universe.} in the left panel of Fig.~\ref{fig:act_match}. The $\sigma$ parameters of the Rayleigh distribution fitting the angular displacement between the ACT SZ sample and \textsc{AMICO} optical detections is $\sigma_r=0.39$ arcmin, $\sigma_r=0.44$ arcmin, and $\sigma_r=0.43$ arcmin for the redshift intervals $0.1<z<0.3$, $0.3<z< 0.6$, and $0.6<z <0.9$, respectively, as shown in central panel of the same figure. The differences in these scatters across redshift intervals are primarily due to small-number statistics rather than an intrinsic redshift dependency. The right panel displays the relationships between the \textsc{AMICO} intrinsic richness $\lambda_*$ and the SZ-derived mass $M_{200}$. Also in this case there is a strong correlation at all redshifts but we prefer not to derive a scaling relation the ACT sample covers a rather small mass interval.

After visually inspecting all unmatched objects, we identified the following cases:
\begin{itemize}
\item one ACT detection with coordinates just outside a large masked region. While this cluster was retained in the ACT catalog, a significant fraction of its members are masked, then explaining the missing match.
\item four ACT detections with no visible galaxies in the images, despite their relatively low redshift.
\item twelve ACT detections that are clearly possible matches but they were not automatically matched because their redshift falls right outside the matching threshold.
\end{itemize}

We also visually inspected the 30 \textsc{AMICO} detections with the highest intrinsic richness $\lambda_*$ (covering the interval $86<\lambda_*$<122) that lacked a corresponding ACT counterpart. Among these:
\begin{itemize}
 \item twenty-eight are due to clear galaxy overdensities.
 \item one is of dubious origin, being located in one of the tiles with significant artifacts.
 \item one is a spurious detection at the very edge of the survey, caused by a bright star that was not masked in the input data. 
\end{itemize}

\section{Conclusions}

We used the Adaptive Matched Identifier of Clustered Objects (\textsc{AMICO}) to detect $23\,965$ galaxy clusters with a signal-to-noise ratio $S/N>3.5$ in the redshift range $0.1 \leq z \leq 0.9$ across the effective area of $839$ deg$^2$ covered by the KiDS-DR4 data set. This cluster sample is accompanied by a probabilistic membership association to the \textsc{AMICO} clusters of all KiDS-DR4 galaxies within the magnitude range $15<r'<24$. All analysis has been optimized for cosmological studies, prioritizing data homogeneity over redshift range or depth.

For each cluster detection, we provide basic information including celestial coordinates, redshift, amplitude, apparent richness, intrinsic richness, two mass estimates based on mass-proxy scaling relations obrained using eRASS1 X-ray data. \new{Mass scaling relations based on a subset of clusters based on the KiDS weak gravitational lensing data will be presented by \cite{lesci25b}.} The redshift estimates are calibrated using the GAMA spectroscopic data set. When available, spectroscopic redshifts are associated with the detections, alongside quality flags and indicators to characterize the reliability of these spectroscopic estimates.
These indicators include:
the number of galaxies with \textsc{AMICO} membership and spectroscopic redshifts,
the weighted average of spectroscopic redshifts using \textsc{AMICO} membership probabilities as weights,
the median spectroscopic redshift, and
the RMS of the weighted average.
We provide cluster redshift uncertainties based on photo-$z$s as well as on the estimated scatter with respect to the spectroscopic redshifts.

We conducted an extensive evaluation of detections near tile borders and masks to identify potential issues. We also enhanced \textsc{AMICO} with additional features, including:
a probability redshift distribution, $P_{det}(z)$, for each detection, from which we derive the $67$th percentile redshift uncertainties and ODDS values analogous to the BPZ galaxy photo-$z$s;
a flag distinguishing detections near tile edges with or without a neighboring tile;
and a flag identifying detections in tiles with potentially problematic photometry.

The sample’s purity, completeness, and uncertainties in observable quantities (redshift, angular position, amplitude, apparent richness, and intrinsic richness) were evaluated using the Selection Function Extractor (\textsc{SinFoniA}). \textsc{SinFoniA} is a data-driven method that constructs mock galaxy catalogs while preserving the statistical properties of the original data set. These mocks are then processed with \textsc{AMICO} to evaluate the aforementioned quantities. In this study, we further enhanced \textsc{SinFoniA} by incorporating mock clusters with ellipsoidal shapes, an improved mask-treatment scheme, and probabilistic weighting to better represent the likelihood of detections being real clusters in the mock generation process. This latter feature eliminates any arbitrariness associated with signal-to-noise ratio cutoffs when selecting detections to produce the mock clusters.
By deriving sample purity and completeness without relying on numerical simulations, we mitigate biases and model dependencies in astrophysical and cosmological analyses of the cluster sample. Additionally, we implemented a blinding scheme for the selection function, introducing perturbations to shift cosmological parameter estimates based on cluster counts in a controlled way. Three realizations of the perturbed selection function were stored, and the key identifying the original, unperturbed version \new{has been released} following the completion of the cosmological analysis of this sample.

Additionally, we cross-matched the \textsc{AMICO}-KiDS-DR4 cluster sample with external data sets, namely the RedMaPPer, X-ray eRASS1, and ACT-POL cluster samples, considering only detections falling within the KiDS-DR4 effective area.
For RedMaPPer, limited to $0.1<z<0.6$, we found a $88\%$ matching rate with \textsc{AMICO} clusters. Angular displacements between matched clusters show a "compact core" with a tail extending to larger separations, likely due to differing center definitions: the RedMaPPer centers are in fact based on individual bright galaxies, while the \textsc{AMICO} ones on the overall associated members. Richness estimates correlate strongly, with \textsc{AMICO}’s $\lambda_*$ aligning well with RedMaPPer’s values, as both estimates are driven by the brightest cluster galaxies.
In the cross-match with the X-ray eRASS1 "primary" cluster sample, we identified $321$ matches ($78\%$ of the sample) within the KiDS-DR4 effective area. The eRASS1 sample includes $M_{500}$ estimates, which strongly correlate with the \textsc{AMICO} intrinsic richness $\lambda_*$. We provide a mass-proxy scaling relation based on $\lambda_*$ and conducted a visual inspection of the brightest eRASS1 sources without an \textsc{AMICO} counterpart. This inspection indicates that three unmatched sources may be X-ray point sources (AGNs) misclassified as clusters rather than galaxy clusters, two cases are due to bimodal galaxy distributions causing different centers, two detections have AMICO counterparts just outside the matching radius, six are missed by AMICO because next to masked areas, and seven are missed due to their low redshifts ($0.12 < z < 0.17$) where photo-z quality is poor. In other cases, no galaxy overdensity is visible in the images. A similar analysis was performed for the ACT-DR5 SZ cluster sample, yielding a positive match for $88\%$ of its clusters across the redshift range covered  by KiDS-DR4 and $93\%$ over the range $0.3<z<0.7$, where the AMICO efficiency is the highest.

This new KiDS-DR4 galaxy cluster sample represents a valuable resource for investigating the fundamental properties of galaxy clusters and advancing cosmological studies.

\section*{Acknowledgements}

\new{The KiDS-1000 results in this paper are based on data products from observations made with ESO Telescopes at the La Silla Paranal Observatory under programme IDs 177.A-3016, 177.A-3017 and 177.A-3018, and on data products produced by Target/OmegaCEN, INAF-OACN, INAF-OAPD, and the KiDS production team, on behalf of the KiDS consortium.}
LM acknowledges the financial contribution from the PRIN-MUR 2022 20227RNLY3 grant “The concordance cosmological model: stress-tests with galaxy clusters” supported by Next Generation EU and from the grant ASI n. 2024-10-HH.0 “Attivit\`a scientifiche per la missione Euclid – fase E”.
GC  acknowledges the support from the Next Generation EU funds within the National Recovery and Resilience Plan (PNRR), Mission 4 - Education and Research, Component 2 - From Research to Business (M4C2), Investment Line 3.1 - Strengthening and creation of Research Infrastructures, Project IR0000012 – “CTA+ - Cherenkov Telescope Array Plus”.
MR acknowledges financial support from the INAF mini-grant 2022 “GALCLOCK”.
BG acknowledges support from the UKRI Stephen Hawking Fellowship (grant reference EP/Y017137/1).
HH is supported by a DFG Heisenberg grant (Hi 1495/5-1), the DFG Collaborative Research Center SFB1491, an ERC Consolidator Grant (No. 770935), and the DLR project 50QE2305.
SJ acknowledges the Dennis Sciama Fellowship at the University of Portsmouth and the Ramón y Cajal Fellowship from the Spanish Ministry of Science.



\bibliographystyle{aa}
\bibliography{master}

\begin{thebibliography}{74}
\expandafter\ifx\csname natexlab\endcsname\relax\def\natexlab#1{#1}\fi

\bibitem[{{Abbott} {et~al.}(2018){Abbott}, {Abdalla}, {Alarcon}, {Aleksi{\'c}},
  {Allam}, {Allen}, {Amara}, {Annis}, {Asorey}, {Avila}, {Bacon}, {Balbinot},
  {Banerji}, {Banik}, {Barkhouse}, {Baumer}, {Baxter}, {Bechtol}, {Becker},
  {Benoit-L{\'e}vy}, {Benson}, {Bernstein}, {Bertin}, {Blazek}, {Bridle},
  {Brooks}, {Brout}, {Buckley-Geer}, {Burke}, {Busha}, {Campos}, {Capozzi},
  {Carnero Rosell}, {Carrasco Kind}, {Carretero}, {Castander}, {Cawthon},
  {Chang}, {Chen}, {Childress}, {Choi}, {Conselice}, {Crittenden}, {Crocce},
  {Cunha}, {D'Andrea}, {da Costa}, {Das}, {Davis}, {Davis}, {De Vicente},
  {DePoy}, {DeRose}, {Desai}, {Diehl}, {Dietrich}, {Dodelson}, {Doel},
  {Drlica-Wagner}, {Eifler}, {Elliott}, {Elsner}, {Elvin-Poole}, {Estrada},
  {Evrard}, {Fang}, {Fernandez}, {Fert{\'e}}, {Finley}, {Flaugher}, {Fosalba},
  {Friedrich}, {Frieman}, {Garc{\'\i}a-Bellido}, {Garcia-Fernandez}, {Gatti},
  {Gaztanaga}, {Gerdes}, {Giannantonio}, {Gill}, {Glazebrook}, {Goldstein},
  {Gruen}, {Gruendl}, {Gschwend}, {Gutierrez}, {Hamilton}, {Hartley}, {Hinton},
  {Honscheid}, {Hoyle}, {Huterer}, {Jain}, {James}, {Jarvis}, {Jeltema},
  {Johnson}, {Johnson}, {Kacprzak}, {Kent}, {Kim}, {King}, {Kirk}, {Kokron},
  {Kovacs}, {Krause}, {Krawiec}, {Kremin}, {Kuehn}, {Kuhlmann}, {Kuropatkin},
  {Lacasa}, {Lahav}, {Li}, {Liddle}, {Lidman}, {Lima}, {Lin}, {MacCrann},
  {Maia}, {Makler}, {Manera}, {March}, {Marshall}, {Martini}, {McMahon},
  {Melchior}, {Menanteau}, {Miquel}, {Miranda}, {Mudd}, {Muir}, {M{\"o}ller},
  {Neilsen}, {Nichol}, {Nord}, {Nugent}, {Ogando}, {Palmese}, {Peacock},
  {Peiris}, {Peoples}, {Percival}, {Petravick}, {Plazas}, {Porredon}, {Prat},
  {Pujol}, {Rau}, {Refregier}, {Ricker}, {Roe}, {Rollins}, {Romer}, {Roodman},
  {Rosenfeld}, {Ross}, {Rozo}, {Rykoff}, {Sako}, {Salvador}, {Samuroff},
  {S{\'a}nchez}, {Sanchez}, {Santiago}, {Scarpine}, {Schindler}, {Scolnic},
  {Secco}, {Serrano}, {Sevilla-Noarbe}, {Sheldon}, {Smith}, {Smith}, {Smith},
  {Soares-Santos}, {Sobreira}, {Suchyta}, {Tarle}, {Thomas}, {Troxel},
  {Tucker}, {Tucker}, {Uddin}, {Varga}, {Vielzeuf}, {Vikram}, {Vivas},
  {Walker}, {Wang}, {Wechsler}, {Weller}, {Wester}, {Wolf}, {Yanny}, {Yuan},
  {Zenteno}, {Zhang}, {Zhang}, {Zuntz}, \& {Dark Energy Survey
  Collaboration}}]{abbott18}
{Abbott}, T.~M.~C., {Abdalla}, F.~B., {Alarcon}, A., {et~al.} 2018, \prd, 98,
  043526

\bibitem[{{Abbott} {et~al.}(2020){Abbott}, {Aguena}, {Alarcon}, {Allam},
  {Allen}, {Annis}, {Avila}, {Bacon}, {Bechtol}, {Bermeo}, {Bernstein},
  {Bertin}, {Bhargava}, {Bocquet}, {Brooks}, {Brout}, {Buckley-Geer}, {Burke},
  {Carnero Rosell}, {Carrasco Kind}, {Carretero}, {Castander}, {Cawthon},
  {Chang}, {Chen}, {Choi}, {Costanzi}, {Crocce}, {da Costa}, {Davis}, {De
  Vicente}, {DeRose}, {Desai}, {Diehl}, {Dietrich}, {Dodelson}, {Doel},
  {Drlica-Wagner}, {Eckert}, {Eifler}, {Elvin-Poole}, {Estrada}, {Everett},
  {Evrard}, {Farahi}, {Ferrero}, {Flaugher}, {Fosalba}, {Frieman},
  {Garc{\'\i}a-Bellido}, {Gatti}, {Gaztanaga}, {Gerdes}, {Giannantonio},
  {Giles}, {Grandis}, {Gruen}, {Gruendl}, {Gschwend}, {Gutierrez}, {Hartley},
  {Hinton}, {Hollowood}, {Honscheid}, {Hoyle}, {Huterer}, {James}, {Jarvis},
  {Jeltema}, {Johnson}, {Johnson}, {Kent}, {Krause}, {Kron}, {Kuehn},
  {Kuropatkin}, {Lahav}, {Li}, {Lidman}, {Lima}, {Lin}, {MacCrann}, {Maia},
  {Mantz}, {Marshall}, {Martini}, {Mayers}, {Melchior}, {Mena-Fern{\'a}ndez},
  {Menanteau}, {Miquel}, {Mohr}, {Nichol}, {Nord}, {Ogando}, {Palmese},
  {Paz-Chinch{\'o}n}, {Plazas}, {Prat}, {Rau}, {Romer}, {Roodman}, {Rooney},
  {Rozo}, {Rykoff}, {Sako}, {Samuroff}, {S{\'a}nchez}, {Sanchez}, {Saro},
  {Scarpine}, {Schubnell}, {Scolnic}, {Serrano}, {Sevilla-Noarbe}, {Sheldon},
  {Smith}, {Smith}, {Suchyta}, {Swanson}, {Tarle}, {Thomas}, {To}, {Troxel},
  {Tucker}, {Varga}, {von der Linden}, {Walker}, {Wechsler}, {Weller},
  {Wilkinson}, {Wu}, {Yanny}, {Zhang}, {Zhang}, {Zuntz}, \& {DES
  Collaboration}}]{abbott20}
{Abbott}, T.~M.~C., {Aguena}, M., {Alarcon}, A., {et~al.} 2020, \prd, 102,
  023509

\bibitem[{{Artis} {et~al.}(2025){Artis}, {Bulbul}, {Grandis}, {Ghirardini},
  {Clerc}, {Seppi}, {Comparat}, {Cataneo}, {von der Linden}, {Bahar}, {Balzer},
  {Chiu}, {Gruen}, {Kleinebreil}, {Kluge}, {Krippendorf}, {Li}, {Liu},
  {Malavasi}, {Merloni}, {Miyatake}, {Miyazaki}, {Nandra}, {Okabe}, {Pacaud},
  {Predehl}, {Ramos-Ceja}, {Reiprich}, {Sanders}, {Schrabback}, {Zelmer}, \&
  {Zhang}}]{artis25}
{Artis}, E., {Bulbul}, E., {Grandis}, S., {et~al.} 2025, \aap, 696, A5

\bibitem[{{Bellagamba} {et~al.}(2018){Bellagamba}, {Roncarelli}, {Maturi}, \&
  {Moscardini}}]{Bellagamba18}
{Bellagamba}, F., {Roncarelli}, M., {Maturi}, M., \& {Moscardini}, L. 2018,
  \mnras, 473, 5221

\bibitem[{{Bellagamba} {et~al.}(2019){Bellagamba}, {Sereno}, {Roncarelli},
  {Maturi}, {Radovich}, {Bardelli}, {Puddu}, {Moscardini}, {Getman},
  {Hildebrandt}, \& {Napolitano}}]{bellagamba18b}
{Bellagamba}, F., {Sereno}, M., {Roncarelli}, M., {et~al.} 2019, \mnras, 484,
  1598

\bibitem[{{Bellstedt} {et~al.}(2020){Bellstedt}, {Driver}, {Robotham},
  {Davies}, {Bogue}, {Cook}, {Hashemizadeh}, {Koushan}, {Taylor}, {Thorne},
  {Turner}, \& {Wright}}]{Bellstedt2019}
{Bellstedt}, S., {Driver}, S.~P., {Robotham}, A. S.~G., {et~al.} 2020, \mnras,
  496, 3235

\bibitem[{{Ben{\'\i}tez}(2000)}]{benitez00}
{Ben{\'\i}tez}, N. 2000, \apj, 536, 571

\bibitem[{{Blake} {et~al.}(2016){Blake}, {Joudaki}, {Heymans}, {Choi}, {Erben},
  {Harnois-Deraps}, {Hildebrandt}, {Joachimi}, {Nakajima}, {van Waerbeke}, \&
  {Viola}}]{blake16}
{Blake}, C., {Joudaki}, S., {Heymans}, C., {et~al.} 2016, \mnras, 456, 2806

\bibitem[{{Bleem} {et~al.}(2015){Bleem}, {Stalder}, {de Haan}, {Aird}, {Allen},
  {Applegate}, {Ashby}, {Bautz}, {Bayliss}, {Benson}, {Bocquet}, {Brodwin},
  {Carlstrom}, {Chang}, {Chiu}, {Cho}, {Clocchiatti}, {Crawford}, {Crites},
  {Desai}, {Dietrich}, {Dobbs}, {Foley}, {Forman}, {George}, {Gladders},
  {Gonzalez}, {Halverson}, {Hennig}, {Hoekstra}, {Holder}, {Holzapfel},
  {Hrubes}, {Jones}, {Keisler}, {Knox}, {Lee}, {Leitch}, {Liu}, {Lueker},
  {Luong-Van}, {Mantz}, {Marrone}, {McDonald}, {McMahon}, {Meyer}, {Mocanu},
  {Mohr}, {Murray}, {Padin}, {Pryke}, {Reichardt}, {Rest}, {Ruel}, {Ruhl},
  {Saliwanchik}, {Saro}, {Sayre}, {Schaffer}, {Schrabback}, {Shirokoff},
  {Song}, {Spieler}, {Stanford}, {Staniszewski}, {Stark}, {Story}, {Stubbs},
  {Vanderlinde}, {Vieira}, {Vikhlinin}, {Williamson}, {Zahn}, \&
  {Zenteno}}]{bleem15}
{Bleem}, L.~E., {Stalder}, B., {de Haan}, T., {et~al.} 2015, \apjs, 216, 27

\bibitem[{{Bocquet} {et~al.}(2019){Bocquet}, {Dietrich}, {Schrabback}, {Bleem},
  {Klein}, {Allen}, {Applegate}, {Ashby}, {Bautz}, {Bayliss}, {Benson},
  {Brodwin}, {Bulbul}, {Canning}, {Capasso}, {Carlstrom}, {Chang}, {Chiu},
  {Cho}, {Clocchiatti}, {Crawford}, {Crites}, {de Haan}, {Desai}, {Dobbs},
  {Foley}, {Forman}, {Garmire}, {George}, {Gladders}, {Gonzalez}, {Grandis},
  {Gupta}, {Halverson}, {Hlavacek-Larrondo}, {Hoekstra}, {Holder}, {Holzapfel},
  {Hou}, {Hrubes}, {Huang}, {Jones}, {Khullar}, {Knox}, {Kraft}, {Lee}, {von
  der Linden}, {Luong-Van}, {Mantz}, {Marrone}, {McDonald}, {McMahon}, {Meyer},
  {Mocanu}, {Mohr}, {Morris}, {Padin}, {Patil}, {Pryke}, {Rapetti},
  {Reichardt}, {Rest}, {Ruhl}, {Saliwanchik}, {Saro}, {Sayre}, {Schaffer},
  {Shirokoff}, {Stalder}, {Stanford}, {Staniszewski}, {Stark}, {Story},
  {Strazzullo}, {Stubbs}, {Vanderlinde}, {Vieira}, {Vikhlinin}, {Williamson},
  \& {Zenteno}}]{bocquet19}
{Bocquet}, S., {Dietrich}, J.~P., {Schrabback}, T., {et~al.} 2019, \apj, 878,
  55

\bibitem[{{Bulbul} {et~al.}(2024){Bulbul}, {Liu}, {Kluge}, {Zhang}, {Sanders},
  {Bahar}, {Ghirardini}, {Artis}, {Seppi}, {Garrel}, {Ramos-Ceja}, {Comparat},
  {Balzer}, {B{\"o}ckmann}, {Br{\"u}ggen}, {Clerc}, {Dennerl}, {Dolag},
  {Freyberg}, {Grandis}, {Gruen}, {Kleinebreil}, {Krippendorf}, {Lamer},
  {Merloni}, {Migkas}, {Nandra}, {Pacaud}, {Predehl}, {Reiprich}, {Schrabback},
  {Veronica}, {Weller}, \& {Zelmer}}]{bulbul24}
{Bulbul}, E., {Liu}, A., {Kluge}, M., {et~al.} 2024, \aap, 685, A106

\bibitem[{{Castignani} {et~al.}(2022){Castignani}, {Radovich}, {Combes},
  {Salom{\'e}}, {Maturi}, {Moscardini}, {Bardelli}, {Giocoli}, {Lesci},
  {Marulli}, {Puddu}, \& {Sereno}}]{Castignani2022}
{Castignani}, G., {Radovich}, M., {Combes}, F., {et~al.} 2022, \aap, 667, A52

\bibitem[{{Castignani} {et~al.}(2023){Castignani}, {Radovich}, {Combes},
  {Salom{\'e}}, {Moscardini}, {Bardelli}, {Giocoli}, {Lesci}, {Marulli},
  {Maturi}, {Puddu}, {Sereno}, \& {Tramonte}}]{Castignani2023}
{Castignani}, G., {Radovich}, M., {Combes}, F., {et~al.} 2023, \aap, 672, A139

\bibitem[{{Conley} {et~al.}(2006){Conley}, {Goldhaber}, {Wang}, {Aldering},
  {Amanullah}, {Commins}, {Fadeyev}, {Folatelli}, {Garavini}, {Gibbons},
  {Goobar}, {Groom}, {Hook}, {Howell}, {Kim}, {Knop}, {Kowalski}, {Kuznetsova},
  {Lidman}, {Nobili}, {Nugent}, {Pain}, {Perlmutter}, {Smith}, {Spadafora},
  {Stanishev}, {Strovink}, {Thomas}, {Wood-Vasey}, \& {Supernova Cosmology
  Project}}]{conley06}
{Conley}, A., {Goldhaber}, G., {Wang}, L., {et~al.} 2006, \apj, 644, 1

\bibitem[{{Costanzi} {et~al.}(2021){Costanzi}, {Saro}, {Bocquet}, {Abbott},
  {Aguena}, {Allam}, {Amara}, {Annis}, {Avila}, {Bacon}, {Benson}, {Bhargava},
  {Brooks}, {Buckley-Geer}, {Burke}, {Carnero Rosell}, {Carrasco Kind},
  {Carretero}, {Choi}, {da Costa}, {Pereira}, {De Vicente}, {Desai}, {Diehl},
  {Dietrich}, {Doel}, {Eifler}, {Everett}, {Ferrero}, {Fert{\'e}}, {Flaugher},
  {Fosalba}, {Frieman}, {Garc{\'\i}a-Bellido}, {Gaztanaga}, {Gerdes},
  {Giannantonio}, {Giles}, {Grandis}, {Gruen}, {Gruendl}, {Gupta}, {Gutierrez},
  {Hartley}, {Hinton}, {Hollowood}, {Honscheid}, {James}, {Jeltema}, {Krause},
  {Kuehn}, {Kuropatkin}, {Lahav}, {Lima}, {MacCrann}, {Maia}, {Marshall},
  {Menanteau}, {Miquel}, {Mohr}, {Morgan}, {Myles}, {Ogando}, {Palmese},
  {Paz-Chinch{\'o}n}, {Plazas}, {Rapetti}, {Reichardt}, {Romer}, {Roodman},
  {Ruppin}, {Salvati}, {Samuroff}, {Sanchez}, {Scarpine}, {Serrano},
  {Sevilla-Noarbe}, {Singh}, {Smith}, {Soares-Santos}, {Stark}, {Suchyta},
  {Swanson}, {Tarle}, {Thomas}, {To}, {Tucker}, {Varga}, {Wechsler}, {Zhang},
  {DES}, \& {SPT Collaborations}}]{costanzi21}
{Costanzi}, M., {Saro}, A., {Bocquet}, S., {et~al.} 2021, \prd, 103, 043522

\bibitem[{{Costanzi} {et~al.}(2013){Costanzi}, {Villaescusa-Navarro}, {Viel},
  {Xia}, {Borgani}, {Castorina}, \& {Sefusatti}}]{Costanzi13}
{Costanzi}, M., {Villaescusa-Navarro}, F., {Viel}, M., {et~al.} 2013, \jcap,
  12, 012

\bibitem[{{de Jong} {et~al.}(2015){de Jong}, {Verdoes Kleijn}, {Boxhoorn},
  {Buddelmeijer}, {Capaccioli}, {Getman}, {Grado}, {Helmich}, {Huang},
  {Irisarri}, {Kuijken}, {La Barbera}, {McFarland}, {Napolitano}, {Radovich},
  {Sikkema}, {Valentijn}, {Begeman}, {Brescia}, {Cavuoti}, {Choi}, {Cordes},
  {Covone}, {Dall'Ora}, {Hildebrandt}, {Longo}, {Nakajima}, {Paolillo},
  {Puddu}, {Rifatto}, {Tortora}, {van Uitert}, {Buddendiek},
  {Harnois-D{\'e}raps}, {Erben}, {Eriksen}, {Heymans}, {Hoekstra}, {Joachimi},
  {Kitching}, {Klaes}, {Koopmans}, {K{\"o}hlinger}, {Roy}, {Sif{\'o}n},
  {Schneider}, {Sutherland}, {Viola}, \& {Vriend}}]{deJong15}
{de Jong}, J.~T.~A., {Verdoes Kleijn}, G.~A., {Boxhoorn}, D.~R., {et~al.} 2015,
  \aap, 582, A62

\bibitem[{{Despali} {et~al.}(2017){Despali}, {Giocoli}, {Bonamigo}, {Limousin},
  \& {Tormen}}]{despali17}
{Despali}, G., {Giocoli}, C., {Bonamigo}, M., {Limousin}, M., \& {Tormen}, G.
  2017, \mnras, 466, 181

\bibitem[{{Driver} {et~al.}(2022){Driver}, {Bellstedt}, {Robotham}, {Baldry},
  {Davies}, {Liske}, {Obreschkow}, {Taylor}, {Wright}, {Alpaslan}, {Bamford},
  {Bauer}, {Bland-Hawthorn}, {Bilicki}, {Bravo}, {Brough}, {Casura}, {Cluver},
  {Colless}, {Conselice}, {Croom}, {de Jong}, {D'Eugenio}, {De Propris},
  {Dogruel}, {Drinkwater}, {Dvornik}, {Farrow}, {Frenk}, {Giblin}, {Graham},
  {Grootes}, {Gunawardhana}, {Hashemizadeh}, {H{\"a}u{\ss}ler}, {Heymans},
  {Hildebrandt}, {Holwerda}, {Hopkins}, {Jarrett}, {Heath Jones}, {Kelvin},
  {Koushan}, {Kuijken}, {Lara-L{\'o}pez}, {Lange}, {L{\'o}pez-S{\'a}nchez},
  {Loveday}, {Mahajan}, {Meyer}, {Moffett}, {Napolitano}, {Norberg}, {Owers},
  {Radovich}, {Raouf}, {Peacock}, {Phillipps}, {Pimbblet}, {Popescu}, {Said},
  {Sansom}, {Seibert}, {Sutherland}, {Thorne}, {Tuffs}, {Turner}, {van der
  Wel}, {van Kampen}, \& {Wilkins}}]{Driver2022}
{Driver}, S.~P., {Bellstedt}, S., {Robotham}, A. S.~G., {et~al.} 2022, \mnras,
  513, 439

\bibitem[{{Driver} {et~al.}(2011){Driver}, {Hill}, {Kelvin}, {Robotham},
  {Liske}, {Norberg}, {Baldry}, {Bamford}, {Hopkins}, {Loveday}, {Peacock},
  {Andrae}, {Bland-Hawthorn}, {Brough}, {Brown}, {Cameron}, {Ching}, {Colless},
  {Conselice}, {Croom}, {Cross}, {de Propris}, {Dye}, {Drinkwater}, {Ellis},
  {Graham}, {Grootes}, {Gunawardhana}, {Jones}, {van Kampen}, {Maraston},
  {Nichol}, {Parkinson}, {Phillipps}, {Pimbblet}, {Popescu}, {Prescott},
  {Roseboom}, {Sadler}, {Sansom}, {Sharp}, {Smith}, {Taylor}, {Thomas},
  {Tuffs}, {Wijesinghe}, {Dunne}, {Frenk}, {Jarvis}, {Madore}, {Meyer},
  {Seibert}, {Staveley-Smith}, {Sutherland}, \& {Warren}}]{Driver2011}
{Driver}, S.~P., {Hill}, D.~T., {Kelvin}, L.~S., {et~al.} 2011, \mnras, 413,
  971

\bibitem[{{Edge} {et~al.}(2013){Edge}, {Sutherland}, {Kuijken}, {Driver},
  {McMahon}, {Eales}, \& {Emerson}}]{edge13}
{Edge}, A., {Sutherland}, W., {Kuijken}, K., {et~al.} 2013, The Messenger, 154,
  32

\bibitem[{{Euclid Collaboration} {et~al.}(2019){Euclid Collaboration}, {Adam},
  {Vannier}, {Maurogordato}, {Biviano}, {Adami}, {Ascaso}, {Bellagamba},
  {Benoist}, {Cappi}, {D{\'\i}az-S{\'a}nchez}, {Durret}, {Farrens}, {Gonzalez},
  {Iovino}, {Licitra}, {Maturi}, {Mei}, {Merson}, {Munari}, {Pell{\'o}},
  {Ricci}, {Rocci}, {Roncarelli}, {Sarron}, {Amoura}, {Andreon}, {Apostolakos},
  {Arnaud}, {Bardelli}, {Bartlett}, {Baugh}, {Borgani}, {Brodwin}, {Castander},
  {Castignani}, {Cucciati}, {De Lucia}, {Dubath}, {Fosalba}, {Giocoli},
  {Hoekstra}, {Mamon}, {Melin}, {Moscardini}, {Paltani}, {Radovich},
  {Sartoris}, {Schultheis}, {Sereno}, {Weller}, {Burigana}, {Carvalho},
  {Corcione}, {Kurki-Suonio}, {Lilje}, {Sirri}, {Toledo-Moreo}, \&
  {Zamorani}}]{adam19}
{Euclid Collaboration}, {Adam}, R., {Vannier}, M., {et~al.} 2019, \aap, 627,
  A23

\bibitem[{{Garrel} {et~al.}(2022){Garrel}, {Pierre}, {Valageas}, {Eckert},
  {Marulli}, {Veropalumbo}, {Pacaud}, {Clerc}, {Sereno}, {Umetsu},
  {Moscardini}, {Bhargava}, {Adami}, {Chiappetti}, {Gastaldello},
  {Koulouridis}, {Le Fevre}, \& {Plionis}}]{garrel22}
{Garrel}, C., {Pierre}, M., {Valageas}, P., {et~al.} 2022, \aap, 663, A3

\bibitem[{{Ghirardini} {et~al.}(2024){Ghirardini}, {Bulbul}, {Artis}, {Clerc},
  {Garrel}, {Grandis}, {Kluge}, {Liu}, {Bahar}, {Balzer}, {Chiu}, {Comparat},
  {Gruen}, {Kleinebreil}, {Krippendorf}, {Merloni}, {Nandra}, {Okabe},
  {Pacaud}, {Predehl}, {Ramos-Ceja}, {Reiprich}, {Sanders}, {Schrabback},
  {Seppi}, {Zelmer}, {Zhang}, {Bornemann}, {Brunner}, {Burwitz}, {Coutinho},
  {Dennerl}, {Freyberg}, {Friedrich}, {Gaida}, {Gueguen}, {Haberl}, {Kink},
  {Lamer}, {Li}, {Liu}, {Maitra}, {Meidinger}, {Mueller}, {Miyatake},
  {Miyazaki}, {Robrade}, {Schwope}, \& {Stewart}}]{ghirardini24}
{Ghirardini}, V., {Bulbul}, E., {Artis}, E., {et~al.} 2024, \aap, 689, A298

\bibitem[{{Giocoli} {et~al.}(2021){Giocoli}, {Marulli}, {Moscardini}, {Sereno},
  {Veropalumbo}, {Gigante}, {Maturi}, {Radovich}, {Bellagamba}, {Roncarelli},
  {Bardelli}, {Contarini}, {Covone}, {Harnois-D{\'e}raps}, {Ingoglia}, {Lesci},
  {Nanni}, \& {Puddu}}]{giocoli21}
{Giocoli}, C., {Marulli}, F., {Moscardini}, L., {et~al.} 2021, \aap, 653, A19

\bibitem[{{Giocoli} {et~al.}(2024){Giocoli}, {Palmucci}, {Lesci}, {Moscardini},
  {Despali}, {Marulli}, {Maturi}, {Radovich}, {Sereno}, {Bardelli},
  {Castignani}, {Covone}, {Ingoglia}, {Romanello}, {Roncarelli}, \&
  {Puddu}}]{giocoli24}
{Giocoli}, C., {Palmucci}, L., {Lesci}, G.~F., {et~al.} 2024, \aap, 687, A79

\bibitem[{{Hennig} {et~al.}(2017){Hennig}, {Mohr}, {Zenteno}, {Desai},
  {Dietrich}, {Bocquet}, {Strazzullo}, {Saro}, {Abbott}, {Abdalla}, {Bayliss},
  {Benoit-L{\'e}vy}, {Bernstein}, {Bertin}, {Brooks}, {Capasso}, {Capozzi},
  {Carnero}, {Carrasco Kind}, {Carretero}, {Chiu}, {D'Andrea}, {daCosta},
  {Diehl}, {Doel}, {Eifler}, {Evrard}, {Fausti-Neto}, {Fosalba}, {Frieman},
  {Gangkofner}, {Gonzalez}, {Gruen}, {Gruendl}, {Gupta}, {Gutierrez},
  {Honscheid}, {Hlavacek-Larrondo}, {James}, {Kuehn}, {Kuropatkin}, {Lahav},
  {March}, {Marshall}, {Martini}, {McDonald}, {Melchior}, {Miller}, {Miquel},
  {Neilsen}, {Nord}, {Ogando}, {Plazas}, {Reichardt}, {Romer}, {Rozo},
  {Rykoff}, {Sanchez}, {Santiago}, {Schubnell}, {Sevilla-Noarbe}, {Smith},
  {Soares-Santos}, {Sobreira}, {Stalder}, {Stanford}, {Suchyta}, {Swanson},
  {Tarle}, {Thomas}, {Vikram}, {Walker}, \& {Zhang}}]{Hennig17}
{Hennig}, C., {Mohr}, J.~J., {Zenteno}, A., {et~al.} 2017, \mnras, 467, 4015

\bibitem[{{Heymans} {et~al.}(2012){Heymans}, {Van Waerbeke}, {Miller}, {Erben},
  {Hildebrandt}, {Hoekstra}, {Kitching}, {Mellier}, {Simon}, {Bonnett},
  {Coupon}, {Fu}, {Harnois D{\'e}raps}, {Hudson}, {Kilbinger}, {Kuijken},
  {Rowe}, {Schrabback}, {Semboloni}, {van Uitert}, {Vafaei}, \&
  {Velander}}]{heymans12}
{Heymans}, C., {Van Waerbeke}, L., {Miller}, L., {et~al.} 2012, \mnras, 427,
  146

\bibitem[{{Hildebrandt} {et~al.}(2017){Hildebrandt}, {Viola}, {Heymans},
  {Joudaki}, {Kuijken}, {Blake}, {Erben}, {Joachimi}, {Klaes}, {Miller},
  {Morrison}, {Nakajima}, {Verdoes Kleijn}, {Amon}, {Choi}, {Covone}, {de
  Jong}, {Dvornik}, {Fenech Conti}, {Grado}, {Harnois-D{\'e}raps}, {Herbonnet},
  {Hoekstra}, {K{\"o}hlinger}, {McFarland}, {Mead}, {Merten}, {Napolitano},
  {Peacock}, {Radovich}, {Schneider}, {Simon}, {Valentijn}, {van den Busch},
  {van Uitert}, \& {Van Waerbeke}}]{hildebrandt17}
{Hildebrandt}, H., {Viola}, M., {Heymans}, C., {et~al.} 2017, \mnras, 465, 1454

\bibitem[{{Hilton} {et~al.}(2021){Hilton}, {Sif{\'o}n}, {Naess},
  {Madhavacheril}, {Oguri}, {Rozo}, {Rykoff}, {Abbott}, {Adhikari}, {Aguena},
  {Aiola}, {Allam}, {Amodeo}, {Amon}, {Annis}, {Ansarinejad}, {Aros-Bunster},
  {Austermann}, {Avila}, {Bacon}, {Battaglia}, {Beall}, {Becker}, {Bernstein},
  {Bertin}, {Bhandarkar}, {Bhargava}, {Bond}, {Brooks}, {Burke}, {Calabrese},
  {Carrasco Kind}, {Carretero}, {Choi}, {Choi}, {Conselice}, {da Costa},
  {Costanzi}, {Crichton}, {Crowley}, {D{\"u}nner}, {Denison}, {Devlin},
  {Dicker}, {Diehl}, {Dietrich}, {Doel}, {Duff}, {Duivenvoorden}, {Dunkley},
  {Everett}, {Ferraro}, {Ferrero}, {Fert{\'e}}, {Flaugher}, {Frieman},
  {Gallardo}, {Garc{\'\i}a-Bellido}, {Gaztanaga}, {Gerdes}, {Giles}, {Golec},
  {Gralla}, {Grandis}, {Gruen}, {Gruendl}, {Gschwend}, {Gutierrez}, {Han},
  {Hartley}, {Hasselfield}, {Hill}, {Hilton}, {Hincks}, {Hinton}, {Ho},
  {Honscheid}, {Hoyle}, {Hubmayr}, {Huffenberger}, {Hughes}, {Jaelani}, {Jain},
  {James}, {Jeltema}, {Kent}, {Knowles}, {Koopman}, {Kuehn}, {Lahav}, {Lima},
  {Lin}, {Lokken}, {Loubser}, {MacCrann}, {Maia}, {Marriage}, {Martin},
  {McMahon}, {Melchior}, {Menanteau}, {Miquel}, {Miyatake}, {Moodley},
  {Morgan}, {Mroczkowski}, {Nati}, {Newburgh}, {Niemack}, {Nishizawa},
  {Ogando}, {Orlowski-Scherer}, {Page}, {Palmese}, {Partridge},
  {Paz-Chinch{\'o}n}, {Phakathi}, {Plazas}, {Robertson}, {Romer}, {Carnero
  Rosell}, {Salatino}, {Sanchez}, {Schaan}, {Schillaci}, {Sehgal}, {Serrano},
  {Shin}, {Simon}, {Smith}, {Soares-Santos}, {Spergel}, {Staggs}, {Storer},
  {Suchyta}, {Swanson}, {Tarle}, {Thomas}, {To}, {Trac}, {Ullom}, {Vale}, {Van
  Lanen}, {Vavagiakis}, {De Vicente}, {Wilkinson}, {Wollack}, {Xu}, \&
  {Zhang}}]{hilton20}
{Hilton}, M., {Sif{\'o}n}, C., {Naess}, S., {et~al.} 2021, \apjs, 253, 3

\bibitem[{{Ingoglia} {et~al.}(2022){Ingoglia}, {Covone}, {Sereno}, {Giocoli},
  {Bardelli}, {Bellagamba}, {Castignani}, {Farrens}, {Hildebrandt}, {Joudaki},
  {Jullo}, {Lanzieri}, {Lesci}, {Marulli}, {Maturi}, {Moscardini}, {Nanni},
  {Puddu}, {Radovich}, {Roncarelli}, {Sapio}, \& {Schimd}}]{ingoglia22}
{Ingoglia}, L., {Covone}, G., {Sereno}, M., {et~al.} 2022, \mnras, 511, 1484

\bibitem[{{Jenkins} {et~al.}(2001){Jenkins}, {Frenk}, {White}, {Colberg},
  {Cole}, {Evrard}, {Couchman}, \& {Yoshida}}]{jenkins01}
{Jenkins}, A., {Frenk}, C.~S., {White}, S.~D.~M., {et~al.} 2001, \mnras, 321,
  372

\bibitem[{{Kuijken}(2011)}]{Kuijken11}
{Kuijken}, K. 2011, The Messenger, 146, 8

\bibitem[{{Kuijken} {et~al.}(2019){Kuijken}, {Heymans}, {Dvornik},
  {Hildebrandt}, {de Jong}, {Wright}, {Erben}, {Bilicki}, {Giblin}, {Shan},
  {Getman}, {Grado}, {Hoekstra}, {Miller}, {Napolitano}, {Paolilo}, {Radovich},
  {Schneider}, {Sutherland}, {Tewes}, {Tortora}, {Valentijn}, \& {Verdoes
  Kleijn}}]{Kuijken2019}
{Kuijken}, K., {Heymans}, C., {Dvornik}, A., {et~al.} 2019, \aap, 625, A2

\bibitem[{{Kuijken} {et~al.}(2015){Kuijken}, {Heymans}, {Hildebrandt},
  {Nakajima}, {Erben}, {de Jong}, {Viola}, {Choi}, {Hoekstra}, {Miller}, {van
  Uitert}, {Amon}, {Blake}, {Brouwer}, {Buddendiek}, {Conti}, {Eriksen},
  {Grado}, {Harnois-D{\'e}raps}, {Helmich}, {Herbonnet}, {Irisarri},
  {Kitching}, {Klaes}, {La Barbera}, {Napolitano}, {Radovich}, {Schneider},
  {Sif{\'o}n}, {Sikkema}, {Simon}, {Tudorica}, {Valentijn}, {Verdoes Kleijn},
  \& {van Waerbeke}}]{kuijken15}
{Kuijken}, K., {Heymans}, C., {Hildebrandt}, H., {et~al.} 2015, \mnras, 454,
  3500

\bibitem[{{Lesci} {et~al.}(2025{\natexlab{a}}){Lesci}, {Marulli}, {Moscardini},
  {Maturi}, {Sereno}, {Radovich}, {Romanello}, {Giocoli}, {Wright}, {Bardelli},
  {Bilicki}, {Castignani}, {Hildebrandt}, {Kannawadi}, {Ingoglia}, {Joudaki},
  \& {Puddu}}]{lesci25b}
{Lesci}, G.~F., {Marulli}, F., {Moscardini}, L., {et~al.} 2025{\natexlab{a}},
  arXiv e-prints, arXiv:2507.14285

\bibitem[{{Lesci} {et~al.}(2022{\natexlab{a}}){Lesci}, {Marulli}, {Moscardini},
  {Sereno}, {Veropalumbo}, {Maturi}, {Giocoli}, {Radovich}, {Bellagamba},
  {Roncarelli}, {Bardelli}, {Contarini}, {Covone}, {Ingoglia}, {Nanni}, \&
  {Puddu}}]{lesci22}
{Lesci}, G.~F., {Marulli}, F., {Moscardini}, L., {et~al.} 2022{\natexlab{a}},
  \aap, 659, A88

\bibitem[{{Lesci} {et~al.}(2025{\natexlab{b}}){Lesci}, {Marulli},
  L.~{Moscardini}, M.~{Maturi}, {Sereno}, {Radovich}, {Romanello}, {Giocoli},
  {Wright}, {Bardelli}, {Castignani}, \& {Puddu}}]{lesci25}
{Lesci}, G.~F., {Marulli}, G., L.~{Moscardini}, L., {et~al.}
  2025{\natexlab{b}}, in prep.

\bibitem[{{Lesci} {et~al.}(2022{\natexlab{b}}){Lesci}, {Nanni}, {Marulli},
  {Moscardini}, {Veropalumbo}, {Maturi}, {Sereno}, {Radovich}, {Bellagamba},
  {Roncarelli}, {Bardelli}, {Castignani}, {Covone}, {Giocoli}, {Ingoglia}, \&
  {Puddu}}]{lesci22b}
{Lesci}, G.~F., {Nanni}, L., {Marulli}, F., {et~al.} 2022{\natexlab{b}}, \aap,
  665, A100

\bibitem[{{Liske} {et~al.}(2015){Liske}, {Baldry}, {Driver}, {Tuffs},
  {Alpaslan}, {Andrae}, {Brough}, {Cluver}, {Grootes}, {Gunawardhana},
  {Kelvin}, {Loveday}, {Robotham}, {Taylor}, {Bamford}, {Bland-Hawthorn},
  {Brown}, {Drinkwater}, {Hopkins}, {Meyer}, {Norberg}, {Peacock}, {Agius},
  {Andrews}, {Bauer}, {Ching}, {Colless}, {Conselice}, {Croom}, {Davies}, {De
  Propris}, {Dunne}, {Eardley}, {Ellis}, {Foster}, {Frenk}, {H{\"a}u{\ss}ler},
  {Holwerda}, {Howlett}, {Ibarra}, {Jarvis}, {Jones}, {Kafle}, {Lacey},
  {Lange}, {Lara-L{\'o}pez}, {L{\'o}pez-S{\'a}nchez}, {Maddox}, {Madore},
  {McNaught-Roberts}, {Moffett}, {Nichol}, {Owers}, {Palamara}, {Penny},
  {Phillipps}, {Pimbblet}, {Popescu}, {Prescott}, {Proctor}, {Sadler},
  {Sansom}, {Seibert}, {Sharp}, {Sutherland}, {V{\'a}zquez-Mata}, {van Kampen},
  {Wilkins}, {Williams}, \& {Wright}}]{Liske2015}
{Liske}, J., {Baldry}, I.~K., {Driver}, S.~P., {et~al.} 2015, \mnras, 452, 2087

\bibitem[{{Maccoun} \& {Perlmutter}(2015)}]{maccoun15}
{Maccoun}, R. \& {Perlmutter}, S. 2015, \nat, 526, 187

\bibitem[{{Maturi} {et~al.}(2019){Maturi}, {Bellagamba}, {Radovich},
  {Roncarelli}, {Sereno}, {Moscardini}, {Bardelli}, \& {Puddu}}]{maturi19}
{Maturi}, M., {Bellagamba}, F., {Radovich}, M., {et~al.} 2019, \mnras, 485, 498

\bibitem[{{Maturi} {et~al.}(2023){Maturi}, {Finoguenov}, {Lopes}, {Gonz{\'a}lez
  Delgado}, {Dupke}, {Cypriano}, {Carrasco}, {Diego}, {Penna-Lima}, {Doubrawa},
  {V{\'\i}lchez}, {Moscardini}, {Marra}, {Bonoli},
  {Rodr{\'\i}guez-Mart{\'\i}n}, {Zitrin}, {M{\'a}rquez},
  {Hern{\'a}n-Caballero}, {Jim{\'e}nez-Teja}, {Abramo}, {Alcaniz}, {Benitez},
  {Carneiro}, {Cenarro}, {Crist{\'o}bal-Hornillos}, {Ederoclite},
  {L{\'o}pez-Sanjuan}, {Mar{\'\i}n-Franch}, {Mendes de Oliveira}, {Moles},
  {Sodr{\'e}}, {Taylor}, {Varela}, {V{\'a}zquez Rami{\'o}}, \&
  {Fern{\'a}ndez-Ontiveros}}]{maturi23}
{Maturi}, M., {Finoguenov}, A., {Lopes}, P.~A.~A., {et~al.} 2023, \aap, 678,
  A145

\bibitem[{{Maturi} {et~al.}(2005){Maturi}, {Meneghetti}, {Bartelmann}, {Dolag},
  \& {Moscardini}}]{2005A&A...442..851M}
{Maturi}, M., {Meneghetti}, M., {Bartelmann}, M., {Dolag}, K., \& {Moscardini},
  L. 2005, \aap, 442, 851

\bibitem[{{Merloni} {et~al.}(2024){Merloni}, {Lamer}, {Liu}, {Ramos-Ceja},
  {Brunner}, {Bulbul}, {Dennerl}, {Doroshenko}, {Freyberg}, {Friedrich},
  {Gatuzz}, {Georgakakis}, {Haberl}, {Igo}, {Kreykenbohm}, {Liu}, {Maitra},
  {Malyali}, {Mayer}, {Nandra}, {Predehl}, {Robrade}, {Salvato}, {Sanders},
  {Stewart}, {Tub{\'\i}n-Arenas}, {Weber}, {Wilms}, {Arcodia}, {Artis},
  {Aschersleben}, {Avakyan}, {Aydar}, {Bahar}, {Balzer}, {Becker}, {Berger},
  {Boller}, {Bornemann}, {Br{\"u}ggen}, {Brusa}, {Buchner}, {Burwitz},
  {Camilloni}, {Clerc}, {Comparat}, {Coutinho}, {Czesla}, {Dannhauer},
  {Dauner}, {Dauser}, {Dietl}, {Dolag}, {Dwelly}, {Egg}, {Ehl}, {Freund},
  {Friedrich}, {Gaida}, {Garrel}, {Ghirardini}, {Gokus}, {Gr{\"u}nwald},
  {Grandis}, {Grotova}, {Gruen}, {Gueguen}, {H{\"a}mmerich}, {Hamaus},
  {Hasinger}, {Haubner}, {Homan}, {Ider Chitham}, {Joseph}, {Joyce},
  {K{\"o}nig}, {Kaltenbrunner}, {Khokhriakova}, {Kink}, {Kirsch}, {Kluge},
  {Knies}, {Krippendorf}, {Krumpe}, {Kurpas}, {Li}, {Liu}, {Locatelli},
  {Lorenz}, {M{\"u}ller}, {Magaudda}, {Mannes}, {McCall}, {Meidinger},
  {Michailidis}, {Migkas}, {Mu{\~n}oz-Giraldo}, {Musiimenta}, {Nguyen-Dang},
  {Ni}, {Olechowska}, {Ota}, {Pacaud}, {Pasini}, {Perinati}, {Pires},
  {Pommranz}, {Ponti}, {Poppenhaeger}, {P{\"u}hlhofer}, {Rau}, {Reh},
  {Reiprich}, {Roster}, {Saeedi}, {Santangelo}, {Sasaki}, {Schmitt},
  {Schneider}, {Schrabback}, {Schuster}, {Schwope}, {Seppi}, {Serim},
  {Shreeram}, {Sokolova-Lapa}, {Starck}, {Stelzer}, {Stierhof}, {Suleimanov},
  {Tenzer}, {Traulsen}, {Tr{\"u}mper}, {Tsuge}, {Urrutia}, {Veronica},
  {Waddell}, {Willer}, {Wolf}, {Yeung}, {Zainab}, {Zangrandi}, {Zhang},
  {Zhang}, \& {Zheng}}]{merloni24}
{Merloni}, A., {Lamer}, G., {Liu}, T., {et~al.} 2024, \aap, 682, A34

\bibitem[{{Muir} {et~al.}(2020){Muir}, {Bernstein}, {Huterer}, {Elsner},
  {Krause}, {Roodman}, {Allam}, {Annis}, {Avila}, {Bechtol}, {Bertin},
  {Brooks}, {Buckley-Geer}, {Burke}, {Carnero Rosell}, {Carrasco Kind},
  {Carretero}, {Cawthon}, {Costanzi}, {da Costa}, {De Vicente}, {Desai},
  {Dietrich}, {Doel}, {Eifler}, {Everett}, {Fosalba}, {Frieman},
  {Garc{\'\i}a-Bellido}, {Gerdes}, {Gruen}, {Gruendl}, {Gschwend}, {Hartley},
  {Hollowood}, {James}, {Jarvis}, {Kuehn}, {Kuropatkin}, {Lahav}, {March},
  {Marshall}, {Melchior}, {Menanteau}, {Miquel}, {Ogando}, {Palmese},
  {Paz-Chinch{\'o}n}, {Plazas}, {Romer}, {Sanchez}, {Scarpine}, {Schubnell},
  {Serrano}, {Sevilla-Noarbe}, {Smith}, {Suchyta}, {Tarle}, {Thomas}, {Troxel},
  {Walker}, {Weller}, {Wester}, {Zuntz}, \& {DES Collaboration}}]{muir20}
{Muir}, J., {Bernstein}, G.~M., {Huterer}, D., {et~al.} 2020, \mnras, 494, 4454

\bibitem[{{Navarro} {et~al.}(1997){Navarro}, {Frenk}, \&
  {White}}]{1997ApJ...490..493N}
{Navarro}, J.~F., {Frenk}, C.~S., \& {White}, S.~D.~M. 1997, \apj, 490, 493

\bibitem[{{Ota} {et~al.}(2020){Ota}, {Mitsuishi}, {Babazaki}, {Akamatsu},
  {Ichinohe}, {Ueda}, {Okabe}, {Oguri}, {Fujimoto}, {Hamana}, {Miyaoka},
  {Miyazaki}, {Otani}, {Tanaka}, {Tsuji}, \& {Yoshida}}]{ota20}
{Ota}, N., {Mitsuishi}, I., {Babazaki}, Y., {et~al.} 2020, \pasj, 72, 1

\bibitem[{{Planck Collaboration} {et~al.}(2016){Planck Collaboration}, {Ade},
  {Aghanim}, {Arnaud}, {Ashdown}, {Aumont}, {Baccigalupi}, {Banday},
  {Barreiro}, {Barrena}, \& et~al.}]{planckSZ16}
{Planck Collaboration}, {Ade}, P.~A.~R., {Aghanim}, N., {et~al.} 2016, \aap,
  594, A27

\bibitem[{{Predehl} {et~al.}(2021){Predehl}, {Andritschke}, {Arefiev},
  {Babyshkin}, {Batanov}, {Becker}, {B{\"o}hringer}, {Bogomolov}, {Boller},
  {Borm}, {Bornemann}, {Br{\"a}uninger}, {Br{\"u}ggen}, {Brunner}, {Brusa},
  {Bulbul}, {Buntov}, {Burwitz}, {Burkert}, {Clerc}, {Churazov}, {Coutinho},
  {Dauser}, {Dennerl}, {Doroshenko}, {Eder}, {Emberger}, {Eraerds},
  {Finoguenov}, {Freyberg}, {Friedrich}, {Friedrich}, {F{\"u}rmetz},
  {Georgakakis}, {Gilfanov}, {Granato}, {Grossberger}, {Gueguen}, {Gureev},
  {Haberl}, {H{\"a}lker}, {Hartner}, {Hasinger}, {Huber}, {Ji}, {Kienlin},
  {Kink}, {Korotkov}, {Kreykenbohm}, {Lamer}, {Lomakin}, {Lapshov}, {Liu},
  {Maitra}, {Meidinger}, {Menz}, {Merloni}, {Mernik}, {Mican}, {Mohr},
  {M{\"u}ller}, {Nandra}, {Nazarov}, {Pacaud}, {Pavlinsky}, {Perinati},
  {Pfeffermann}, {Pietschner}, {Ramos-Ceja}, {Rau}, {Reiffers}, {Reiprich},
  {Robrade}, {Salvato}, {Sanders}, {Santangelo}, {Sasaki}, {Scheuerle},
  {Schmid}, {Schmitt}, {Schwope}, {Shirshakov}, {Steinmetz}, {Stewart},
  {Str{\"u}der}, {Sunyaev}, {Tenzer}, {Tiedemann}, {Tr{\"u}mper}, {Voron},
  {Weber}, {Wilms}, \& {Yaroshenko}}]{predehl21}
{Predehl}, P., {Andritschke}, R., {Arefiev}, V., {et~al.} 2021, \aap, 647, A1

\bibitem[{{Puddu} {et~al.}(2021){Puddu}, {Radovich}, {Sereno}, {Bardelli},
  {Maturi}, {Moscardini}, {Bellagamba}, {Giocoli}, {Marulli}, \&
  {Roncarelli}}]{puddu21}
{Puddu}, E., {Radovich}, M., {Sereno}, M., {et~al.} 2021, \aap, 645, A9

\bibitem[{{Radovich+}(2025)}]{radovich25}
{Radovich+}, M. 2025, in prep.

\bibitem[{{Radovich} {et~al.}(2020){Radovich}, {Tortora}, {Bellagamba},
  {Maturi}, {Moscardini}, {Puddu}, {Roncarelli}, {Roy}, {Bardelli}, {Marulli},
  {Sereno}, {Getman}, \& {Napolitano}}]{radovich20}
{Radovich}, M., {Tortora}, C., {Bellagamba}, F., {et~al.} 2020, \mnras, 498,
  4303

\bibitem[{{Romanello} {et~al.}(2024){Romanello}, {Marulli}, {Moscardini},
  {Lesci}, {Sartoris}, {Contarini}, {Giocoli}, {Bardelli}, {Busillo},
  {Castignani}, {Covone}, {Ingoglia}, {Maturi}, {Puddu}, {Radovich},
  {Roncarelli}, \& {Sereno}}]{romanello24}
{Romanello}, M., {Marulli}, F., {Moscardini}, L., {et~al.} 2024, \aap, 682, A72

\bibitem[{{Rykoff} {et~al.}(2014){Rykoff}, {Rozo}, {Busha}, {Cunha},
  {Finoguenov}, {Evrard}, {Hao}, {Koester}, {Leauthaud}, {Nord}, {Pierre},
  {Reddick}, {Sadibekova}, {Sheldon}, \& {Wechsler}}]{Rykoff14}
{Rykoff}, E.~S., {Rozo}, E., {Busha}, M.~T., {et~al.} 2014, \apj, 785, 104

\bibitem[{{Rykoff} {et~al.}(2016){Rykoff}, {Rozo}, {Hollowood},
  {Bermeo-Hernandez}, {Jeltema}, {Mayers}, {Romer}, {Rooney}, {Saro}, {Vergara
  Cervantes}, {Wechsler}, {Wilcox}, {Abbott}, {Abdalla}, {Allam}, {Annis},
  {Benoit-L{\'e}vy}, {Bernstein}, {Bertin}, {Brooks}, {Burke}, {Capozzi},
  {Carnero Rosell}, {Carrasco Kind}, {Castander}, {Childress}, {Collins},
  {Cunha}, {D'Andrea}, {da Costa}, {Davis}, {Desai}, {Diehl}, {Dietrich},
  {Doel}, {Evrard}, {Finley}, {Flaugher}, {Fosalba}, {Frieman}, {Glazebrook},
  {Goldstein}, {Gruen}, {Gruendl}, {Gutierrez}, {Hilton}, {Honscheid}, {Hoyle},
  {James}, {Kay}, {Kuehn}, {Kuropatkin}, {Lahav}, {Lewis}, {Lidman}, {Lima},
  {Maia}, {Mann}, {Marshall}, {Martini}, {Melchior}, {Miller}, {Miquel},
  {Mohr}, {Nichol}, {Nord}, {Ogando}, {Plazas}, {Reil}, {Sahl{\'e}n},
  {Sanchez}, {Santiago}, {Scarpine}, {Schubnell}, {Sevilla-Noarbe}, {Smith},
  {Soares-Santos}, {Sobreira}, {Stott}, {Suchyta}, {Swanson}, {Tarle},
  {Thomas}, {Tucker}, {Uddin}, {Viana}, {Vikram}, {Walker}, {Zhang}, \& {DES
  Collaboration}}]{rykoff16}
{Rykoff}, E.~S., {Rozo}, E., {Hollowood}, D., {et~al.} 2016, \apjs, 224, 1

\bibitem[{{Schechter}(1976)}]{schechter76}
{Schechter}, P. 1976, \apj, 203, 297

\bibitem[{{Seppi} {et~al.}(2024){Seppi}, {Comparat}, {Ghirardini}, {Garrel},
  {Artis}, {S{\'a}nchez}, {Liu}, {Clerc}, {Bulbul}, {Grandis}, {Kluge},
  {Reiprich}, {Merloni}, {Zhang}, {Bahar}, {Shreeram}, {Sanders}, {Ramos-Ceja},
  \& {Krumpe}}]{seppi24}
{Seppi}, R., {Comparat}, J., {Ghirardini}, V., {et~al.} 2024, \aap, 686, A196

\bibitem[{{Sereno} {et~al.}(2020){Sereno}, {Ettori}, {Lesci}, {Marulli},
  {Maturi}, {Moscardini}, {Radovich}, {Bellagamba}, \& {Roncarelli}}]{sereno20}
{Sereno}, M., {Ettori}, S., {Lesci}, G.~F., {et~al.} 2020, \mnras, 497, 894

\bibitem[{{Smit} {et~al.}(2022){Smit}, {Dvornik}, {Radovich}, {Kuijken},
  {Maturi}, {Moscardini}, \& {Sereno}}]{smit22}
{Smit}, M., {Dvornik}, A., {Radovich}, M., {et~al.} 2022, \aap, 659, A195

\bibitem[{{Suyu} {et~al.}(2017){Suyu}, {Bonvin}, {Courbin}, {Fassnacht},
  {Rusu}, {Sluse}, {Treu}, {Wong}, {Auger}, {Ding}, {Hilbert}, {Marshall},
  {Rumbaugh}, {Sonnenfeld}, {Tewes}, {Tihhonova}, {Agnello}, {Blandford},
  {Chen}, {Collett}, {Koopmans}, {Liao}, {Meylan}, \& {Spiniello}}]{suyu17}
{Suyu}, S.~H., {Bonvin}, V., {Courbin}, F., {et~al.} 2017, \mnras, 468, 2590

\bibitem[{{Swetz} {et~al.}(2011){Swetz}, {Ade}, {Amiri}, {Appel},
  {Battistelli}, {Burger}, {Chervenak}, {Devlin}, {Dicker}, {Doriese},
  {D{\"u}nner}, {Essinger-Hileman}, {Fisher}, {Fowler}, {Halpern},
  {Hasselfield}, {Hilton}, {Hincks}, {Irwin}, {Jarosik}, {Kaul}, {Klein},
  {Lau}, {Limon}, {Marriage}, {Marsden}, {Martocci}, {Mauskopf}, {Moseley},
  {Netterfield}, {Niemack}, {Nolta}, {Page}, {Parker}, {Staggs}, {Stryzak},
  {Switzer}, {Thornton}, {Tucker}, {Wollack}, \& {Zhao}}]{swetz11}
{Swetz}, D.~S., {Ade}, P.~A.~R., {Amiri}, M., {et~al.} 2011, \apjs, 194, 41

\bibitem[{{Thornton} {et~al.}(2016){Thornton}, {Ade}, {Aiola}, {Angil{\`e}},
  {Amiri}, {Beall}, {Becker}, {Cho}, {Choi}, {Corlies}, {Coughlin}, {Datta},
  {Devlin}, {Dicker}, {D{\"u}nner}, {Fowler}, {Fox}, {Gallardo}, {Gao},
  {Grace}, {Halpern}, {Hasselfield}, {Henderson}, {Hilton}, {Hincks}, {Ho},
  {Hubmayr}, {Irwin}, {Klein}, {Koopman}, {Li}, {Louis}, {Lungu}, {Maurin},
  {McMahon}, {Munson}, {Naess}, {Nati}, {Newburgh}, {Nibarger}, {Niemack},
  {Niraula}, {Nolta}, {Page}, {Pappas}, {Schillaci}, {Schmitt}, {Sehgal},
  {Sievers}, {Simon}, {Staggs}, {Tucker}, {Uehara}, {van Lanen}, {Ward}, \&
  {Wollack}}]{thornton16}
{Thornton}, R.~J., {Ade}, P.~A.~R., {Aiola}, S., {et~al.} 2016, \apjs, 227, 21

\bibitem[{{Toni} {et~al.}(2024){Toni}, {Maturi}, {Finoguenov}, {Moscardini}, \&
  {Castignani}}]{toni24}
{Toni}, G., {Maturi}, M., {Finoguenov}, A., {Moscardini}, L., \& {Castignani},
  G. 2024, \aap, 687, A56

\bibitem[{{Troxel} {et~al.}(2018){Troxel}, {MacCrann}, {Zuntz}, {Eifler},
  {Krause}, {Dodelson}, {Gruen}, {Blazek}, {Friedrich}, {Samuroff}, {Prat},
  {Secco}, {Davis}, {Fert{\'e}}, {DeRose}, {Alarcon}, {Amara}, {Baxter},
  {Becker}, {Bernstein}, {Bridle}, {Cawthon}, {Chang}, {Choi}, {De Vicente},
  {Drlica-Wagner}, {Elvin-Poole}, {Frieman}, {Gatti}, {Hartley}, {Honscheid},
  {Hoyle}, {Huff}, {Huterer}, {Jain}, {Jarvis}, {Kacprzak}, {Kirk}, {Kokron},
  {Krawiec}, {Lahav}, {Liddle}, {Peacock}, {Rau}, {Refregier}, {Rollins},
  {Rozo}, {Rykoff}, {S{\'a}nchez}, {Sevilla-Noarbe}, {Sheldon}, {Stebbins},
  {Varga}, {Vielzeuf}, {Wang}, {Wechsler}, {Yanny}, {Abbott}, {Abdalla},
  {Allam}, {Annis}, {Bechtol}, {Benoit-L{\'e}vy}, {Bertin}, {Brooks},
  {Buckley-Geer}, {Burke}, {Carnero Rosell}, {Carrasco Kind}, {Carretero},
  {Castander}, {Crocce}, {Cunha}, {D'Andrea}, {da Costa}, {DePoy}, {Desai},
  {Diehl}, {Dietrich}, {Doel}, {Fernandez}, {Flaugher}, {Fosalba},
  {Garc{\'\i}a-Bellido}, {Gaztanaga}, {Gerdes}, {Giannantonio}, {Goldstein},
  {Gruendl}, {Gschwend}, {Gutierrez}, {James}, {Jeltema}, {Johnson}, {Johnson},
  {Kent}, {Kuehn}, {Kuhlmann}, {Kuropatkin}, {Li}, {Lima}, {Lin}, {Maia},
  {March}, {Marshall}, {Martini}, {Melchior}, {Menanteau}, {Miquel}, {Mohr},
  {Neilsen}, {Nichol}, {Nord}, {Petravick}, {Plazas}, {Romer}, {Roodman},
  {Sako}, {Sanchez}, {Scarpine}, {Schindler}, {Schubnell}, {Smith}, {Smith},
  {Soares-Santos}, {Sobreira}, {Suchyta}, {Swanson}, {Tarle}, {Thomas},
  {Tucker}, {Vikram}, {Walker}, {Weller}, {Zhang}, \& {DES
  Collaboration}}]{troxel18}
{Troxel}, M.~A., {MacCrann}, N., {Zuntz}, J., {et~al.} 2018, \prd, 98, 043528

\bibitem[{{van den Busch} {et~al.}(2022){van den Busch}, {Wright},
  {Hildebrandt}, {Bilicki}, {Asgari}, {Joudaki}, {Blake}, {Heymans},
  {Kannawadi}, {Shan}, \& {Tr{\"o}ster}}]{vandenbusch22}
{van den Busch}, J.~L., {Wright}, A.~H., {Hildebrandt}, H., {et~al.} 2022,
  \aap, 664, A170

\bibitem[{{von der Linden} {et~al.}(2014){von der Linden}, {Mantz}, {Allen},
  {Applegate}, {Kelly}, {Morris}, {Wright}, {Allen}, {Burchat}, {Burke},
  {Donovan}, \& {Ebeling}}]{vonDerLinden14}
{von der Linden}, A., {Mantz}, A., {Allen}, S.~W., {et~al.} 2014, \mnras, 443,
  1973

\bibitem[{{Wen} \& {Han}(2015)}]{WH15}
{Wen}, Z.~L. \& {Han}, J.~L. 2015, \apj, 807, 178

\bibitem[{{Wen} \& {Han}(2021)}]{WH21}
{Wen}, Z.~L. \& {Han}, J.~L. 2021, \mnras, 500, 1003

\bibitem[{{Wen} \& {Han}(2022)}]{WH22}
{Wen}, Z.~L. \& {Han}, J.~L. 2022, \mnras, 513, 3946

\bibitem[{{Wen} \& {Han}(2024)}]{WH24}
{Wen}, Z.~L. \& {Han}, J.~L. 2024, \apjs, 272, 39

\bibitem[{{Wright} {et~al.}(2019){Wright}, {Hildebrandt}, {Kuijken}, {Erben},
  {Blake}, {Buddelmeijer}, {Choi}, {Cross}, {de Jong}, {Edge},
  {Gonzalez-Fernandez}, {Gonz{\'a}lez Solares}, {Grado}, {Heymans}, {Irwin},
  {Kupcu Yoldas}, {Lewis}, {Mann}, {Napolitano}, {Radovich}, {Schneider},
  {Sif{\'o}n}, {Sutherland}, {Sutorius}, \& {Verdoes Kleijn}}]{wright19}
{Wright}, A.~H., {Hildebrandt}, H., {Kuijken}, K., {et~al.} 2019, \aap, 632,
  A34

\bibitem[{{Wright} {et~al.}(2024){Wright}, {Kuijken}, {Hildebrandt},
  {Radovich}, {Bilicki}, {Dvornik}, {Getman}, {Heymans}, {Hoekstra}, {Li},
  {Miller}, {Napolitano}, {Xia}, {Asgari}, {Brescia}, {Buddelmeijer}, {Burger},
  {Castignani}, {Cavuoti}, {de Jong}, {Edge}, {Giblin}, {Giocoli},
  {Harnois-D{\'e}raps}, {Jalan}, {Joachimi}, {John William}, {Joudaki},
  {Kannawadi}, {Kaur}, {La Barbera}, {Linke}, {Mahony}, {Maturi}, {Moscardini},
  {Nakoneczny}, {Paolillo}, {Porth}, {Puddu}, {Reischke}, {Schneider},
  {Sereno}, {Shan}, {Sif{\'o}n}, {St{\"o}lzner}, {Tr{\"o}ster}, {Valentijn},
  {van den Busch}, {Verdoes Kleijn}, {Wittje}, {Yan}, {Yao}, {Yoon}, \&
  {Zhang}}]{wright24}
{Wright}, A.~H., {Kuijken}, K., {Hildebrandt}, H., {et~al.} 2024, \aap, 686,
  A170

\bibitem[{{Zenteno} {et~al.}(2016){Zenteno}, {Mohr}, {Desai}, {Stalder},
  {Saro}, {Dietrich}, {Bayliss}, {Bocquet}, {Chiu}, {Gonzalez}, {Gangkofner},
  {Gupta}, {Hlavacek-Larrondo}, {McDonald}, {Reichardt}, \& {Rest}}]{Zenteno16}
{Zenteno}, A., {Mohr}, J.~J., {Desai}, S., {et~al.} 2016, \mnras, 462, 830

\end{thebibliography}






\appendix

\section{Examples of newly discovered rich clusters}\label{sec:gallery}

\begin{figure*}
\centering
\includegraphics[width=0.48\textwidth]{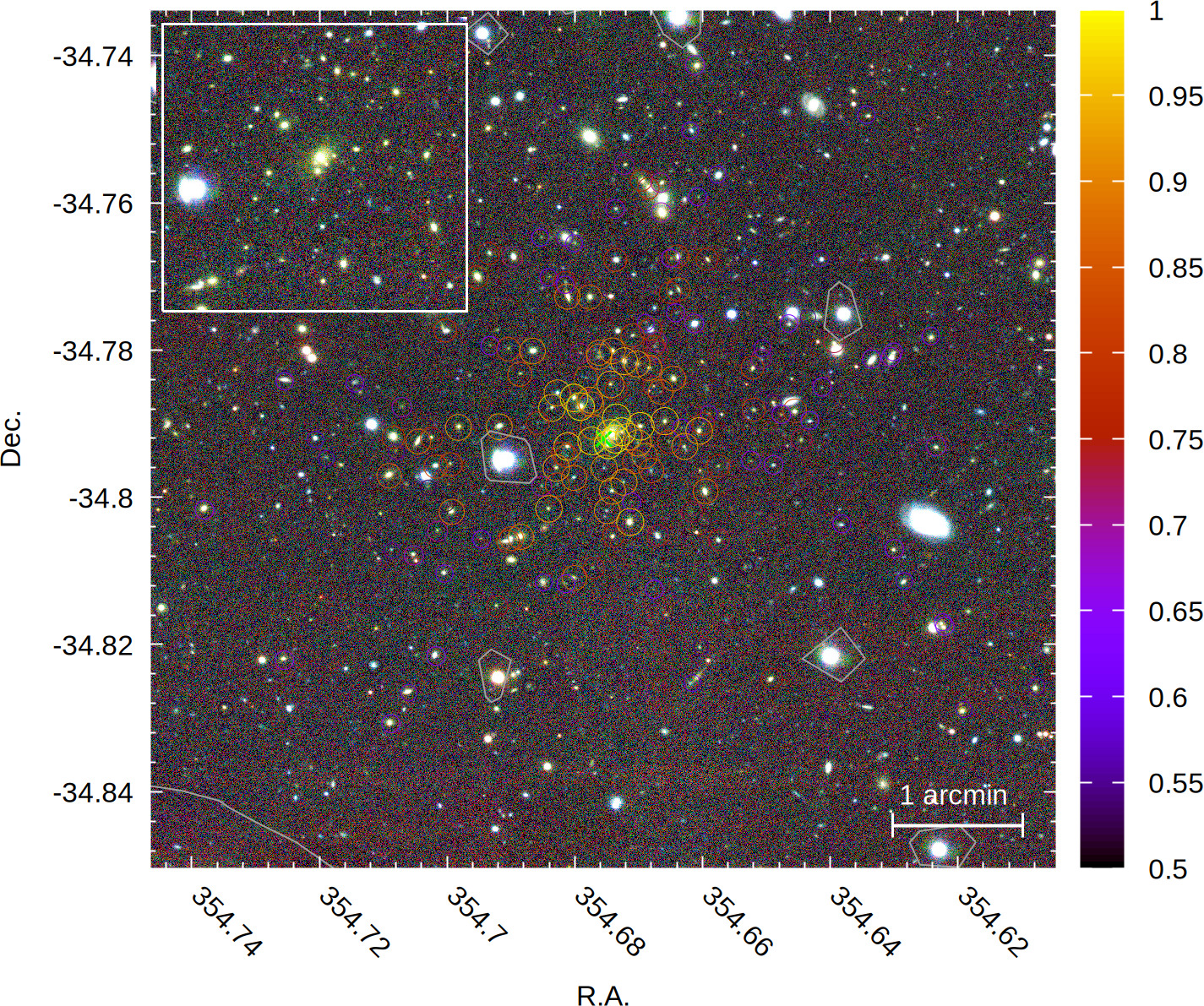}
~
\includegraphics[width=0.48\textwidth]{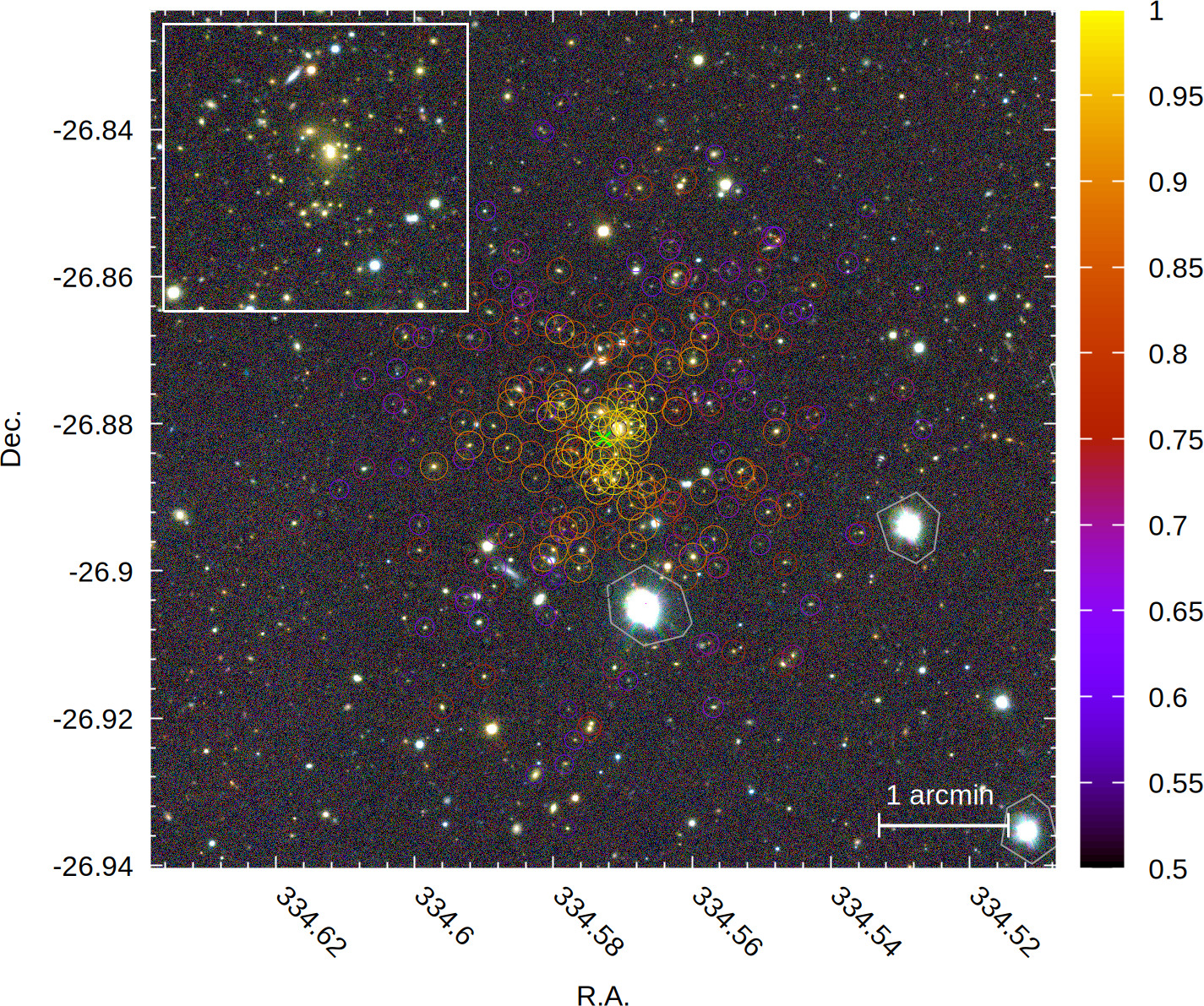}
~
\includegraphics[width=0.48\textwidth]{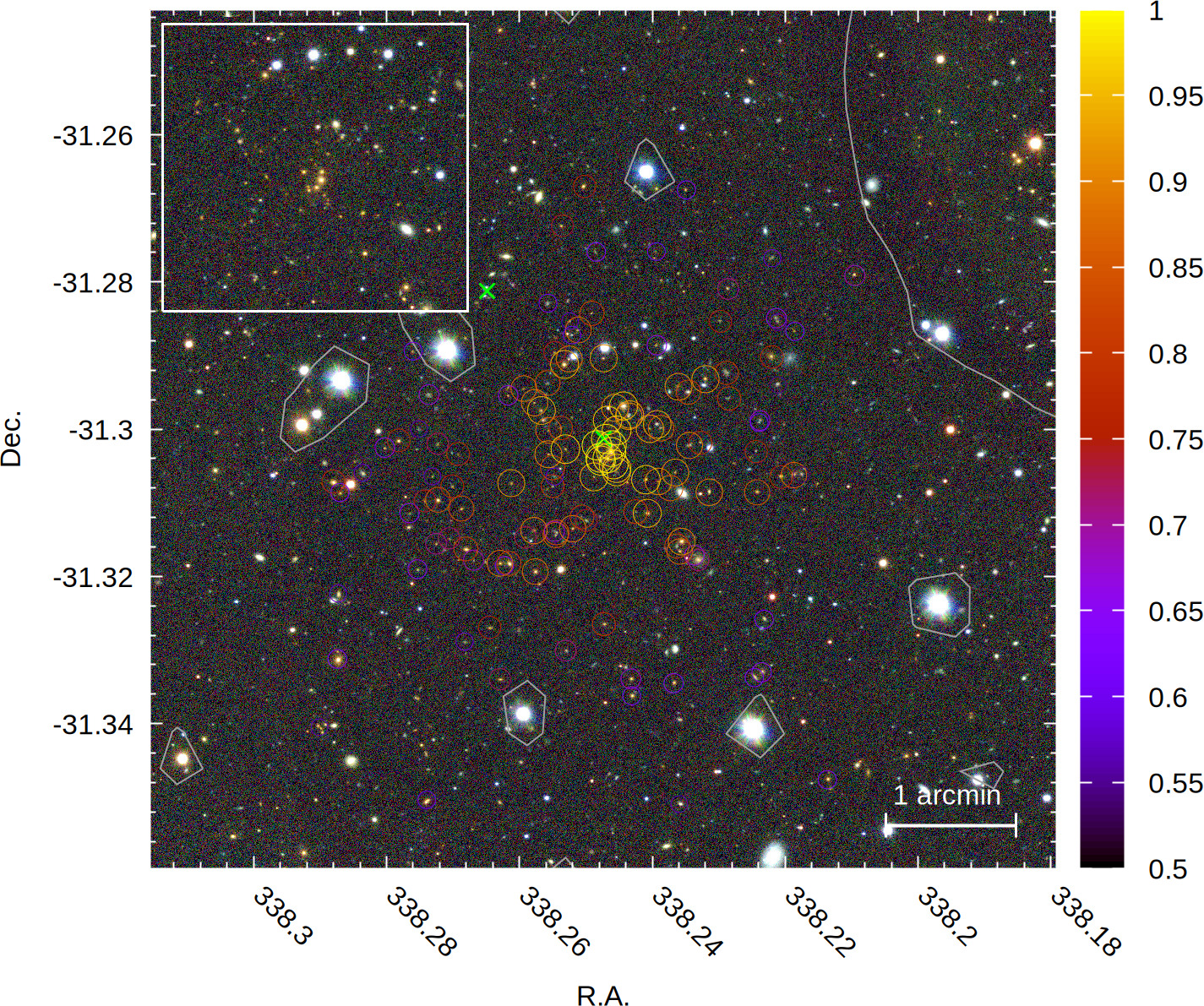}
~
\includegraphics[width=0.48\textwidth]{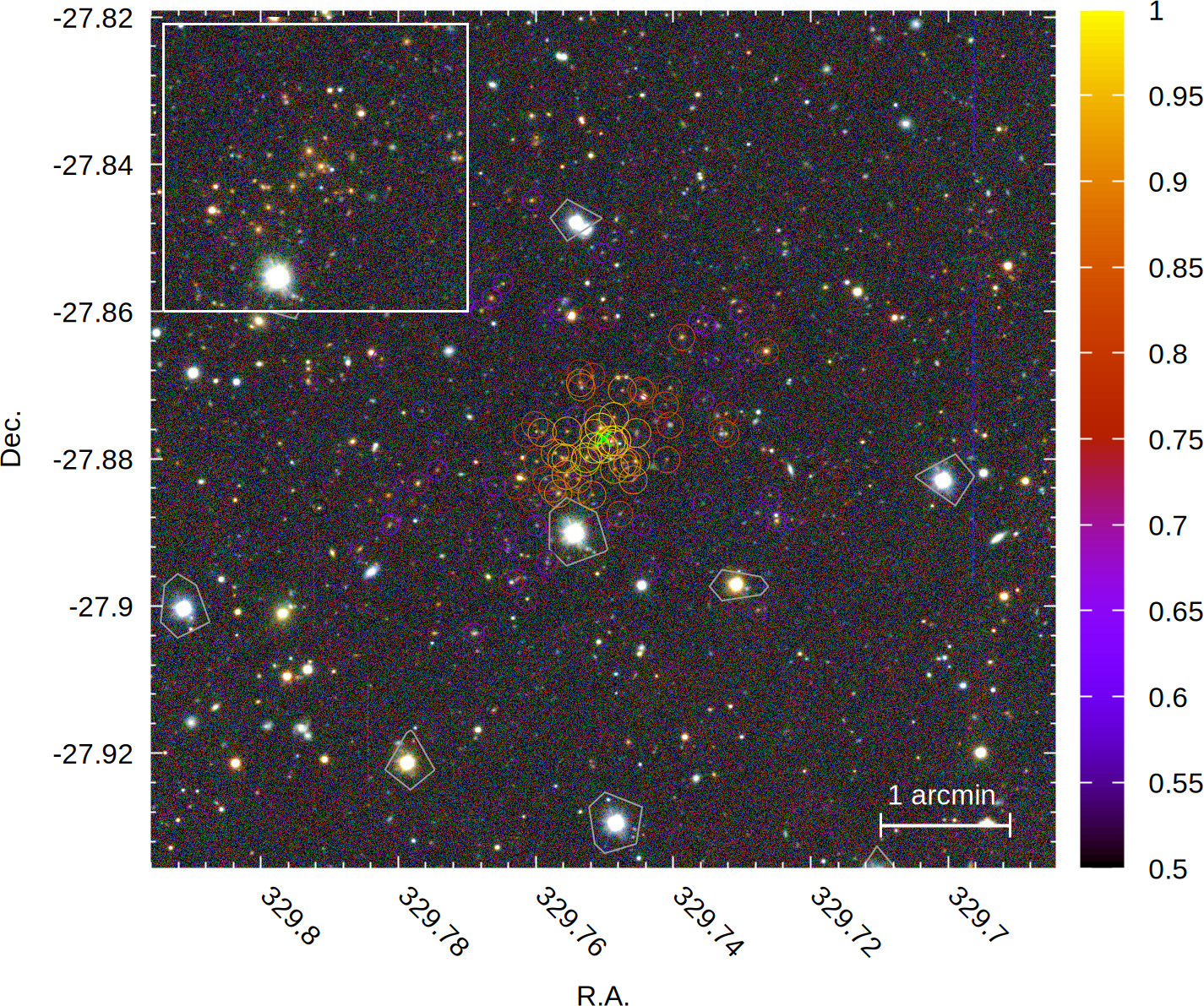}
~
\includegraphics[width=0.48\textwidth]{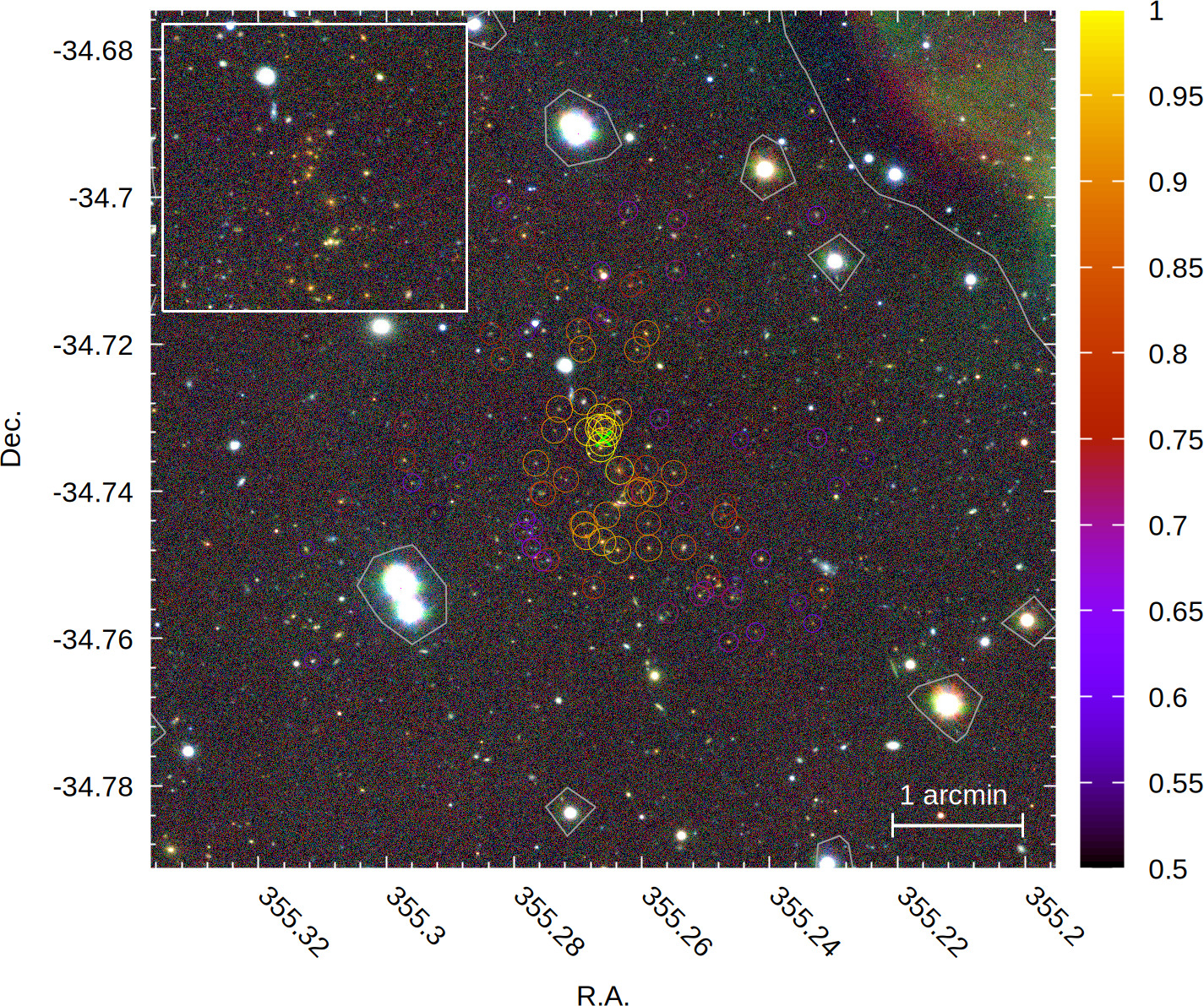}
~
\includegraphics[width=0.48\textwidth]{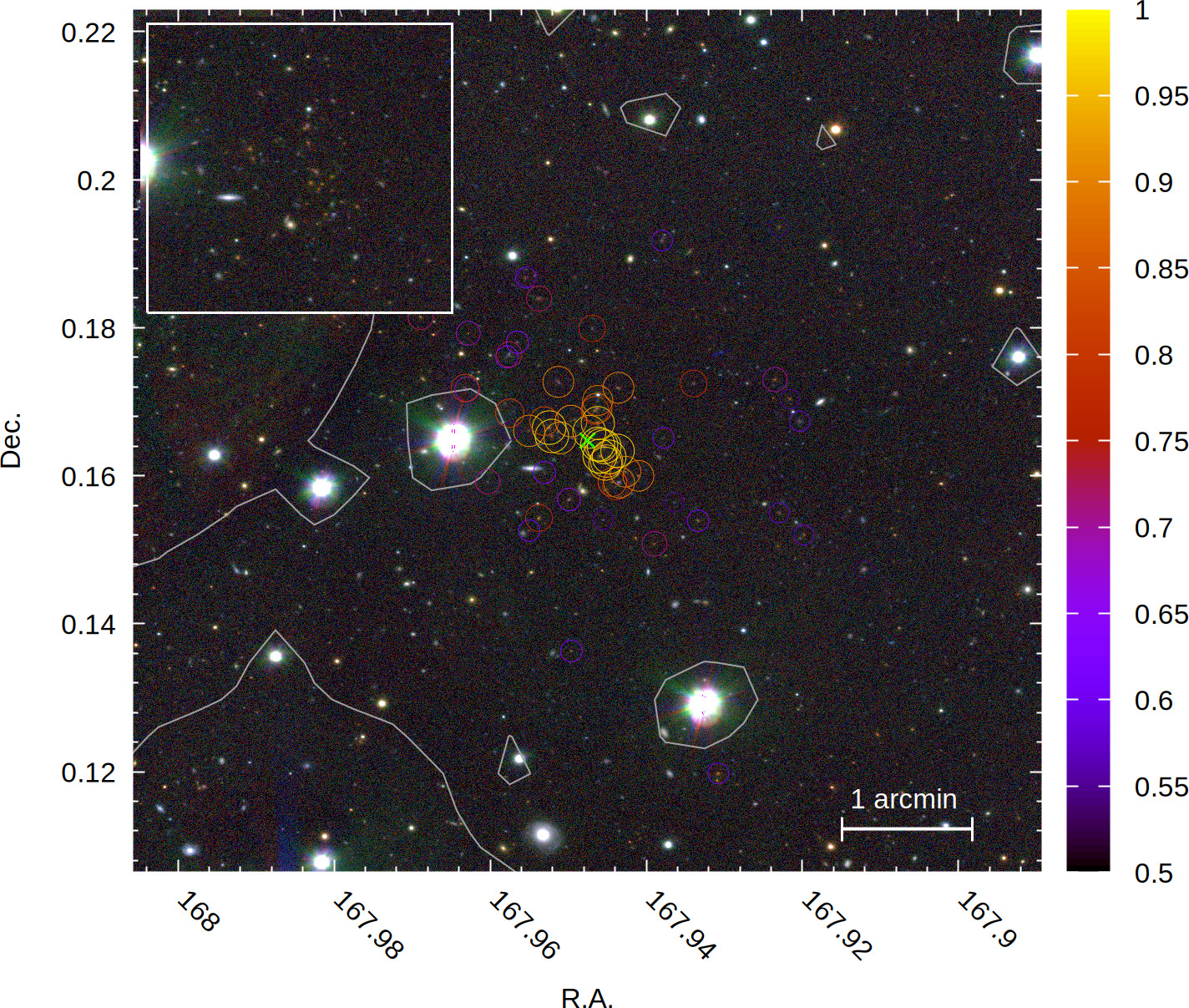}
  \caption{Gallery showcasing a selection of newly discovered clusters with $\lambda_* > 60$. The images are arranged from left to right and top to bottom, following the order of the entries in Table~\ref{tab:gallery}. The cluster center is marked with a green ``x'', while cluster members are highlighted with circles, with colors indicating their probabilistic membership. The top-left sub-panels provide a zoomed-in view of the central region without overlays. Each postage stamp image covers an area of $7$ arcmin per side.}
  \label{fig:gallery}
\end{figure*}

To provide a visual impression of the data quality, Fig.~\ref{fig:gallery} presents $gri$-color composite images of newly discovered clusters with intrinsic richness $\lambda_*>60$, spanning a broad redshift range.  The main properties of these clusters are summarized in Table~\ref{tab:gallery}. To assess whether these clusters were not previously detected in other surveys, we searched for already known clusters within 10\arcmin\ from the AMICO center in the {\em SIMBAD} database\footnote{\url{https://simbad.u-strasbg.fr/simbad/}}. We further verified that the clusters were not contained in the cluster catalogs
\footnote{\url{http://zmtt.bao.ac.cn/galaxy_clusters/catalogs.html}} based on the DESI \citep{WH24}, Dark Energy and unWISE \citep{WH22}, Subaru Hyper-SuprimeCam and unWISE \citep{WH21}, SDSS \citep{WH15} surveys, eRASS1 and ACT. All of them are outside the areas covered by these catalogs, with the exception of 6835: this cluster is within the SDSS footprint,  but due to the high redshift ($z \sim 0.8$) it was not detected by \citet{WH15}.

\begin{table*}
  \centering
  \caption {Main properties of the clusters shown in Fig.~\ref{fig:gallery}.}
  \begin{tabular}{lllllllll}
  \hline
  UID & TILE & RA & DEC & $Z_{fix}$ & SN & $A$ & $\lambda_*$ & $\lambda$ \\
  33493 & \texttt{\detokenize{KiDS_DR4.0_355.1_-35.1}} & 354.67554 & -34.792076 & 0.363141 & 8.1271238 & 3.469 & 70.396614 & 475.92517\\
  51198 & \texttt{\detokenize{KiDS_DR4.1_334.2_-27.2}} & 334.57281 & -26.882067 & 0.443988 & 9.34056 & 4.73776 & 89.256004 & 665.80652\\
  27965 & \texttt{\detokenize{KiDS_DR4.0_337.9_-31.2}} & 338.24738 & -31.301323 & 0.616797 & 7.0126705 & 4.26233 & 71.79953 & 340.97748\\
  50760 & \texttt{\detokenize{KiDS_DR4.1_329.5_-28.2}} & 329.75018 & -27.877317 & 0.64674 & 6.6932201 & 4.23948 & 82.536217 & 367.30576\\
  33483 & \texttt{\detokenize{KiDS_DR4.0_355.1_-35.1}} & 355.26602 & -34.73288 & 0.716602 & 6.3742614 & 4.68677 & 69.213089 & 270.9325\\
  6835 & \texttt{\detokenize{KiDS_DR4.0_168.0_0.5}} & 167.94754 & 0.1647398 & 0.796444 & 4.6347418 & 3.36811 & 67.456497 & 153.05356\\
  \hline
  \end{tabular}
  \label{tab:gallery}
\end{table*}

\section{Blinding for cosmological applications}\label{sec:blinding}

Blinding is a fundamental practice in cosmological analyses, and in physics more broadly, to mitigate potential observer biases arising from unavoidable subjective choices, such as thresholds for data selection, data binning, etc..\citep{conley06,heymans12,vonDerLinden14,maccoun15,kuijken15,blake16,suyu17,hildebrandt17,troxel18,abbott18,ota20,muir20}. In this work, we implement a blinding strategy based on controlled perturbations of the selection function derived from the matching of true and detected clusters in the \textsc{SinFoniA} mock catalogs (see Sect.~\ref{sec:pur-sel}). The selection function plays a pivotal role in modeling cluster counts, which represent one of the most powerful probes to constrain the main cosmological parameters using galaxy clusters.

Our blinding scheme for cosmological analysis, based on cluster counts, involves these key steps:
\begin{itemize}
\item {\it Halo Ratio Computation}: Using the \cite{jenkins01} mass function, we compute the ratio of the number of halos as a function of redshift and mass between a reference flat cosmology ($h=0.6736$, $\Omega_m=0.3$, $S_8=0.78$, $\sigma_8=0.86$) and another flat cosmology with $\Omega_m=0.3$ and $S_8$ randomly drawn from a uniform distribution in the range $0.65 < S_8 < 0.91$.
\item {\it Mapping to mass proxy}: The mass dependence of this "perturbation ratio" is mapped onto the amplitude mass proxy, $A$, using the scaling relation from \cite{bellagamba18b}, recalibrated for KiDS-DR4 data.
\item {\it Perturbation of the matching}: The original matching of detected clusters with $S/N<6.5$ is perturbed by applying the "perturbation ratio" in each redshift and amplitude bin, artificially modifying the number of matches. Entries with $S/N>6.5$ are left unaffected to preserve the statistical integrity of high signal-to-noise detections, ensuring that an excessive presence or absence of such clusters does not inadvertently reveal the blinding status.
\item {\it Selection function evaluation}: We calculate the selection function for the original, unperturbed matching catalog (the "true" selection function) and for two perturbed, blinded matching catalogs (the "blinds"). To mitigate shot noise in the matching counts, particularly for extreme values of richness, we interpolate the selection functions using Chebyshev polynomials of order one and window size of three.
\end{itemize}

\begin{figure*}
    \includegraphics[width=0.33\textwidth]{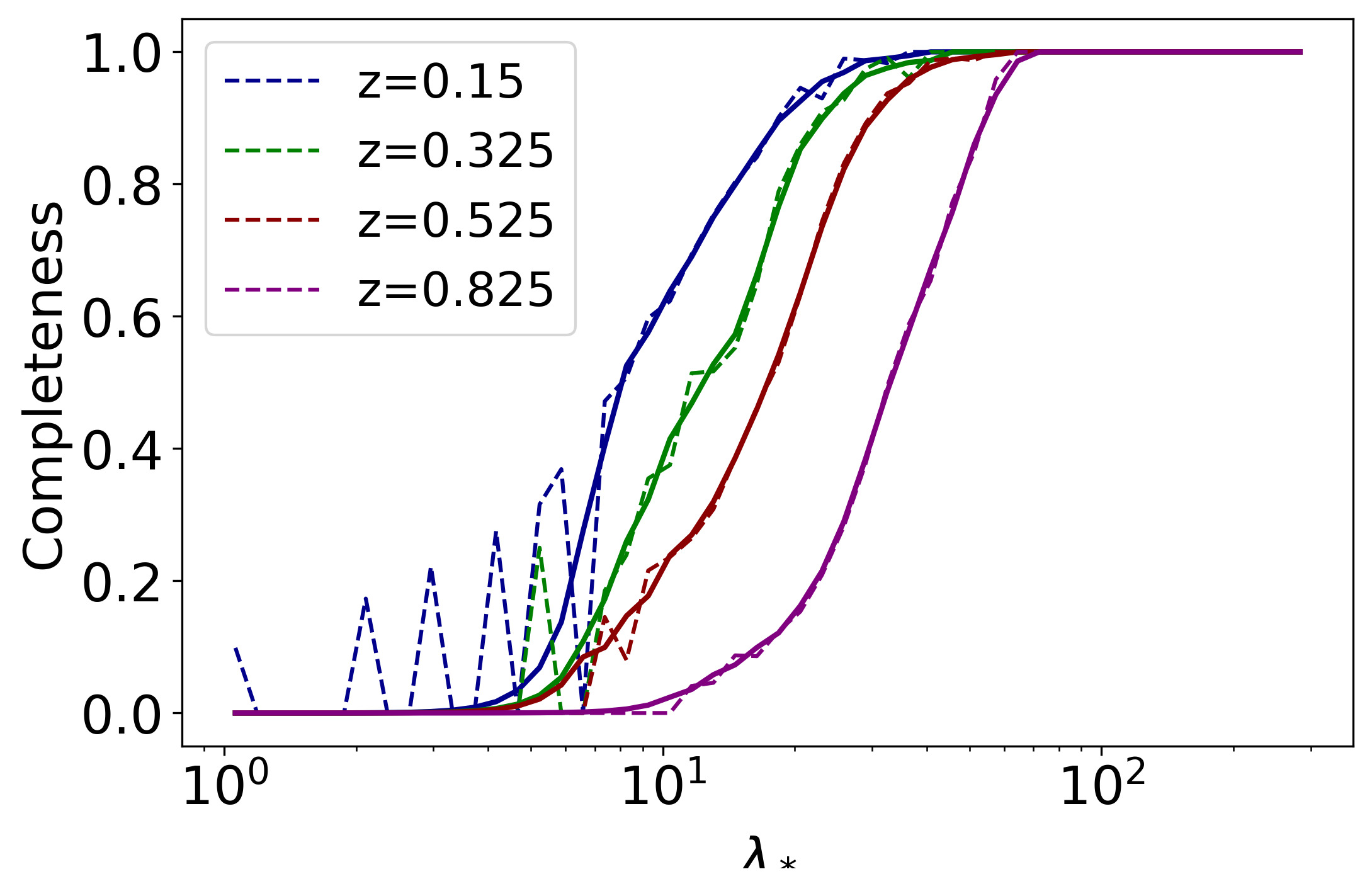}
    \includegraphics[width=0.33\textwidth]{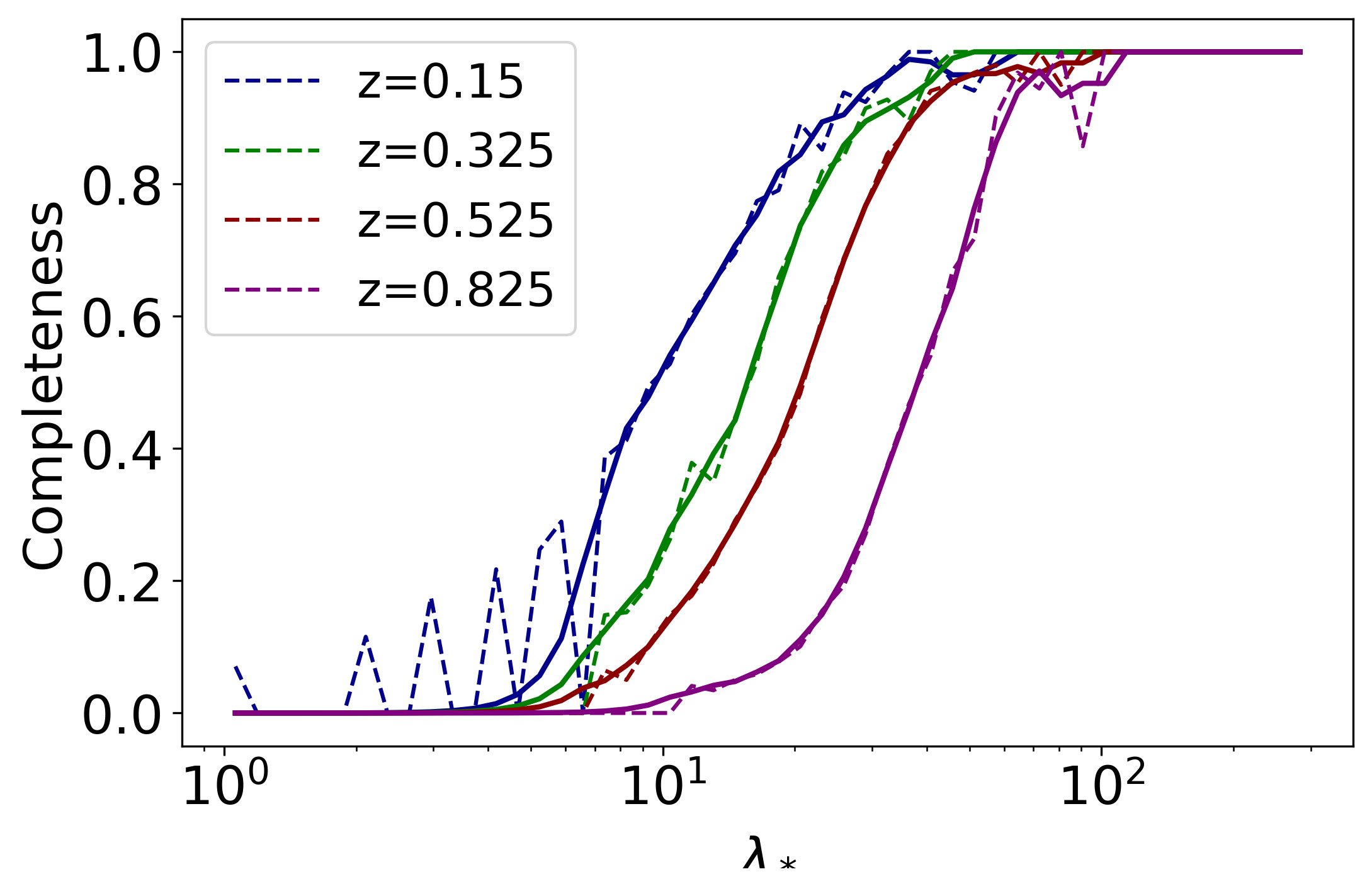}
    \includegraphics[width=0.33\textwidth]{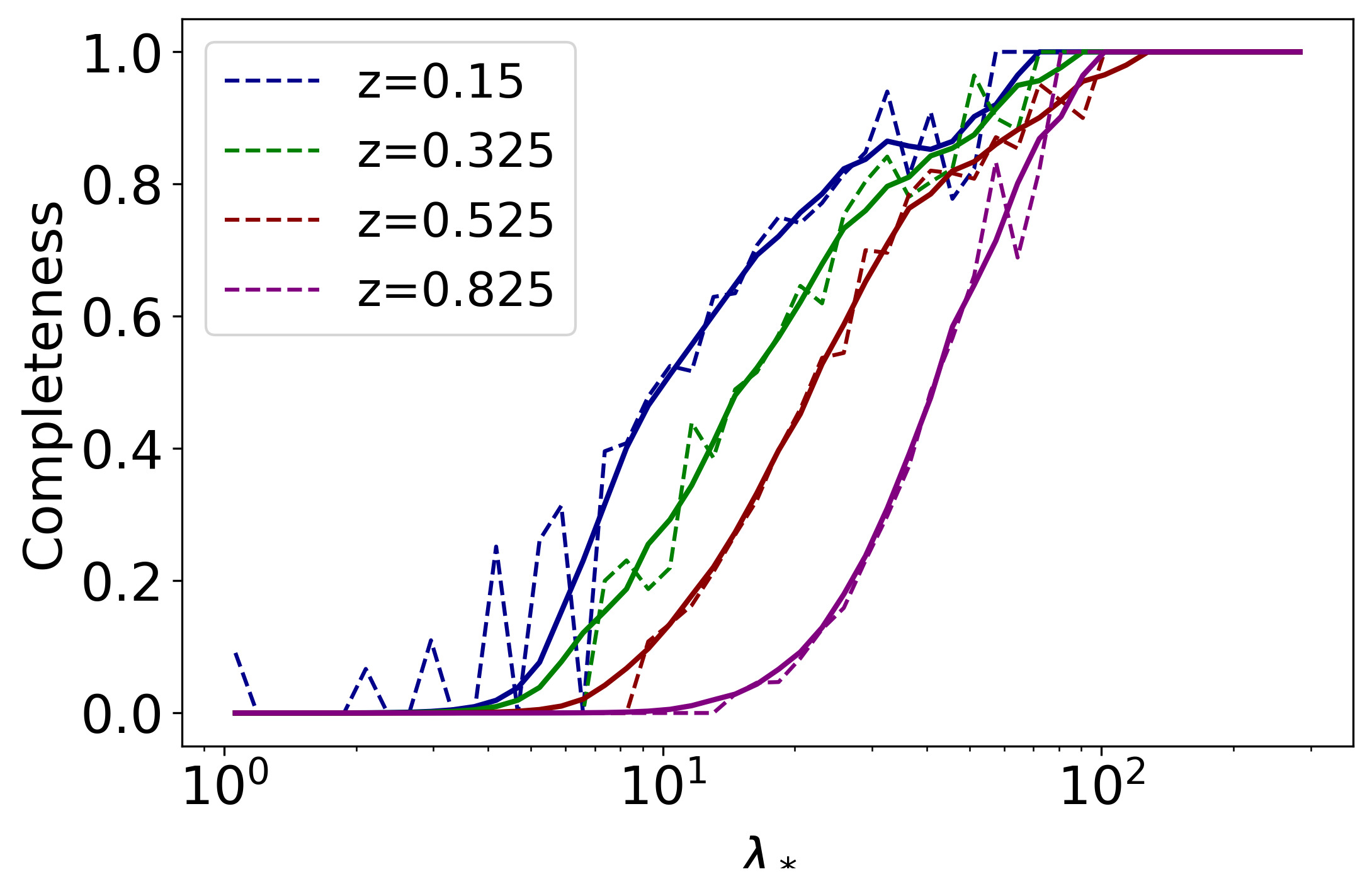}  
    \caption{The selection function of the \textsc{AMICO} catalog as a function of the intrinsic richness $\lambda_*$ computed in 4 different redshift intervals, centered in the values reported in the legend. The solid lines represent the smoothed versions derived from Chebyshev polynomials, while the dashed lines refer to the unsmoothed versions. The three panels include the original \new{(central panel)} and two blinded versions of the selection function.}
    \label{fig:blind_completeness}
\end{figure*}

In Fig.~\ref{fig:blind_completeness}, we compare the true, unperturbed selection function  with two blinded versions. The solid lines represent the smoothed selection functions, obtained using Chebyshev polynomials and used in the cosmological analysis, while the dashed lines show the corresponding unsmoothed versions. We considered seven redshift intervals spanning the range $0.1<z<0.9$, but display only four of them not to overcrowd the plots.  We considered seven redshift intervals running from $0.1$ to $0.9$, but we display here only 4 of them not to overcrowd the plots. The selection function is shown as a function of $\lambda_*$, the mass proxy used in our cosmological analysis.

\new{ The true selection function was kept concealed until the completion of the analysis and is now publicly available (central panel of Fig.~\ref{fig:blind_completeness}).}

\section{Data Availability}

The photometric galaxies catalog and the masks of the KiDS-DR4 survey are publicly available as explained in the KiDS web portal (\url{https://kids.strw.leidenuniv.nl/DR4}).

The \textsc{AMICO}-KIDS-DR4 cluster sample, probabilistic memberships, sample purity, and selection function will be made publicly available on the KiDS data products web portal but can already be obtained on request, after evaluation. Please contact the lead author

\end{document}